\documentclass[1p]{myelsarticle}
\usepackage{latexsym}
\usepackage{amssymb}
\usepackage{amsfonts}
\usepackage{amsmath}
\usepackage{cuted}
\usepackage{multirow}
\usepackage{hyperref}
\usepackage{epsf} 
\usepackage{subfigure}
\usepackage{colortbl}
\usepackage{array}
\usepackage{slashed}
\usepackage{wasysym}
\usepackage{dcolumn}
\usepackage{units}
\usepackage{physics}
\usepackage[euler]{textgreek}
\usepackage[normalem]{ulem}
\definecolor{jocol}{rgb}{0.0,0.1,0.6}

\newcommand{\sqrts}{$\sqrt{s}$}
\newcommand{\sqrtsnn}{$\sqrt{s_{\mathrm{NN}}}$}

\newcommand{\sqsna}[1]{$\sqrt{s}\, = #1 \, A \, \mathrm{G}e\mathrm{V}$}
\newcommand{\sqsnat}[1]{$\sqrt{s}\, = #1 \, A \, \mathrm{T}e\mathrm{V}$}
\newcommand{\sqsnn}[1]{$\sqrt{s_{\mathrm{NN}}}\, = #1 \, \mathrm{G}e\mathrm{V}$}
\newcommand{\sqsnnt}[1]{$\sqrt{s_{\mathrm{NN}}}\, = #1 \, \mathrm{T}e\mathrm{V}$}

\newcommand{\mub}{$\mu_B$}

\newcommand{\piztodal}{\piz$\to$\textgamma\epem}
\newcommand{\etatodal}{\texteta$\to$\textgamma\epem}

\newcommand{\omegatodal}{\textomega$\to$\piz\epem}
\newcommand{\phitodal}{\straightphi$\to$\newcommand{\omegatodalg}{\textomega$\to$\piz\textgamma}
\piz\epem}
\newcommand{\rhoa}{\textrho --a$_1$}

\newcommand{\epem}{$\mathrm{e}^+\mathrm{e}^-$}

\newcommand{\piz}{\textpi$^0$}
\newcommand{\pip}{\textpi$^+$}

\newcommand{\pim}{\textpi$^-$}
\newcommand{\mumu}{\textmu$^+$\textmu$^-$}
\newcommand{\DD}{D$\bar{\text{D}}$}
\newcommand{\pA}{p--A}
\newcommand{\cc}{C--C}

\newcommand{\arkcl}{Ar--KCl}
\newcommand{\tata}{Ta--Ta}
\newcommand{\auau}{Au--Au}
\newcommand{\pbpb}{Pb--Pb}
\newcommand{\pbau}{Pb--Au}
\newcommand{\inin}{In--In}
\newcommand{\nnnn}{N--N}
\newcommand{\pnb}{p--Nb}
\newcommand{\np}{n--p}
\newcommand{\pn}{p--n}
\newcommand{\nn}{n--n}
\newcommand{\pp}{p--p}
\newcommand{\xdp}{p--d}

\newcommand{\mrN}{\mathrm{N}}
\newcommand{\mrn}{\mathrm{n}}
\newcommand{\mrp}{\mathrm{p}}
\newcommand{\mree}{\mathrm{e}^+\mathrm{e}^-}
\newcommand{\mreen}{\mathrm{e}\,\mathrm{e}}
\newcommand{\mrpp}{\mathrm{p}\,\mathrm{p}}

\newcommand{\mrC}{\mathrm{C}}

\newcommand{\mrAu}{\mathrm{Au}}

\newcommand{\gevcc}{\,G$e$V/$c^{2}$}
\newcommand{\mevcc}{\,M$e$V/$c^{2}$}
\newcommand{\gevc}{\,G$e$V/$c$}
\newcommand{\mevc}{\,M$e$V/$c$}
\newcommand{\gev}{\,G$e$V}
\newcommand{\mev}{\,M$e$V}
\newcommand{\tev}{\,T$e$V}
\newcommand{\atev}{\,$A$ T$e$V}
\newcommand{\agev}{$\,A$ G$e$V}
\newcommand{\amev}{$\,A$ M$e$V}
\newcommand{\myie}{\textit{i.e.}~}
\newcommand{\mycf}{\textit{cf.}~}
\newcommand{\myeg}{\textit{e.g.}~}
\newcommand{\mywrt}{\textit{w.r.t.}~}
\begin{document}
\begin{frontmatter}
%
\title{Dilepton Radiation from Strongly Interacting Systems}
\author[ju]{P. Salabura} 
\author[gu,gsi,hfhf]{J. Stroth} 
\address[ju]{M.\ Smoluchowski Instute of Physics, Jagiellonian University, Cracow, Poland}
\address[gu]{Goethe University, Frankfurt am Main, Germany}
\address[gsi]{GSI Helmholtzzentrum f{\"u}r Schwerionenforschung GmbH, Darmstadt, Germany}
\address[hfhf]{Helmholtz Research Academy for FAIR, Campus Frankfurt, Goethe University Frankfurt am Main, Germany}

\begin{abstract}
We review the current understanding of time-like virtual photon emission from QCD matter. 
The phenomenology of dilepton emission is discussed and basic theoretical concepts are introduced. 
The experimental findings are presented, grouped into production of lepton pairs in elementary processes, production off cold nuclear matter and emission from heavy-ion collisions. 
The review emphasizes the role of dilepton emission as tool for studying exotic phases of QCD matter. 
Open questions and a route to probe the QCD phase diagram with dileptons are outlined.  
\end{abstract}
\end{frontmatter}
%
\section{Introduction}
 States of QCD matter at extreme temperatures and densities can be created in the laboratory by colliding heavy ions at (ultra-)relativistic energies. 
The strong interaction provides sufficient stopping to convert all or a large fraction of the kinetic beam energy into excitation and deposit it around the center-of-mass of the collision system. 
Once the energy density in the collision system exceeds a critical value $\epsilon_c \approx 1\,\mathrm{G}e\mathrm{V}/\mathrm{fm}^3$ deconfined matter occurs which can be characterized as a liquid of strongly coupled quarks and gluons.
Due to the high pressure such ``nuclear fireballs'' expand rapidly and the system crosses over from partonic degrees of freedom to a system of hadrons.
This evolution resembles features of the universe about 10 $\mu$s after its birth.
While the universe was cooling down to temperatures of $kT \sim 160$~\mev\ little droplets of the matter were formed building color-neutral states (hadrons). 
Hence, the interior of the nucleons contains the ``hot soup'' of the early universe as the energy density in the interior of a nucleon is of the order of $\epsilon_c$.

Neutron stars (NS) can be considered giant nuclei\footnote{Despite the fact that neutron star matter is electrically neutral and that the neutron density exceeds the proton density substantially.} with radii of the order of $10^4$~m and core densities factors higher than nuclear ground state density. 
The high energy density in the NS core is represented by mass and governed by the immense gravitational forces pulling the matter inward.
It is not known if under such high pressures the hadrons (neutrons) still survive as quasi-particles or if they are squeezed into each other forming a cold partonic phase. 
Similar states of QCD matter are expected to be realized in heavy-ion collisions at a few G$e$V per nucleon pair.
The reaction is that of a bunch of nucleons interacting with a bunch of nucleons which, during the course of a central collision, are almost stopped. 
In contrast, at ultra-relativistic energies, the interaction is rather dominated by the violent interaction of two giant clouds of very low-x gluons, while the``stripped'' valence quarks mostly escape the interaction zone along the collision axis. 
That explains why the net-baryon density, \myie\ $\rho_B-\rho_{\bar{B}}$, is high in case of low collision energies and essentially close to zero at mid-rapidity for collision energies in excess of~\sqsnn{100}. 
Common to both situations is that the matter formed in the collisions are transient states with lifetimes of a few fm/$c$ only. 

The experimental challenge is to measure with sufficient precision probes which are sensitive to the properties of the high-density and high-temperature phase. 
Virtual photons are such a case. 
They occur as intermediary objects which couple lepton pairs (\epem, \mumu) to the electromagnetic current of strongly interacting systems. 
An important advantage of dileptons over hadronic probes is their immediate decoupling from the strongly interacting system allowing them to escape nearly undisturbed during all stages of the collision.
Moreover, they carry rich information since the four-momentum of the intermediary virtual photon and its spin orientation can be reconstructed. 
This is an important difference to real photons. 
But at the same time dileptons also pose challenges: 
First, the signal is rare and has to be reconstructed with sufficient purity. 
Second, experiments integrate the radiation emitted throughout the evolution of the four-volume of the transient states. 
To disentangle the cocktail of all sources contributing to the signal finally involves also the inspection of the hadronic final state of the collision. 
Last but not least, pair spectroscopy gives rise to combinatorial background which has to be determined with high precision and accuracy. 

Dilepton spectroscopy in relativistic heavy-ion collisions has been pioneered at SPS (CERN) in the mid eighties by NA34-HELIOS and NA38 and, soon later, also investigated at much lower beam energies provided by the BEVALAC (LBNL) using the two-arm spectrometer DLS.
With the availability of lead beams at the CERN SPS, dedicated second generation large acceptance experiments started operation. 
CERES was designed to measure electrons with two cylindrical RICH detectors and NA50, the successor experiment of NA38, used a 4~m hadron absorber in front of a cylindrical spectrometer for muon detection. 
The dilepton program at SPS has been temporarily concluded with a high statistics run with the NA60 dimuon spectrometer, an upgraded version of NA50. 
Meanwhile, dilepton spectroscopy is pursued at almost all operational heavy-ion facilities, from relativistic energies with beams of $\sim 1\,$\gev\ per nucleon on stationary targets with HADES at the SIS18 of GSI, to ultra-relativistic energies as provided by the colliders RHIC using the detectors PHENIX and STAR and at LHC by CMS, ATLAS and ALICE. 
At LHC all the three experiments with central detection are in principle able to investigate continuum dilepton radiation, although continuum radiation in the invariant mass range below one \gevcc\ and with small transverse momentum is mostly in the focus of the ALICE experiment.

In this paper, we review the current status of dilepton spectroscopy in relativistic heavy-ion collisions. 
We in particular address continuum radiation as an observable for QCD matter under conditions of high temperature and energy density as formed in (ultra-)relativistic collisions of heavy ions. 
An important aspect for the understanding of such continuum radiation is the coupling of virtual photons to hadrons via intermediary vector mesons which is also discussed in some detail.
To cover the current experimental situation we report recent measurements concluded at GSI/SIS18, BNL/RHIC and CERN/SPS/LHC. 
Dileptons are also the proper tool to study medium modifications of hadrons in cold (\myie nuclear ground state) matter. 
The extraction of medium-modifications requires a detailed knowledge of the coupling of virtual photons to hadrons in vacuum. 
This aspect is related to electromagnetic transition form factors of hadrons in the time-like region, which governs the electromagnetic decay of excited hadronic states.
We will also touch this topic and will discuss the relevance/connection to the heavy-ion program.  
In the following section we will start with a short survey about the phenomenology of relativistic heavy-ion collisions. 
We will introduce the so-called ``standard model'' of heavy-ion collisions and characterize various stages of the reaction. 

In Sec.\,\ref{sec:pheno} we will address experimental aspects of dilepton spectroscopy and discuss the difference between dimuon and dielectron detection. 
An introduction to the theoretical foundations of dilepton spectroscopy is presented in Sec.\,\ref{sec:theory}. 
Sec.\,\ref{sec:exp}, finally, is devoted to experimental results and comparisons to model calculations. 
We finish with a status and outlook. 

%
\section{QCD matter under extreme conditions}
\label{sec:stages}
High-energy heavy-ion collisions are characterized by the collision energy defined by the total center of mass energy per nucleon-nucleon pair \sqrtsnn  , the ``size'' of the collision system, determined by the combination of ions brought to collision and the centrality of the collision estimated \myie by the event activity.
The latter is typically evaluated by the multiplicity of charged particles reconstructed for a given event.
The heaviest systems routinely chosen are the symmetric systems \auau\ or \pbpb\ with around 400 nucleons participating in a central collision. 

In an experiment at the BEVALAC utilizing a medium--heavy collision system (\tata) at 400\amev\ it was observed for the first time that the particles expelled from the reaction zone exhibit collective behavior~\cite{Gustafsson:1984ka}.
In semi-central collisions protons and fragments showed a distinct emission pattern, relative to a plane spanned by the beam axis and the impact parameter then called side-splash or bounce-off and squeeze~\cite{Stoecker:1986ci}.
Observed was a preferred in-plane flow in forward (near projectile) rapidity while at mid-rapidity the flow was directed perpendicular to the reaction plane and symmetrically back-to-back. 
Such a pattern is characteristic for a system in which pressure is built up and the constituents move along the gradient of the local density profile.
Flow patterns are nowadays characterized by the Fourier components $v_n$ of the modulating function 
\begin{equation}
F(\phi)=F_0
\left(
1+\sum_{n=1}^{N}2 v_n cos(n\phi) 
\right)
\label{eq:higher-harmonics}
\end{equation}
which can be investigated in multi-differential manner, like for different particle species, and as a function of the centrality, the rapidity and the transverse momentum.
Here $\phi$ measures the azimuth relative to the event plane.
The second component $v_2$ (``squeeze'') is particularly interesting. 
As a function of beam energy the sign of $v_2$ changes two times. 
At low beam energies, in the fusion--fission or multi-fragmentation regime, fragment and nucleon emission are preferred in plane and caused by attraction and resulting angular momentum of the combined nuclear system.
For BEVALAC, SIS18, AGS energies, $v_2$ turns negative (out of plane) because the projectile and target like residues do not separate fast enough from the fireball and block the in-plane emission of particles from the hot and dense zone.
At ultra-relativistic energies, $v_2$ is positive again and a direct measure for the pressure of the initial, almond-shaped QGP droplet~\cite{Sorensen:2009cz}.
The change of $v_2$ as a function of the collision energy reveals collective behavior of the matter formed in heavy-ion collisions at any energy between \sqsnn{2} and \sqsnnt{5}.
 Moreover, in \cite{Nara:2016phs} it has been conjectured that a strong reduction of $v_1$ around midrapidity would signal a ``softest point'' if QCD features a first order phase transition to a deconfined medium in the region of high net-baryon density. 
 
 At the same time $v_4$ would exhibit a maximum~\cite{Nara:2018ijw}.
Yet, the quantification of the pressure is model dependent~\cite{Heinz:2013th} and the quest for the precise knowledge of the initial condition is a key program of the heavy-ion community.
We will come back to the consequences of collective effects for the emission of dileptons in Sec.\,\ref{sec:sps} and \ref{sec:outlook}.

For the discussion of dilepton emission the course of a heavy-ion reaction is conveniently subdivided into three stages: 
An initial stage, which is characterized by the violent first encounter of the constituents of the nuclei, an intermediate stage when the system has build up transverse motion and which by large can be understood as an extreme state of matter, and a late stage characterized as dilute gas of excited hadrons which changes its chemical composition solely by decays. 
In the first stage the system's energy density is determined by the degree of stopping, then the system converts pressure into collective motion while it is evolving as a locally thermalized state, and finally the system decouples and the chemical composition and phase space-distribution of the particles traversing the detectors is determined.
This approach is indeed a qualitative picture and in reality the different stages cannot be strictly separated. 
Moreover, the assumption that the fireball has homogeneous properties at a given stage  is not necessarily justified. 
Rather, from the Glauber picture for colliding nuclei it follows that less stopping occurs in the periphery.
In phenomenological descriptions of such reactions it has been tried to accommodate this fact by separating the fireball into a core and a corona~\cite{Werner:2007bf}).
In the following, we will discuss in more detail the three stages of the collision with emphasis on the relevance for dilepton production and how an increasing collision energy draws on the microscopic structure and dynamics of the matter. 
%
\subsection{The Initial Stage}
\label{sec:initial}
%
Evidently, this stage is most affected by the collision energy. 
The nucleon wavelength in the \nnnn\ rest frame shrinks from a size of the order of the nucleon itself at SIS18 energies to dimensions where the softest partons of the nucleons are resolved at LHC. 
At the same time also the penetration time, \myie the instant it takes until both nuclei overlap completely, changes substantially.
Compared to the typical hadron formation time (or a baryon resonance lifetime $\tau_{\mathrm{R}})$ the penetration takes almost ten times longer at SIS18 energies down to a fraction of only a few per mill at LHC energies.
At near relativistic collisions (SIS18), which we would like to consider first, the initial collisions are of ``nucleon nucleon type'' and proceed elastically and inelastically through resonant pion production via intermediate baryonic resonances (mostly $\Delta(1232)$'s). 
Since the life times of hadronic 
resonances are substantially shorter than the time it takes until the fireball freezes out,
they can be absorbed and recreated in subsequent collision processes. 
Indeed, several generations of the resonances can be created until complete overlap is achieved in central collision of heavy nuclei. 
The built-up of the pressure is going in-line with the excitation and decay of resonances and hence the separation between the initial and intermediate stages is not clearly defined. 
While the first \nnnn\ collisions have already taken place, more and more nucleons stream into the interaction region. 
Hence, the in-streaming nucleons not only impinge on the counter-streaming nucleons of the collision partner but also collide with (excited) baryons which already ``piled up'' at mid-rapidity.
This situation is referred to as the formation of resonance matter featuring secondary collisions including baryonic resonances~\cite{Metag:1992jh}.
With increasing beam energy, higher-lying baryonic resonances are excited and the ``frequency'' of \nnnn\ collisions increases due to the growing Lorentz contraction. 
Once full overlap is reached, a substantial fraction of the beam energy is converted to excitation and pressure.
The latter is composed of thermal as well as static pressure due to the nuclear mean field.
At beam energies of around 10\,\agev~(top AGS), the time for full penetration is already as short as the lifetime of a baryonic resonance.
In the central region of such collisions baryons might collide subsequently so frequently that the resonance states are not even fully established between subsequent scatterings.

Dilepton radiation from this stage can in general be described as multiple \nnnn\ brems-strahlung (\mycf Sec.\,\ref{sec:N-N-brems}). 
In most general terms, the emission process of a bremsstrahlung photon includes amplitudes for quasi-elastic as well as inelastic collisions involving intermediary baryonic resonances.
One way to theoretically describe this stage is to study the time evolution of the nucleon collisions by means of microscopic transport models (\mycf Sec.\,\ref{sec:dense}). 
However, the baryon densities can be high and the baryon meson-clouds start to overlap substantially.
This situation is expected to give rise to strong modifications of the baryonic resonance properties or even might constitute limiting conditions for a resonance gas.  
A consistent numerical treatment of in-medium effects governing the colliding baryons and mesons is based on kinetic theory and has been formulated by Kadanoff and Baym in the 60s~\cite{KadanoffBaym:1962}. 
Several microscopic transport models have been developed which use approximations of the Kadanoff--Baym equations of various sophistication (\mycf Sec.\,\ref{sec:mass-transport}).  
In cascade mode, \myie with many-body effects essentially turned off, the nucleons flow on straight trajectories between collisions and the individual scattering processes are described on the basis of known, extrapolated or calculated cross sections. 
This is a good description as long as the average time between collisions 
$\tau \sim \left(\sigma\rho\beta\right)^{-1}\;$
is longer than the time it takes until the scattered hadron has reached its asymptotic state~\cite{Danielewicz:1995ay}.

Towards higher collision energies the situation is fundamentally different. 
Even a \pp\ collision can turn into a complex process since parton-parton interactions are at action. 
The high energy and strong interaction give rise to a vast range of possible final states. 
Moreover, due to the Lorentz contraction essentially all parton-parton interactions happen at the same time. 
Parton-parton collisions range from the very soft (non-perturbative) to the hard (perturbative) regime. 
The discussion of the many facets of such calculations carried out to describe proton--proton collisions at collider energies goes way beyond the scope of this paper. 
We will focus on observations which are relevant for dilepton production. 

Hard initial state parton-parton collisions are responsible for the production of charm and heavier flavor and to direct virtual photon production through quark anti-quark annihilation (the Drell-Yan process).
Both processes give rise to hard virtual photon production dominating, depending on beam energy, different regions of the dilepton invariant mass spectrum. 
It is commonly understood that these contributions can be calculated on the basis of perturbative QCD taking into account next to leading order graphs (\mycf\ Sec.\,\ref{sec:hard-processes}). 
Perturbative methods, however, cannot straightforwardly be applied to beam energies below $\sqrt{s_{NN}} \lesssim 10$\gev . 
The calculation of charm production down to threshold energies might become possible, though, long distance dynamics in the hard process described by so-called Sudakov corrections seems to dominate over the higher order corrections~\cite{Kidonakis:2002vj}. 
In contrast, the large energy density generated in the initial stage of the heavy-ion collision is by large due to multiple collision of soft partons, \myie mostly gluons. 
In this context it is interesting to note that pomeron exchange has recently found interest again to describe the non-perturbative (soft) hadron-hadron interactions~\cite{Drescher:2000ha}.
Pomerons can be understood as coherent multi-gluon states carrying the quantum numbers of the vacuum.
In particular they do not carry charge and consequently do not radiate (virtual) photons.  
In the NEXUS model, the multi-parton collisions are separated into hard collisions (those appearing at large momentum transfer) and de-localized, soft collisions treated in an effective manner via string formation~\cite{Drescher:2000ec}. 
The stochastic treatment of the multi-parton collisions allows to study also the energy density variations over the initial fireball volume.
These distributions, which fluctuate event by event, are in particular responsible for the higher order angular modulations observed in the particle emission~\cite{Gale:2013da}. 

While many observables in heavy-ion collision scale with the number of nucleons participating in the collision, hard processes do not trivially scale with the the number of hard binary collisions due to shadowing effects~\cite{Eskola:2020lee}. 
Nevertheless, the number of participating nucleons is a key quantity to classify heavy-ion reactions at high energies.
Centrality allows a good classification of the transverse energy production, which in turn is a good measure of the entropy of the system. 
As the centrality increases, more and more nucleons take part in the formation of the nuclear fireball.
In experiment, the number of charged particles in the acceptance of the detector system is usually used to approximate the centrality of collision.
Glauber Monte Carlo simulations can be used to relate a certain multiplicity class to the respective centrality class defined by a mean number of participating (primordial) nucleons $\langle A_{part}\rangle$~\cite{Miller:2007ri}.
Its has been shown, that this strategy can be applied for experiments with beams of a few \agev\ on stationary targets~\cite{Adamczewski-Musch:2017sdk} up to a few \atev\ energies at the LHC~\cite{Abelev:2013qoq}.
At ultra-relativistic energies the initial stage of the collision does not contribute substantially to the yield of soft (virtual) photons. 
However, the high pressure in the initial state and the flow generated out of it has some effect on the momentum distribution of the (virtual) photons (\mycf Sec.\,\ref{sec:MultiDiff}). 
\subsection{The Hot and Dense Phase}
\label{sec:dense}
%
The intermediate stage evolves from a state of maximal energy density. 
The values achieved depend on the collision system and the energy as well as on the impact parameter. 
In the most general case, the energy density is not constant across the interaction volume but will drop towards its periphery and will exhibit fluctuations.
However, there is good evidence that throughout the evolution during this stage the system is locally equilibrated, certainly so at ultra-relativistic energies. 
The true nature of the ``thermalization'' process leading to this situation is still a matter of research.
In particular, at lower beam energies, rapidity distributions of protons suggest that the fireball is not fully equilibrated.  
Yet, it is conceivable that different degrees of freedom, \myeg the baryon excitation spectrum versus the baryon momentum distribution, will have different equilibration times.
Fast thermalization in general can be achieved if three or many-body collisions are at work, or, arguing in the language of quantum mechanics, if a certain degree of entanglement is realized such that the combined wave function of a number of constituents probes the phase space according to the level density available (Fermi's golden rule). 
At the high-energy frontier, recently a pre-equilibrium stage has been introduced, which connects the initial stage characterized by multi-gluon interactions, to a later stage, when the system can be safely treated as hydrodynamic evolution.
This stage describes the evolution of the system to equilibration using kinetic theory for gluons~\cite{Kurkela:2018wud}.

Throughout the ``hot and dense'' stage, the microscopic structure of the matter can primarily be of hadronic or partonic character. 
It is said to have hadronic character if the constituents are good quasi particles identifiable by their pole masses and partonic, if quasi-free quarks and gluons are the relevant degree of freedom.
These are somewhat idealized pictures resembling features of a dilute hadron resonance gas and perturbative QGP, respectively, both not precisely realized in nuclear collisions. 
We would like to give an intuitive picture how a dense hadronic phase could be understood based on the cloudy bag model. 
In a sufficiently dense baryonic system, meson clouds of neighboring baryons will substantially overlap.
If such a system is heated, the average kinetic energy of a baryon increases as does the probability to find baryons in an excited state of growing excitation energy. 
In an excited baryon the energy can be carried as well by the mesonic excitations of the cloud or by the dynamics of the valence quarks. 
In other words, the wave function of an excited baryon can contain contributions of baryon-meson states.    
Such a separation into core and cloud effects has recently been put forward in describing the structure of baryonic resonances probed by electron scattering experiments~\cite{Aznauryan:2012ba}.
Here it has been observed that in particular the soft, \myie long wave length, perturbations are carried by the cloud.
Kinetic energies governing the hadronic interactions in a thermalized system formed in heavy-ion collision are generally of soft character. 
In a baryon-dense medium, however, it will be difficult to attribute the meson cloud to a single baryon core only.
This mechanism might lead to a different spectral structure due to the simultaneous coupling of the meson cloud to several baryonic cores. 
In such a case, one can speak of a de-localization of the mesonic excitation which, however, would be strongly modified due to the coupling to neighboring hadrons. 
As will be discussed in Sec.\,\ref{sec:thermal-radiation}, vector meson states propagating to a dense hadronic medium are believed to be the prime source of dileptons with invariant masses up to $\sim 1\,$\gevcc .

This picture resembles ideas developed by Hagedorn to interpret particle spectra observed in violent hadronic collisions, known as the bootstrap model (see \cite{JohannRafelski:2015xx} for a comprehensive discussion and historical aspects).
The basic idea is that a nuclear fireball contains in itself little fireballs, each populated with densities governed by Boltzmann weights (\mycf Eq.\,\ref{eq:shm_density}).
As a consequence, already in the 60s Hagedorn suggested that a system composed of hadrons cannot be heated beyond a critical temperature since at this temperature mass production would be preferred over the increase of the average kinetic energy.
These constituent fireballs are nowadays interpreted as high-lying hadron resonances, which in turn can have internal structures comprising multi-quark components.
Examples for such types of baryonic resonances at lower excitation energies are \myeg the $\Lambda(1405)$, the Roper resonance $\mathrm{N}^{*}(1440)$, which have strong meson-core components in their wave functions.

At the highest collider energies, the initial energy density reached in the collision zone is well beyond $1\,\mathrm{G}e\mathrm{V}/\mathrm{fm}^3$ and net-baryon density practically zero. 
It is known from lattice QCD calculations that at such energy densities QCD matter resides in a partonic phase. 
Moreover, from the evaluation of the chiral susceptibility~\cite{Bazavov:2011nk}, which is obtained from the second derivative of (the logarithm of) the partition function \mywrt the quark mass, a pseudo-critical temperature of $T_c = 156.5 \pm 5$ \mev\ is found \cite{Borsanyi:2013bia, Borsanyi:2010bp}.  
If quarks were precisely massless, QCD matter would undergo a second order phase transition to a chirally restored phase in the vicinity of this temperature. There is also an evidence \cite{BraunMunzinger:2003zd}, \cite{Andronic:2017pug} that a fireball created at collider energies, will freeze-out into a hadronic medium at a temperature just around the pseudo critical temperature derived from QCD.  
There is still debate, to what extent residual inelastic collisions will still occur in the hadronic system. 
This question is of some importance for the understanding of the dilepton emission and will be addressed later. 

While virtual photon radiation from a QGP can be understood as being due to abundant radiative $q\bar{q}$ annihilation (\mycf \ref{sec:thermal-radiation}), the cloudy bag model is in particular helpful to understand virtual photon radiation expected from a dense, baryon-rich hadronic medium. 
Indeed, the Vector Meson Dominance Model (VDM) suggests that all photons couple to hadrons through intermediate vector mesons, which in turn can be regarded as an excitation of the meson cloud 
carrying the quantum numbers of the photon. 
As it will be discussed in detail in Sec.\,\ref{sec:VDM}, VDM is very successful in the description of electromagnetic transitions of hadrons in vacuum (Dalitz and two-body decays).  Investigating the spectral distribution of virtual photons emitted from a dense hadronic system in the context of VDM means studying in-medium vector meson propagators, most importantly the one of the \textrho\ meson. 
As we will discuss in Sec~\ref{sec:thermal-radiation}, various model calculations demonstrate strong modifications of the in-medium \textrho\ spectral function due to the coupling of the \textrho\ to baryon-resonance hole states in nuclear matter. 
The salient feature of these calculations is the ansatz of VDM for baryon--photon coupling stressing the importance of a sound microscopic understanding of the underlying radiative transitions (\mycf Sec.\,\ref{sec:cocktail-phenom}).       

To extract model predictions for the baryon densities and temperatures reached in the various stages of heavy-ion collisions one can use ``coarse graining'' methods.
For this, multiple simulated heavy-ion collision events, each computed with the same collision energy, system size and impact parameter, are sampled on a 3+1 dimensional space-time grid~\cite{Lang:1991qa,Huovinen:2003sa}. 
The densities of a given space-time cell can then be derived by boosting each cell in its local rest-frame.
The degree of thermalization can be assessed by inspecting the stress-energy tensor in each cell. 
For example, using UrQMD it has been found that the density in central \auau\ collisions at \sqsnn{2.4}, averaged over a central volume of $7^3 \,\textrm{fm}^3$, reaches already $\rho = 0.5\;\mathrm{fm}^{-3} \simeq 3 \, \rho_0$ about $12\;\textrm{fm}/c$ after first contact of the nuclei.
With a similar approach, densities of approximately $1$ and $2\;\mathrm{fm}^{-3}$ where found for \sqsnn{3.5} and \sqsnn{8.8}~\cite{Arsene:2006vf}, respectively. 
Using 3-fluid hydro-simulations densities reach even about 20\% higher values.
It can therefore be conjectured that maximum net-baryon densities 20~times higher than the ground state density are produced in the central region of heavy-ion collisions.
However, the densities derived in this way have to be taken as upper values since the short-range part of the nucleon-nucleon potential is not explicitly treated in most microscopic transport models.
%
\subsection{Freeze-out}
\label{sec:freezeout}
%
%
\begin{figure}[tbh]
\begin{center}
\includegraphics[width=0.8\linewidth]{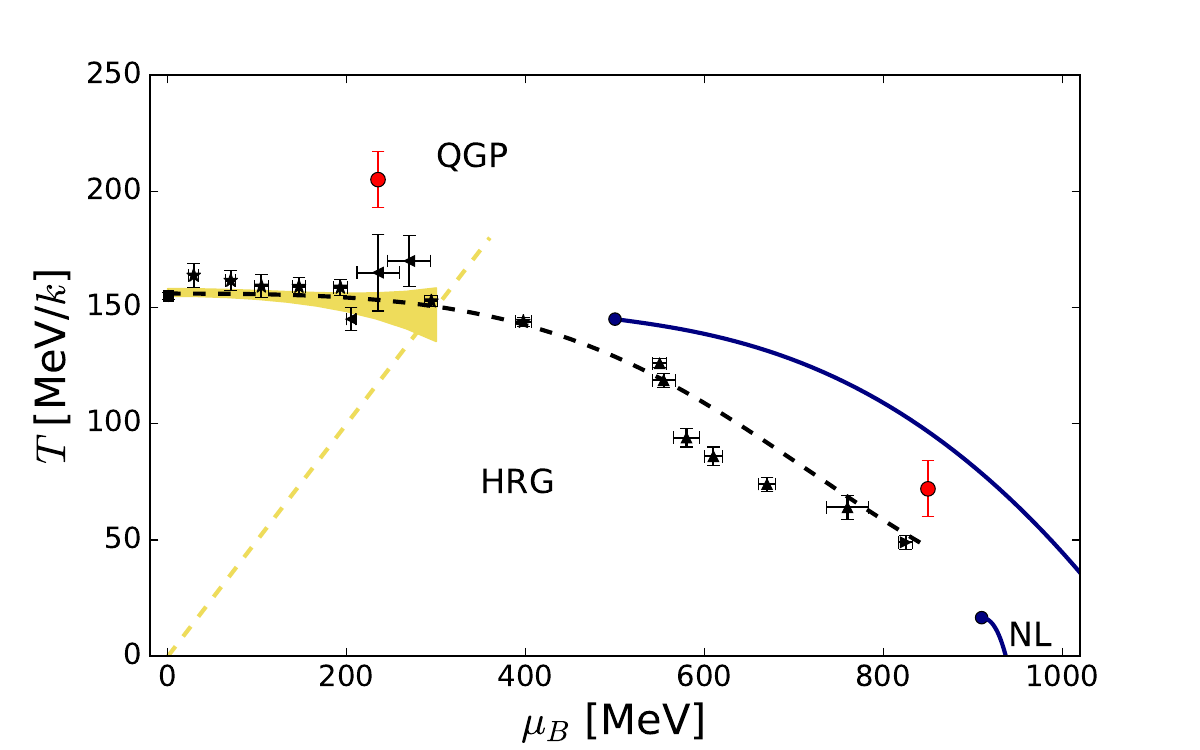}
\caption{ 
The solid blue lines depict conjectured first order phase transition lines ending in critical points. 
The phase regions are the quark gluon plasma (QGP), hadron gas (HG) and nuclear liquid (NL).
The (black) symbols represent freeze-out parameters derived from hadronic final states and are taken from the compilation presented in~\cite{Andronic:2017pug}. 
The (red) circles depict mean fireball temperatures extracted from dilepton excess radiation~\cite{Specht:2010xu,Adamczewski-Musch:2019byl}. 
Results from lattice QCD calculation for the location of the HR-QGP cross over are shown as yellow band and 
the dashed yellow line marks the limit of validity of the Taylor expansion used for deriving thermodynamic parameters from lQCD~\cite{Bazavov:2018mes}. 
}
\label{fig:QCDphasediagram}
\end{center}
\end{figure}

In this conceptional approach to dilepton emission from QCD matter created in heavy-ion collisions it is logical to define a third phase, which spans the time after the chemical composition of particles in the system is frozen and can only be changed by electroweak, or strong decays with ($\tau \gg t_{\mathrm{f.o.}}$). 
The idea behind this classification is to identify contributions to the integrated dilepton yield which can be understood as being due to dilepton  decays of hadrons at freeze-out. 
With this definition, the dilepton yield connected to this stage of the collision can in principle be derived independently if the respective hadron multiplicities of a certain reaction class are measured and the decay processes is well understood. 
Based on observed multiplicities and phase space distributions of hadrons, the expected dilepton yield from this stage can be calculated using the respective branching ratios for decay channels with a dilepton in the final state.
The resulting invariant mass distribution containing dileptons from different hadronic sources is called \textit{hadronic cocktail}, or in brief \textit{cocktail}.
This approach can be verified experimentally by comparing a derived cocktail to the measured dilepton invariant mass distribution in elementary reactions (\mycf  Sec.\,\ref{sec:exp-elementary}), or by inspecting a very peripheral heavy-ion reaction class.

\begin{figure}[tbh]
\begin{center}
\includegraphics[width=26pc]{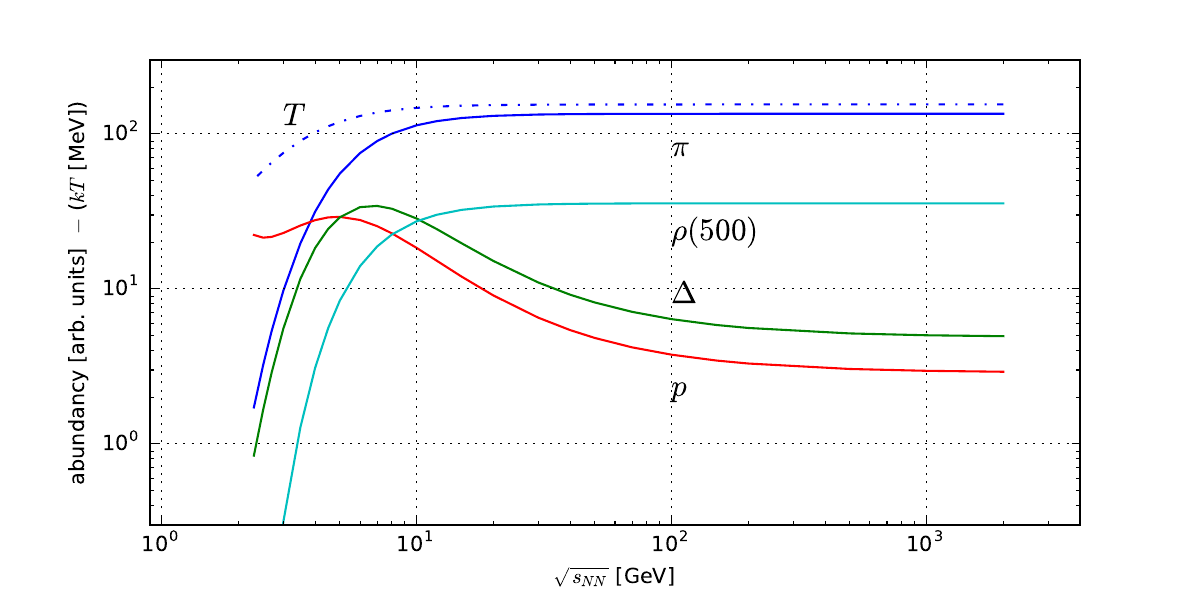}
\caption[QCD phase diagram]{ 
Hadron densities at chemical freeze-out assuming a thermalized hadron resonance gas according to Eqs.\,\ref{eq:shm_T}, \ref{eq:shm_mu} and \ref{eq:shm_density} as function of the collision energy.
For the \textrho\ a hypothetical ``off-shell'' \textrho\ with mass 500\mevcc\ is chosen representing the LMR medium radiation.
Widths of resonances are neglected throughout. 
The dashed--dotted line shows the temperature at chemical freeze-out.
}
\label{fig:freezout_densities}
\end{center}
\end{figure}
%
Moreover, at high collision energies it is well established that the (charged) particle multiplicities can be described by grand canonical partition functions assuming that the system suddenly hadronizes out of a thermalized, deconfined phase (QGP)~\cite{BraunMunzinger:2003zd,Andronic:2017pug}. 
While details are still under discussion (see \myeg \cite{Lo:2017ldt}), this observation gives straight access to particle abundance as they are solely determined by a few thermodynamical parameters like volume, temperature and chemical potentials. 
Indeed, from the inspection of hadron multiplicities observed for various collision energies and spanning from \sqsnat{5.13} (top LHC energies) down to \sqrts $\;=7$\agev\ (lowest RHIC energies in collider mode) it was found that the derived freeze-out temperatures ($T$) and baryo-chemical potentials ($\mu_b$) all line up on a narrow band in the QCD phase diagram (see Fig.\,\ref{fig:QCDphasediagram}).
As an example, in \cite{Andronic:2008gu} a phenomenological functional dependence has been reported for central collision of heavy collision systems (\myie \auau ~or \pbpb ) as 
\begin{equation}
\label{eq:shm_T}
T(\sqrt{s_{\mathrm{NN}}}) = 
\frac{T_{\mathrm{lim}}}
{
\left(
1+\exp
\left(
2.6-\ln
\left(
\sqrt{s_{\mathrm{NN}}}/(0.45 \cdot \mathrm{G}e\mathrm{V})
\right)
\right)
\right)}
\end{equation}
and
\begin{equation}
\label{eq:shm_mu}
\mu_B(\sqrt{s_{\mathrm{NN}}}) = 
1303 \cdot \mathrm{M}e\mathrm{V} \cdot
\left(
1+0.286 \cdot \mathrm{G}e\mathrm{V} \cdot \sqrt{s_{\mathrm{NN}}}
\right)^{-1}.
\end{equation}
The parameter $T_{\mathrm{lim}}$ defines the plateau in temperature for high beam energies observed in experiment (\mycf Fig.\,\ref{fig:QCDphasediagram}). 
Other definitions of a \textit{universal freeze-out line} assume fixed energy per particle or constant entropy density etc.~\cite{Cleymans:2005xv}.
Using the above parametrization (Eqs.\,\ref{eq:shm_T} and~\ref{eq:shm_mu}), the particle multiplicity densities $n_i$ in central collisions can be calculated for a given collision energy using the approximate formula~\cite{Randrup:2006nr}
\begin{eqnarray}
\label{eq:shm_density}
\nonumber
n_i \left( T,\{ \mu \} \right) & = & 
\frac{g_i T^3}{2\pi^2} 
\sum_{n=1}^{\infty}
\frac{\left( \pm \lambda_i \right)^n}{n^3}
\left(\frac{n m_i}{T}\right)^2
K_2\left( \frac{n m_i}{T} \right)
\\
& = & 
\frac{g_i \lambda_i}{2\pi^2}
m_i^2 T K_2\left( \frac{m_i}{T} \right) \pm \mathrm{h.o.}\: .
\end{eqnarray}
Here $n_i$ denotes the number density for the particle of type $i$, $\{ \mu \} = \{ \mu_B, \mu_Q, \mu_S \}$ the chemical potentials of the emitting system connected to the conserved quantum numbers: baryon number ($B$), charge ($Q$) and strangeness ($S$) and $m_i$ the mass of the particle. 
$K_2$ is the Bessel function of second kind.
The minus sign has to be used for fermions (baryons) while the plus sign is for bosons (mesons). 
Moreover, 
\begin{equation}
\label{eq:shm_fugacity}
\lambda_i \left( T,\{ \mu_i \} \right) = 
e^{\mu_B B_i+\mu_Q Q_i+\mu_S S_i}
\end{equation}
is the fugacity factor of particle $i$ with the chemical potentials $\{\mu_i \}$ and 
\begin{equation}
\label{eq:shm_degeneracy}
g_i = 2 J_i + 1
\end{equation}
the spin degeneracy of the particle. 
The formula breaks the series expansion (\ref{eq:shm_density}) after the first term and is precise to within 10\% for not too small $m_i/T$.
The pion to nucleon ratio along the freeze-out line can now readily be calculated (see Fig.\,\ref{fig:freezout_densities}).
As can be seen, according to the thermal description matter created at colliders freezes out as pion dominated hadron gas while at SIS18/BEVALAC energies the system has only about 10\% pions in the final state. 
Generally, a substantial additional amount of pions adds to the final state because of rapid decays of (heavy) mesonic or baryonic resonances. 
In general, such a freeze-out scenario can describe the yields and the soft part of the transverse momentum distributions of measured particles, once in addition collective expansion of the fireball is taken into account. 

As it has already been mentioned, in semi-central collisions a strong $v_2$ signal is observed.
At beam energies of a few \agev\ particle emission in the reaction plane is partially suppressed due to absorption of particles emitted from the collision zone (participant matter) in the bypassing projectile and target (spectator) matter.
The latter effect results in a negative $v_2$.
With increasing beam energy the $v_2$ signal turns positive (to in-plane) with increasing strength. 
At the highest RHIC energies it has been found that $v_2$~\cite{Adare:2006ti} for identified baryons and mesons can be collapsed to a unique curve if $v_2$ is scaled to the number of constituents quarks $n_q$ of the identified particle (\myie $v_q = 2$ for mesons and $v_q = 3$ for baryons) and plotted in bins of transverse kinetic energy per constituent quark $E_T/n_q = \left(m_T - m_0\right)/n_q$, where $m_T$ is the transverse mass and $m_0$ the rest mass of the identified hadron.
This observation is commonly called {\em quark number scaling}.
The success of the hydrodynamical description of the fireball evolution at ultra-relativistic collision energies supports the assumption of a very rapid thermalization in the initial stage. 
In particular the strong flow effects and quark number scaling give very clear evidence for a deconfined phase (partonic) phase of the  matter which is strongly coupled. 
Moreover, the flow develops throughout the whole expansion until the system freezes out. 
Hence, it is to be expected that any penetrating probe, as dileptons, which on average decouple from the system early on, should not follow this flow systematics. 

The validity of this expansion scenario has been studied at RHIC by STAR in the the so-called beam-energy scan~\cite{Adamczyk:2012ku}. 
No major non-monotonic behavior of observables characterizing the fireball evolution has been found down to collision energies of~\sqsna{7.7} so far.
Exceptions are maybe related to the excitation function of anti-hyperons~\cite{Adam:2019koz}, which show indeed a change in the centrality dependence, and to the excitation function of the source radii extracted from two-pion interferometry~\cite{Lacey:2014wqa}.  
The main purpose of this energy scan, which is currently repeated with more statistics and an upgraded detector system, is to scrutinize this search for non-monotonic excitation functions of some observables and/or for the disappearance of signatures evidently attributed to the existence of a partonic phase in this intermediate stage. 

As it was already said, a strict division of heavy-ion reactions into well separated stages is an idealization. 
Yet, it eases the way to the interpretation of dilepton emission  in (ultra-) relativistic heavy-ion collisions. 
At the high energy frontier dileptons emitted in the initial stage (\mycf Sec.\,\ref{sec:initial}) are characterized by partonic (hard) processes. 
Although at a few \agev\ first chance collisions are not hard, in the sense that perturbative QCD would apply, typical values of \sqrtsnn\ (taking the Fermi momentum into account) soon significantly depart from those characteristic for binary \nnnn\ collisions in the thermalized fireball. The contribution from this first chance collisions or pre-equilibrium stage can be experimentally accessed by studying elementary (reference) systems (\mycf Sec.\,\ref{sec:exp-elementary}). 

Finally, one might like to challenge the assumption of a possible thermalization of the fireball at low beam energies.
In that case, microscopic (transport) theory would be the only way to theoretically address the dilepton continuum. 
However, the recent development of "coarse graining" methods (\mycf Sec.\,\ref{sec:initial}) for the description of dilepton production offers yet another possibility merging both approaches. 
We will come back to this point in  the discussion of experimental results~\mycf Sec.\,\ref{sec:status}.   
%

\subsection{The phase diagram of QCD matter}
The present understanding of the phase structure of QCD matter is illustrated in Fig.\,\ref{fig:QCDphasediagram}. 
The properties of QCD matter at vanishing baryo-chemical potential, \myie\ for baryon anti-baryon symmetric matter, is well understood from lattice QCD calculations.
A cross over between a hadron gas at low temperature to a strongly coupled liquid of quarks and gluon has been located around a pseudo-critical temperature of $T_{pc} = 156$~M$e$V \cite{Borsanyi:2013bia,Ding:2019fzc}.
Below the pseudo-critical temperature thermodynamic quantities derived from QCD lattice calculations are in line with respective quantities computed in the hadron resonance gas model~\cite{Vovchenko:2017xad}. 
Above this temperature they deviate and lattice QCD only gradually approaches the properties of an ideal gas of quarks and gluons at very high temperatures.
For temperatures likely reached in heavy-ion collisions, \myie below 300~MeV, the quantities reflect the properties of a strongly couple fluid.
The transition from the chirally broken situation to its full restoration is a smooth cross over and falls approximately in place with the transition from a confined to deconfined state.

In contrast, for matter with large net-baryon density (high $\mu_B$) the deconfinement transition is expected to be of first order.
The region in the phase diagram where the transition turns to be first-order, characterized by the occurrence of a critical point, has not been located yet. 
Thermodynamic properties of strongly interacting matter in the region of finite baryo-chemical potential cannot easily be assessed with lattice QCD. 
The reason is a sign problem which causes a complex Fermi determinant once $\mu_B$ takes finite values.   
Using Taylor expansion techniques it became possible to extend lattice results to $\mu_B \simeq 0.5 T$, showing no indication for critical point indicating that the hadron-QGP transition would turn into first order.
The search for the critical point, or likewise for indications of a first order deconfinement transition, is a key program of relativistic heavy-ion experiments. 
In parallel, theoretical approaches to solve QCD based on effective field theories, like Functional Renormalization Group methods or Schwinger-Dyson calculations, are developed to narrow done the uncertainties~\cite{Fischer:2018sdj,Drews:2016wpi}.   

McLerran and Pisarski have proposed to ``organize'' the phase diagram of QCD matter by inspecting QCD properties in the limit of a large number of colors ($N_c \rightarrow \infty$ limit)~\cite{McLerran:2007qj}.
Obviously, in the hadronic world the number of degrees of freedom do not change with the number of colors since hadrons are color neutral objects and need to contain all colors to neutralize color charge. 
The number of degrees of freedom in the partonic phase would scale like $N_c^2$, because gluons carry combinations of color and anti-color. 
They argued that there should exist a third phase, termed quarkyonic, which would be characterized by a linear scaling of the degrees of freedom with the number of colors. 
It would occur at high net-baryon densities and moderate temperatures and could be considered as a gas of confined quarks surrounded by hadronic excitations and glueballs.  
However, a rigorous derivation of the phase structure in the real world with three colors only is not available.

The quest is to locate ``trajectories'' in the phase diagram along which the collision systems evolve during the intermediate stage of heavy-ion collisions and before the system freezes out. 
Since the abundance and phase space distributions of hadrons are fixed only after the system has frozen out, the determination of such trajectories based on hadron spectra is close to impossible.
Guidance where to locate fireballs formed in heavy-ion collisions probe the QCD phase diagram is given by the so-called freeze-out parameters which are depicted in Fig.\,\ref{fig:QCDphasediagram} as black ``data points'' (\mycf Sec.\,\ref{sec:freezeout}). 
They are obtained from fits of the hadron resonance gas (Eq.\,\ref{eq:shm_density}) to multiplicity for a large number of hadrons measured in the experiments, taking properly into account feed-down form excited hadron states.
The dashed line connecting the data points is defined by Eqs.\,\ref{eq:shm_T} and \ref{eq:shm_mu}.

Electromagnetic probes are currently the only way to obtain direct information from the dense system before freeze-out.
Two such data points are included in Fig.\,\ref{fig:QCDphasediagram} as circles.
They refer to temperatures obtained from the slope of dilepton invariant mass spectra. 
The approximate baryo chemical-potentials are taken from the respective freeze-out point. 
Both lie above the freeze-out curve as it is expected due to the penetrating nature of dileptons. 
The procedure to extract such information from dileptons emitted from heavy-ion collision is a central subject of this review. 
%
%
\section{Experimental aspects of dilepton spectroscopy}
\label{sec:pheno}
The central goal of continuum dilepton spectroscopy is to reconstruct the so-called excess radiation, which is emitted out of the dense and hot stage of a heavy-ion reaction.
This is a formidable task since on the one hand we have to deal with pair reconstruction, which requires to form all possible opposite sign pairs from all leptons (electrons or muons) reconstructed in a given event and hence generates combinatorial pairs not originating from the same virtual photon. 
On the other hand, the expected distribution of the excess  radiation is rather structure-less and has to be extracted as one of all signal sources  adding to a eventually huge background of combinatorial pairs. 
Moreover, lepton-pair decays of hadrons or pair emission from annihilation processes are suppressed of order $\alpha^2$ and hence are truly rare probes. 
The reconstruction procedure for the dilepton signal can be grouped to the following steps:
\begin{enumerate}
\item 
Reconstruction of sufficiently pure lepton track candidates.
\item
Rejection of background leptons on the single track level.
\item
Background rejection on the pair level.
\item
Estimation and subtraction of the combinatorial and correlated background.
\item
Identification of the excess radiation.
\item
Efficiency and acceptance correction of spectra.
\end{enumerate}
In the following we first introduce the dilepton observables. 
Thereafter, we address experimental aspects of dimuon and dielectron signal extraction and then discuss how the unknown combinatorial background can be approximated based on random combinations of leptons. 
At the end, we introduce selected experiments which have been used, or are still in use, for dilepton spectroscopy. 
%
\subsection{The dilepton signal}
\begin{figure}[tbh]
\begin{minipage}[b]{0.58\textwidth}
\includegraphics[width=20pc,height=10pc]{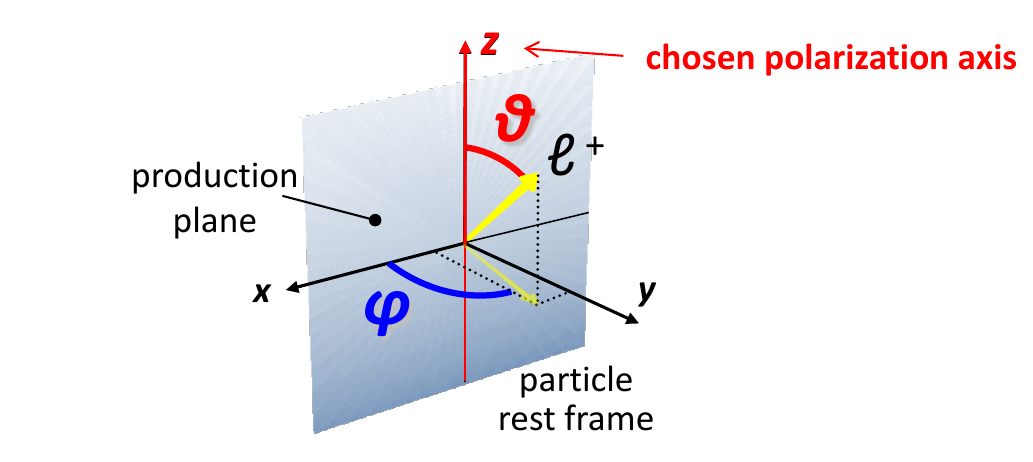}
\end{minipage}
\hfill
\begin{minipage}[b]{0.38\textwidth}
\caption[Decay angles]{
Angles characterizing the decay of virtual photons in its rest frame with the z axis pointing along the photon direction. 
The {\em production plane} is spanned by the beam axis and the momentum vector of the virtual photon in the laboratory frame. 
The definition of the two angles $\theta$ and $\phi$ depends on the polarization axis used. 
}
\label{fig:ref-frame}
\end{minipage}
\end{figure}

%
The 6-fold differential production probability for dileptons in covariant form reads  
%
\begin{equation}
E_+ E_-
\frac{\mathrm{d}^6P}{\mathrm{d}^3p_{+} \mathrm{d}^3p_{-}}
= \frac{\mathrm{d}^6P}{q_{\perp}\,\mathrm{d}M_{ll}\,\mathrm{d}y\, \,\mathrm{d}q_{\perp}\,\mathrm{d}\varphi_{l}\,\mathrm{d}\alpha_h
}\,.
\label{eqn:total-differential}
\end{equation}
Here $E_{+-}$ and $p_{+-}$ denote the lepton energies and momenta, respectively.
A common choice for the six independent coordinates, which characterize the emission of the virtual photon in the beam system, is typically taken to be the invariant mass $M_{\ell\ell}$, the transverse momentum $q_{\perp}$, the rapidity $y$, 
and the azimuthal emission angle $\phi$ of the virtual photon relative to the beam axis,
as well as the helicity angle $\alpha_h$.
Another choice for the angles are the two independent lepton decay angles in the virtual photon rest frame $\vartheta_l$ and $\varphi_l$ (\mycf in Fig.\ref{fig:ref-frame}).
%
%

In case electrons are observed ($m_\ell = m_e \ll M_{\ell\ell}$), the invariant mass can be well approximated using the geometric mean of the lepton momenta as   
\begin{equation}
M_{ee} \simeq 2 \sqrt{|p_{\ell}| \, |p_{\ell}|} \;   
\sin{\left( \frac{\alpha_{\ell\ell}}{2}\right)}
\label{eqn:invariant_mass}
\end{equation}
with $\alpha_{\ell\ell}$ denoting the opening angle of the lepton pair in the laboratory.
This equation demonstrates that dileptons with small invariant mass but sufficient momentum to traverse the detector systems appear with small opening angle. 
This topology can be used to reduce combinatorial background (\mycf Sec.\,\ref{sec:electrons}). 
It may be convenient to use the transverse mass $m_\perp$ instead of the transverse momentum.
For the lepton pair it is defined through
\begin{equation}
m_\perp^2 = M_{\ell\ell}^2 + q_\perp^2 = q_\mu q^\mu + q_\perp^2 \;
\end{equation}
where $q$ is the virtual photon four-momentum.

It is common to distinguish dilepton invariant mass regions according to the sources dominating them: the Low Mass Region (LMR) below $1.1$ \gevcc, \myie including the \textrho\ meson pole mass, the Intermediate Mass Region (IMR) $1.1<M_{l^+l^-}<2.9$ \gevcc , spanning between the low-mass vector mesons and charmonium, and a High Mass Region (HMR) above. 
The relative abundance of pairs in these regions depends primarily on the collision energy per nucleon pair. 
While in experiments at LHC, RHIC and SPS, dileptons are populating all regions, experiments at SIS18/BEVALAC up to today could only explore the LMR.

The polar and azimuthal emission angles of individual leptons are connected to the photon polarization, which in turn depends on the production mechanism. 
The respective angular distributions are given in the rest frame of the virtual photon and can be decomposed in the following way ~(see~\cite{Speranza:2018osi} and references therein):
\begin{align}
\label{eqn:ang-distr-full}
\frac{\mathrm{d}P}{\mathrm{d}\Omega_{\ell}}  = 
1
& +\lambda_{\vartheta} \cos{\vartheta_{\ell}}^2 
+ \lambda_{\varphi} \sin^2\vartheta_{\ell}  \cos{2\varphi_{\ell}} 
+ \lambda_{\vartheta\varphi} \sin{2\vartheta_{\ell}}  \cos{\varphi_{\ell}} \\ \nonumber 
& +\lambda_{\varphi}^{\perp} \sin^2\vartheta_{\ell} \sin{2\varphi_{\ell}}
+\lambda_{\vartheta\varphi}^{\perp} \sin{2\vartheta_{\ell}}  \sin{\varphi_{\ell}}
\,.
\end{align}
The factors $\lambda_{\vartheta}, \lambda_{\varphi}, \lambda_{\vartheta\varphi}, \lambda_{\varphi}^{\perp}$  and $\lambda_{\vartheta\varphi}^{\perp}$ are the anisotropy coefficients where $\lambda_{\varphi}^{\perp}$ and $\lambda_{\vartheta\varphi}^{\perp}$ are non-zero only if the emission is asymmetric with respect to the scattering plane and $\mathrm{d}\Omega_{\ell} = \mathrm{d}(\cos{\vartheta_{\ell}} )\: \mathrm{d}\varphi_{\ell}$.
Depending on the choice of the reference frame, the coefficients can take different values. 
A good choice for continuum radiation is the helicity frame (\mycf Fig.\,\ref{fig:ref-frame}). 
Here, the quantization axis (z) is taken to be along the momentum direction of the virtual photon in the heavy-ion collision (or hadron-hadron) center of mass and $\vartheta_l = \alpha_h$. 
The x-axis is taken in line with the reaction plane.
For the discussion of Drell-Yan radiation or quarkonium production also the Collins--Soper frame is used for which the bisector between the two hadron (heavy-ion) momentum directions in the virtual photon rest frame is taken as the quantization axis. 
The photon polarization and consequently lepton angular distributions can provide essential information about the dilepton production mechanism as well in vacuum as in hot and dense medium. 
This topic is discussed in Sec.\,\ref{sec:theory}.
%
\subsection{Extraction of the dilepton signal}
%
Signal dileptons are pairs which originate from the same virtual photon (mother particle).  
From the physics point of view it does \textit{a priori} not matter if electrons or muons are used to reconstruct the virtual photon spectrum, aside from a different phase space cut-off at low invariant masses due to the by a factor $\simeq 200$ higher rest mass of muons. 
Experimentally, however, there are substantial differences which are related to both, the detection of the signal with sufficient purity as well as the rejection of background.
\subsubsection{Muons}
The detection of muons has advantages over electron detection if the laboratory momenta of the leptons are sufficiently high.
Using hadron absorbers in front of a muon detection system enables that essentially only muons reach these detectors. 
Such absorber technique can, in certain cases, allow to easily trigger on the production of muons or even on muon pairs which is in most cases impossible for electrons.
A disadvantage of the absorber technique is that hadron identification in the acceptance of the absorber requires placing of respective detectors in front of the absorber which enlarges the whole detection system accordingly.
Hadron absorbers, moreover, introduce collisional energy loss and multiple scattering to muons.
Consequently, absorber techniques are advised if the laboratory momenta of the muons are not too small.
This can be assessed by computing the the mean muon energy loss $<dE/dx>$ of an absorber material over one nuclear interaction length $\lambda_I$ in this material: 
\begin{equation} 
 \expval{E_{\mathrm{loss}}}
 = \int_{0}^{\lambda_I} \frac{\mathrm{d}E}{\mathrm{d}x} 
 \mathrm{d}l
 \approx 
 \lambda_I \; 
 \frac{\mathrm{d}E}{\mathrm{d}x}\biggr\rvert_{\mathrm{MIP}} \,.
 \end{equation}
For Iron, using for simplicity the mean energy loss of a minimum ionizing (MIP) muon of $1.45$ \mev\ /(g/cm$^2)$, an energy loss of $\approx 200$~\mev\ can be estimated over a distance, after which the hadron's survival probability drops to $P_{\mathrm{surv.}} = 1/e$.
This is a rough approximation in case of low hadron momenta where hadronic cross sections strongly vary because of resonant scattering. 

The success of such a detection strategy depends on the configuration of the spectrometer.
NA50 used a stand-alone muon spectrometer behind a thick absorber and no magnet in front of it.
As a consequence, the muon tracks had to be extrapolated backward to the target (primary vertex) region through the absorber in order to gain information about the origin of the muons.
Multiple scattering of muons in the absorber, however, blur the vertex resolution of such tracks.
This situation was overcome after the upgrade to the NA60 experiment (\cite{Usai:2005zh} and references therein) when a dipole magnet was added in front of the absorber. 
The region in the dipole field was filled with high-resolution silicon strip and pixel detectors. 
This enabled matching of track segments reconstructed in the muon spectrometer to track candidates obtained from the silicon tracker by matching the candidates in configuration and momentum space.  
The ``double match'' provided sufficient selectivity to reach high purity in the muon reconstruction by avoiding fake matches of tracks produced by different particles in front of and behind the absorber.
 
Moreover, the precision of the track reconstruction right behind the target empowered a partial rejection of off-vertex tracks, which are characteristic for daughter muons escaping after weak decays of charmed hadrons near to the primary vertex. 
A similar approach will also be realized in the upgrade of the ALICE muon forward detector for Run~3 and~4~\cite{Abelevetal:2014cna}. 
A third option is to segment the absorber and to interlace it with tracking detectors.
This allows to substantially reduce the negative effect of multiple scattering by continuously matching the proper track segments to a global track.
Such a solution has been chosen by the CMS experiment at CERN \cite{Chatrchyan:2012xi} which in addition uses the return yoke of the solenoid as hadron absorbers and the reversed field herein as second spectrometer.  
This technique is also foreseen for the muon detection system of the upcoming CBM experiment at FAIR~\cite{Friman:2011zz}, albeit not having a strong magnetic field in the absorber region.
The challenge here is the high track density in the detector behind the first absorber which ``sees'' substantial low-energy secondary particles at a stage where not all primary hadrons are stopped yet. 

A background source of muons, when using absorber techniques, is weak decays of charged pions or kaons in front of the absorber or punch-through hadrons, \myie hadrons which escape the absorber including hadrons produced through interactions in the absorber.
In the case of a weak decay \myeg a pion will decay into a muon by radiating off a neutrino and carrying on most of the pion's momentum with gradual changes in the direction. 
It then will penetrate the hadron absorbers in the same way as signal muons do. 
The fraction of pions (or kaons) decaying before an absorber placed at a distance $d$ away from the interaction point is
\begin{equation} 
P_{\downarrow} = 1-e^{-\frac{d}{\beta\gamma\,c\tau}}
\end{equation}
for a pion (kaon) with laboratory energy $E$, lifetime $\tau$ and $\gamma= \frac{E}{m_{\pi,K}}$.
Evidently, muon detection is advantageous for fixed target experiments at ultra-relativistic energies, in collider experiments at large forward or backward rapidities or for high-$p_\perp$ probes in the central rapidity region due to the large $\gamma$ factors in the laboratory frame.
In all cases signal impurities due to muons from weak decays can additionally be suppressed by active detection of the momentum change to the trajectory of the decaying hadron (both a directional change and a sudden kinetic energy loss) introduced by the escaping neutrino. 
For the latter, the ``mass'' of the tracking system plays an important role since the ability to detect such a kink in the trajectory is again strongly affected by multiple scattering.
Even if the track reconstruction is not able to spot the kink in the trajectory, it is still possible to draw on the Distance of Closest Approach (DCA) of the extrapolated track to the primary interaction vertex. 
%
\subsubsection{Electrons}
\label{sec:electrons}
%
The detection of electrons is not affected by above considerations but faces other challenges.
Below momenta of around $300$\mev , the precise limit depends on the detector geometry and the resolution of the time-of-flight measurement, electrons can effectively be separated from hadrons (and muons) by their high velocity ($v_e \simeq c $).  
The most crucial ``background'' hadron is the pion which, due to its small mass, is the first to reach near $c$ and, moreover, is typically much more abundant than the electron.
The pion rejection power, the fraction of falsely identified pions (adding to the electron sample) over the total number of detected charged pions, should ideally reach at least $10^{-3}$--$10^{-4}$. 
Additional detectors are needed to provide sufficient electron hadron separation for higher momenta.
Commonly used are Ring Imaging Cerenkov (RICH) Detectors based on gaseous radiators, as for example in the CERES~\cite{Baur:1993mw} and the HADES spectrometers~\cite{Agakishiev:2009am}.   
However, RICH detectors need to be placed in regions without magnetic field to guarantee straight trajectories of the electrons inside the radiator which avoids blurring of the imaged ring. 
Hence, the introduction of RICH detectors limits the flexibility in the detector arrangement. 
Depending on the index of refraction of the radiator gas, pions can be effectively discriminated from electrons up to momenta of around of 4-5~\gevc ~\cite{Bohmer:2000jd,Hohne:2005ec}. 
The other option is to combine time-of-flight with a precise dE/dx (energy loss) measurement.
This strategy is successful if Time-Projection-Chambers (TPC) are used for tracking, as demonstrated by STAR experiment at RHIC \cite{Adamczyk:2012yb} and a ALICE at LHC  \cite{Abelevetal:2014cna}.
Continuous tracking in the large active volumes of these detectors provides a large number of quasi-independent energy loss measurements.
As observable a truncated mean is typically used to minimize the effect of fluctuations governed by a highly skewed probability distribution of the energy loss with a long tail~\cite{Olive:2016xmw}.  

Additional electron-to-pion separation at higher momenta is provided by Transition Radiation Detectors (TRD).
These detectors make use of the transition radiation produced by charged particles crossing boundaries between materials with different refractive indices. 
Since the emission probability is small, the particles have to cross many boundaries and consequently low-mass foams or compounds are the material of choice.
The difficulty is to detect the transition radiation, which is emitted sparsely in the soft x-ray domain and typically together with the signal induced by the charged electron traversing the same region of the detector where the x-rays are converted.
Yet another option to identify electrons is to combine momentum (tracking) information with the signal induced in an electromagnetic calorimeter.
At reasonably high energy the resolution is sufficient to identify the leptons by simultaneous measurement of momentum and energy\footnote{The relative energy resolution of a calorimeter improves like $E_e^{-\frac{1}{2}}$}.
Again, precise values depend on the type and quality of the calorimeter. 

To maximize the reconstruction efficiency for leptons in situations where several detectors can contribute, it became common to exploit neural networks in the decision process. 
The input layer to the networks is given by the signals of all detector systems which can contribute to the identification.
This includes also the momenta of the track candidates selected for the decision process. 
The output layer is a single quantity ranging between zero and unity and providing a probabilistic information on the track candidate's nature. 
The advantage is that there are no strict cuts required for each signal response. 
Otherwise, each condition would impose a probability $\epsilon_i$ that the candidate survives this condition.  
For a particle candidate to survive all conditions, a ``survival'' probability of $P = \prod_{i=1}^N \epsilon_i$ would result where $n$ is the number of conditions applied. 
Hence, any fluctuation of an individual signal response out of the the selection range would reject the respective track. 

While weak decays of hadrons are not of relevance as possible background sources in dielectron measurements, external photon conversion is. 
In heavy-ion collisions abundant photons come from \piz\ decay. 
In fixed target experiments, a high conversion probability can already occur in the target.
This is in particular a problem for measurements using high-$Z$ targets as the conversion probability increases quadratic with the atomic number of the target material.
This effect can be mitigated with segmented targets, for which the total material budget is distributed over a number of target slices arranged at distances significantly larger than the diameter of the target. 
In such an arrangement, the average distance a prompt photon emitted at some angle w.r.t beam propagates in the target material is reduced.

There are two important strategies to control the combinatorial background in dielectron spectroscopy.
One is to reconstruct as many as possible conversion and $\pi^0\rightarrow \gamma e^+e^-$ Dalitz-pairs by tracking down to low momentum and applying conditions on invariant mass of the reconstructed pair.
While for the conversion both, the opening angle of the two electron tracks and the invariant mass are close to zero by all practical means, also in case of Dalitz decays a large fraction of the electron pair has opening angles below a few degree.  
The condition can also be placed as 2-dim selection window evaluating independently the mean momentum and the opening angle, \myie the two factors defining the invariant according to  Eq.\,\ref{eqn:invariant_mass}. 
This provides a measure to optimize the signal survival probability for a given background rejection power.  
With some detector configurations a ``close pair rejection'' can already be achieved without tracking by detecting such pairs by respective double-signals in the Cerenkov detector.
This is possible if the radiator volume of the RICH is placed close to the target (primary vertex) and in a field-free region.
However, the separation power between detector signals due to a single electron and electron-positron close pair is critically depending on the total amount of Cerenkov photons reconstructed. 

Indeed, conversion pairs and a good fraction of the Dalitz pairs have opening angles below a few degree owing to the small invariant mass of the intermediate virtual photon.
Such opening angles are less likely for random pairs.
If a combination is found with $\theta^{+-} < \theta^{+-}_{cut}$ a pair is assumed to come from a \piz\ decay (photo conversion or Dalitz), eventually is filled to the histogram but the individual electron and positrons are furthermore not used for other combinations.
It has been found that this strategy can reduce background without substantially reducing signal efficiency. 
The rejection can even be improved if also incompletely reconstructed tracks (track fragment) are included in the search of a close by partner lepton.  
A fully identified electron or positron would then be rejected if a ``track fragment'' is found in close vicinity of it.
The virtue of this method will, however, critically depend on the two-track separation power of the tracking system.
Dielectrons with small opening angle can also be identified if a RICH detector is placed right behind the interaction region.
The Cherenkov photons produced by the the close by track can either be directly detected by counting the fired cells in the ring region or by identifying ring distortions or clear signatures of overlapping double rings.  
The success of such rejection cuts depends on details of the spectrometer configuration and the performance of its detectors, in particular of the RICH. 

The strategy for close-pair rejection pursued by the PHENIX was to introduce the Hadron Blind Detector (HBD).
This detector had been installed nearest to the interaction point in a field free region.
The detector was able to record Cherenkov light emitted from electron and positron tracks traversing its radiator gas prior to the particles entering the tracking section of the spectrometer.
Matching of candidate tracks in the spectrometer with the signals in the HBD enables both, pion rejection (if no Cherenkov signal is seen in the region of interest) and Dalitz (close) pair rejection (if the Cherenkov signal has about twice the strength as the one expected for singles).

A different handle to reduce background contributions is to reject electron tracks which have transverse momenta below a certain threshold. 
Electrons from \piz\ Dalitz decays or from conversion of the \piz\'s decay photons peak at low transverse momentum.
These cuts, however, have to be handled with caution as it also reduces the phase space of the detected signal pairs in the region of small invariant mass. 
\subsection{Determination of the combinatorial background.}
\label{sec:CBdeterm}
Once pure samples of lepton tracks have been obtained, the signal reconstruction is rather similar for muon and electron pair spectroscopy.
Since in heavy-ion experiments we are in particular interested in the dilepton continuum sources, an important task is to find the best approximation for the a priori unknown combinatorial background.
In fact, the true signal distribution is often obtained from subtracting two smooth distributions in situations where the signal eventually amounts to less than a percent of the reconstructed unlike-sign spectrum.
Such situations are typically reached in environments with highest charged particle multiplicity.
Hence, background rejection strategies are key to dilepton-continuum spectroscopy in heavy-ion collision experiments.

Generally, the bulk of the background pairs can be grouped into three categories: (1) wrong combinations of leptons from two fully reconstructed true signal pairs, (2) pairs which contain at least one lepton from an incompletely detected pair or (3) pairs which contain at least one falsely identified hadron. 
To the latter, we count also muons originating from weak decays. 
The term combinatorial is chosen here for pairs which result from ``uncorrelated'' processes, \myie do not originate from the same virtual photon. 
The probability for observing such combinatorial pairs  
\begin{equation}
P(M^+,M^-) = P(M^+) \times P(M^-)
\end{equation}
factorizes,  where $P^+(M^+)$ and $P^-(M^-)$ are the probability density functions to observe $M$ leptons in an event. 
In the most general case, $P^{+/-}$ depends on the detector configuration and the event characteristic like, most importantly, collision energy, system size and centrality.
Moreover, it can be defined as multi-differential probability \mywrt quantities defining the leptons' phase space. 
Indeed, signal spectra have been derived in situations with signal-to-background as low as 1\% \cite{Adare:2015ila}, which then requires a precision in the background determination of a few per mill.
For step (4) of the reconstruction procedure, outlined above, the lepton samples for both polarities, with multiplicities $M^+_{i}$ and $M^-_{i}$ in a given event $i$, are then paired taking all possible combinations which result in 
\begin{equation}
M^{+-} = \sum_i {M^{+-}_i} = \sum_i{\left(M^+_i \times  M^-_i\right)}
\end{equation}
unlike-sign pairs. 
Most pairs are of combinatorial type and their multiplicity 
increases about quadratically with the charged particle multiplicity of the event class.   
This is easily understood from the fact that pions, which are the lightest and hence most abundant hadrons coupling to leptons, contribute to background either through weak decays (muons) or through their Dalitz or conversion of photons. 
The true signal pair distribution $S^{+-}$ is hence obtained from the total pair yield\footnote{The total (unlike-sign same-event) yield is sometimes also called foreground although, in the true sense of the word, foreground does not include the background.} $M^{+-}$ by subtracting the best estimate for the a priori unknown background $B^{+-}$:
\begin{equation}
S^{+-} = M^{+-} - B^{+-} \;.
\end{equation}
To derive the signal with an anticipated statistical precision $\delta S/S$, the background has to be known with an superior precision of 
\begin{equation}
\frac{\delta B^{+-}}{B^{+-}} \approx \frac{\delta S^{+-}}{S^{+-}} \cdot \frac{S^{+-}}{B^{+-}} \;.
\label{eqn:ErrorBackground}
\end{equation}

To minimize the statistical uncertainty, the true background is typically estimated using both, event mixing and like-sign same event methods. 
Either of these methods has their drawbacks and caution has to be taken. 
The strategy for event mixing is to subdivide a given number of events into event classes such that events with similar characteristics like \myeg the number of participant nucleons (centrality) or the location of the event plane relative to the detector are combined to form the event classes (pools). 
Then pairs are generated by combining each a lepton from a different event of the same pool.
In this way it is guaranteed that there are no correlations between the two leptons while they still were reconstructed from events with the same event characteristic. 
The latter is needed to mitigate effects in event mixing as the true combinatorial background is originating from exactly the same event.    
Since a given lepton from one event can be group with oppositely charged leptons of any other event from the same event pool the statistical error can in principle be brought to the required level (\mycf \ref{eqn:ErrorBackground}).  
This is in particular helpful to generate the background in sparsely populated region of the phase space. 
However, it was found  that statistical limitations in the total event sample can still cause remaining structures in the event mixed pair sample~\cite{Voloshin:1984aa}.
Also, the topology of pairs has to be chosen such, that the reconstruction efficiency of the pair can be assumed to be given by the product of the single track reconstruction efficiencies.
This is to exclude ``correlations'' in the lepton reconstruction induced by detector effects like \myeg a dropping reconstruction efficiency if two tracks are propagating too close to each other, in respect to the granularity of the detection system.  
This is not always guaranteed if a pair has a small opening angle.
Depending on the type of tracking detector used the reconstruction efficiency might be affected by the local track density.  

A draw back of this multi-event sampling is that the generated background is a priori not normalized.
Several methods have been developed to provide proper normalization.
Most common is to exploit like-sign same-event pairs which cannot originate from a single virtual photon due to charge conservation. 
In this method, for each event like-sign (charged) pairs $M^{++}$ and $M^{--}$ are formed.
Since in this procedure the same events are used for which the unlike-sign pairs are generated, a proper normalization is in reach.
To increase the statistics and to cope with asymmetries in the detection of leptons with opposite charge the best approximation is obtained from an average of the two charged pair samples. 
However, the exact procedure to normalize this charged pair spectrum depends on the nature of dominant pairs producing the background. 
In case of dimuon spectroscopy, dominant background pairs contain candidates originating from weak decays of pions and kaons. 
If the emitting source is truly thermal, the production of these mesons is uncorrelated and as a consequence also their decay muons.
In dielectron spectroscopy, the dominant contributions stem from photo conversion and \piz\ Dalitz decay. 
Here, the background is mostly produced in pairs.
These are two extreme cases and in practice one can assume that the true background pairs are always a mix of the two.
For example, in dielectron spectroscopy background can also originate from misidentified hadrons which then also come to large extent as uncorrelated singles.
Vice versa, in the dimuon case background can occur due to simultaneous weak decays of two pions or kaons originating from the same mother particle.

For the case of fully uncorrelated background tracks it has been shown that the combinatorial background is given by the geometric mean of the two charged pair same event samples:
\begin{equation}
B^{+-}_{\mathrm{LS}} = 2\sqrt{N^{++} \cdot N^{--}}.
\label{eq:combi}
\end{equation}
This equation holds strictly if the multiplicities show Poissonian probability distributions. 
It has been demonstrated in~\cite{Gazdzicki:2001hy}, however, that different multiplicities for positively and negatively charged background muons can indeed cause deviations of the estimated from the true background. 
Hence, an additional factor has to be multiplied on the right side of Eq.\,\ref{eq:combi} which takes care of this asymmetry.
It can be derived by comparing the mixed-event like-sign with the mixed-event opposite-sign background as 
\begin{equation}
R = \frac{N^{+-}_\mathrm{mixed}}
{2\sqrt{N^{--}_\mathrm{mixed} \cdot N^{++}_\mathrm{mixed}}} \; .
\end{equation}
Moreover, as mentioned above, such like-sign electron pairs include quasi correlated pairs if both leptons originate from each one of the two intermediary photons after internal or external conversion.
Common sources of this kind are \textpi$^0$ or \texteta\ (via two photon or Dalitz decays) and, to a smaller amount, \omegatodal\ decays.
For the latter, there occur three photons as intermediary states due to the dominant \piz\ decay into two photons. 
To avoid that the normalization of the mixed-event background to the same-event sample is hampered by quasi correlated pairs one has to remove tracks from external conversion most efficiently and restrict the normalization to a region where these abundant sources do not contribute.   

At high collision energy, quasi correlated like-sign pairs can also arise from correlated open heavy-flavor mesons~\cite{Crochet:2001qd}. 
For example a pair of charged $D$ mesons produced in a collision can both decay semi-leptonically thus producing a quasi-correlated lepton pair. 
Due to the finite life-time of the $D$ mesons such possible background sources can be mitigated by rejecting lepton tracks not originating from the primary vertex.
Similar background is produced by correlated $B$ meson production. 

Last, it has to be ensured that the background determination is not hampered by selective trigger conditions used to record the events in the experiment.
For example, the reconstruction of proper like-sign pairs is impossible if the trigger was set on detection of at least two leptons with opposite charge as it is often done in two-arm spectrometers~\cite{Sekimoto:2004sy}. 
But care has also to be taken in case the trigger on two leptons is not charge sensitive as pointed out in detail for the NA60 experiment in~\cite{Arnaldi:2008er}. 
%
%
\section{Basic Theoretical Concepts}
\label{sec:theory}
Time-like virtual photons are an ideal probe to study the microscopic properties of strong-interaction matter. 
They couple directly to the electromagnetic current of hadronic processes and probe the spectroscopic properties of the involved hadrons.
Sufficient understanding of the various radiative decay or scattering processes is an important prerequisite to enable the separation of the total dilepton yield obtained in experiments into various contributing sources.
In the following, we will outline the theoretical approaches to model dilepton radiation.
We will focus on soft radiation as this is relevant for dilepton emission of hadronic systems emphasized in this review.
\subsection{Electromagnetic decays of hadrons}
\label{sec:cocktail-phenom}
Leptonic and semi-leptonic two-body and three-body decays of hadrons are the important processes contributing to the hadronic cocktail in the LMR (\mycf Sec.\,\ref{sec:freezeout}).
They can be grouped into three cases, each featuring an intermediary virtual photon decaying into a pair of leptons: 
The first one is the transition of a hadron ($P$) into another hadron ($P'$) with smaller mass and a virtual photon.
A second group is a special case of the first type but with a neutral meson decaying into a real and a virtual photon.  
The last case is the direct transition (annihilation) of vector mesons into a virtual photon.

In the most general case the partial width for a decay of particle ($P$) into a n-body final state can be expressed by~\cite{Olive:2016xmw}
\begin{equation}
       d\Gamma =
       \frac{\left( 2\pi \right)^4}{2M}\,
       \mid\mathcal{T}\mid^2 \,
       d\Phi_n\left(P; p_1 \dots p_n \right) \,,
       \label{eqn:partial_decay_width_general}
\end{equation}   
where $M$ is the mass of the decaying particle and $d\Phi_n$ the $n$-body final state phase space. 
In case of a Dalitz decay of a hadron into a daughter hadron and a lepton pair, the invariant transition amplitude $\mathcal{T}$ is defined by \cite{Landsberg:1986fd} 
\begin{equation}
       \sum_{pol}\mid 
       \mathcal{T}\left(P\rightarrow P'\ell^+\ell^-  \right)\mid ^2\, 
       = e^2\mathcal{W^{\mu\nu}} \frac{1}{q^4} L_{\mu\nu} \, ,
       \label{eqn:mes-matrix-el}
\end{equation}       
where the sum is running over the spin states of the initial and final state particles.
The squared transition matrix element has three parts: (1) the hadronic tensor 
\begin{equation}
\mathcal{W^{\mu\nu}}=\sum_{spins}\mathcal{M}^{\mu}\mathcal{M}^{\nu} \, ,
\label{eqn:hadronic-tensor}
\end{equation}
which describes the $P\rightarrow P'\gamma^*$ transition, (2) the square of the photon propagator $1/q^2$ 
and (3) the lepton tensor 
\begin{equation}
L_{\mu \nu} = \sum_{spins}j_{\mu} j_{\nu} 
\end{equation}
accounting for the photon-dilepton pair conversion. 
The leptonic current
\begin{equation}
j^\mu=\bar{u}(k_1)\gamma^\mu \upsilon(k_2)
\end{equation}
%
is a vector current coupling to photons and contains the respective Dirac wave functions $u(k_1)$ and $\upsilon(k_2)$ of leptons with four momenta $k_1$, $k_2$ and mass $m_l$.
The leptonic tensor can explicitly be calculated as
\begin{equation}
L^{\mu\nu} = 4(k_1^{\mu}k_2^{\nu}+k_2^{\mu}k_1^{\nu} -(k_1\cdot k_2+m_l^2)g^{\mu\nu})\,,
\end{equation}
and also controls the proper infrared cut-off due to the lepton phase space.

The hadronic tensor includes all information about the interaction between the involved hadrons and the coupling to the virtual photon produced in the radiative process. 
The structure of the hadron-photon vertex is encoded in electromagnetic Transition Form Factors (eTFF) which depend on the four-momentum transfer $q$. 
In case of dilepton decays with $q^2>0$ the eTFF's are defined in the time-like region. 
In the space-like region ($q^2<0$ ) eTFF's can be studied via electron scattering experiments. 
In view of dilepton spectroscopy most interesting are time-like eTFF for light neutral mesons, which have been extracted with good precision from dilepton invariant mass distributions of Dalitz decays. 
Similar results on baryon Dalitz decays are very scarce.
The results will be reviewed in the remainder of this section. 

As it has already been stressed in Sec.\,\ref{sec:pheno}, photon polarization affects the lepton angular distributions. 
Indeed, the general form of the transition matrix element $\mathcal{T}$ depends on the polarization of the virtual photon and  includes non-trivial angular dependencies of the emitted leptons.  
To better visualize this dependence one can rewrite the matrix element (Eq.\,\ref{eqn:mes-matrix-el}) by introducing spin density matrix elements~\cite{Choi:1989yf,Speranza:2016tcg} as
\begin{equation}
 \sum_{pol} \mid \mathcal{T} \mid^2 =\frac{e^2}{q^4}\sum_{\lambda,\lambda'}\rho^{had}_{\lambda,\lambda'}\rho^{lep}_{\lambda,\lambda'}\, ,
 \label{eqn:spin-density}
\end{equation} 
where the hadron $\rho^{had}_{\lambda,\lambda'}$ and the lepton $\rho^{lep}_{\lambda,\lambda'}$ spin density matrix elements are defined in similar fashion as the hadronic and leptonic tensors.
They can be written using the four-vector of the virtual photon $\epsilon^\mu (q,\lambda)$ with the helicities $\lambda=\pm 1$, $\lambda=0$ and the four-momentum $q$ as 
\begin{eqnarray}
\rho^{lep}_{\lambda,\lambda'} &=& \epsilon^\mu (q,\lambda) \, L_{\mu,\nu} \, \epsilon^\nu (q,\lambda')^{*}\, , \\
\rho^{had}_{\lambda,\lambda'} &=& \epsilon^\mu (q,\lambda) \, \mathcal{M}_{\mu,\nu} \, \epsilon^\nu (q,\lambda')^{*}\,.
\end{eqnarray}
The asymmetry coefficients given in Eq.\,\ref{eqn:ang-distr-full} can be expressed as combinations of the spin density parameters (see \myeg \cite{Speranza:2016tcg}).
While the hadronic part ($\rho^{had}_{\lambda,\lambda'}$) depends in general on the reaction dynamics and on the type of transition (spin and parity), the leptonic part can be calculated explicitly in QED. 
In the virtual photon rest frame (leptons are emitted back-to-back) it is given by~\cite{Speranza:2016tcg}
\begin{equation}
\label{eqn:ang-distr-hel}
\rho^{lep}_{\lambda,\lambda'} =
4\mid \mathbf{k} \mid ^2 \times
 \left( 
 \begin{array}{ccc}
1+\cos^2 \vartheta_e + a & -\sqrt{2} \cos\vartheta_e \sin \vartheta\, e^{-i\varphi_e} & \sin^2\vartheta_e\, e^{-2i\varphi_e} \\
-\sqrt{2}\cos\vartheta_e \sin\vartheta\, e^{i\varphi_e} & 2(1-\cos^2\vartheta)+a &\sqrt{2}\cos\vartheta_e \sin\vartheta\, e^{-i\varphi_e} \\
 \sin^2\vartheta_e\, e^{2i\varphi_e} & \sqrt{2} \cos\vartheta_e \sin\vartheta e^{i\varphi_e}& 1+\cos^2\vartheta_e + a \,,
 \end{array} 
 \right)
\end{equation} 
where $a=2m_\ell^2 / \mid\mathbf{k}\mid^2$ and $\mathbf{k}$ is the momentum vector of one of the two leptons.
The polar ($\vartheta_e$) and the azimuthal ($\varphi_e$) angles are given in the rest frame of the virtual photon with respect to a quantization axis (\mycf  Fig.\,\ref{fig:ref-frame}). 
In this review we will use the helicity reference frame defined such that the $z$ axis points along the direction of the virtual photon in the CM frame .
The diagonal elements $\rho_{11}=\rho_{-1-1}$ 
correspond to transversely polarized photons ($\sim 1+cos^2\vartheta_e$), while the $\rho_{00}$ ($\sim sin^2\vartheta_e$) to the longitudinally polarized ones. 
As we shall see below, these components appear in the expressions for the angular distributions of electrons emitted in Dalitz decays.
\subsubsection{Vector Meson Dominance}
\label{sec:VDM}
Successful theoretical approaches describe the radiative hadron decays by applying the Vector Meson Dominance Model (VDM), which was already mentioned as an important concept for the description of thermal radiation of QCD matter~({\mycf Sec.\,\ref{sec:dense}}).
Vector Meson Dominance has been suggested by Sakurai already in the sixties~\cite{Sakurai:1960ju} and provides a very intuitive picture of the hadron-photon interaction. 
The conjecture is that the coupling is actually mediated by the light vector mesons \textrho, \textomega\ and \textphi, which carry the same spin and parity as the photon and play the role of an interpolating field between strong and electromagnetic interactions.   
With their intrinsic quark structure they act as the microscopic electric dipole to which the photon couples, very much like nucleon-hole excitations in the description of the giant iso-vector dipole resonance of nuclei.
According to VDM the hadronic electromagnetic current can  straightforwardly be written as  
\begin{equation}
\mathcal{M}^\mu=J^\mu = e \sum_V \frac{m_V^2}{g_{V}} V^{\mu}\,,
\label{eqn:vdm}
\end{equation}
where $V^{\mu}$ are the vector meson fields \textrho, \textomega, \textphi) and $g_{V}$ and $e^2 = 4\pi \alpha$ ($\hbar, c = 1$) are the vector meson and the electromagnetic coupling constants, respectively. 
This current operator is acting on the QCD vacuum and creates antiquark-quark pairs overlapping with the vector meson states. 
An important conjecture of VDM is that the strong interaction between vector mesons and other hadrons is proportional to exactly the same coupling constant $g_V$ as defined in Eq.\,\ref{eqn:vdm}. 

Another important aspect of VDM is interferences between the amplitudes for different intermediary vector meson states. 
Besides the well known effect of the $\rho/\omega$ interferences in the \epem$\rightarrow$\pip\pim\ channel entering the pion electromagnetic form factor~\cite{Wolfe:2009ts}, much stronger effects due to intermediary baryonic resonances have been predicted for the dilepton invariant masses below the vector meson poles for pion induced reactions ~\cite{Lutz:2001mi,Lutz:2005yv,Kampfer:2002ze}. 
The latter one is closely connected to Dalitz decays of low-mass $N^*$, $I=1/2$, $\Delta,I=3/2$ baryon resonances as will be discussed below~(Sec.\,\ref{sec:dalitz-decays}).
In heavy-ion collisions at beam energies of a few~\agev\ temperatures reached in the fireballs are about a factor two lower than at SPS or collider energies.
Consequently such effects may play an important role since baryon resonances may dominate over mesonic excitations (\mycf Fig.\,\ref{fig:freezout_densities}).  
%
\subsubsection{Two body decays of vector mesons}
\label{sec:two-body}
The partial decay width $d\Gamma$ for a two body decay of a particle with mass $M$ into particles $a$ and $b$ is given as \cite{Olive:2016xmw}
\begin{equation}
d\Gamma = \frac{1}{32\pi^2}\mid \mathcal{T}(V\rightarrow \ell^+\ell^-)\mid^2 \frac{p}{M^2} d\Omega
\label{eqn:2body-gamma}
\end{equation}
where
\begin{equation}
p = \frac{\left((M^2 -(m_a+m_b)^2)(M^2-(m_a-m_b)^2)\right)^{1/2}}{2M}\,
\label{eqn:mom}
\end{equation}
is the electron momentum in the CM frame and  $(p/M^2)d\Omega$ is the phase space factor including the solid angle $ d\Omega = d\phi \, d(\cos\theta)$ of one of the decay particles.
The partial decay width of the vector mesons \textrho,\textomega and \textphi\ into a pair of leptons with masses $m_{\ell^+}, m_{\ell^-}$ is calculated as two-step process of the meson-virtual photon transition and its subsequent conversion to a dilepton pair. 
Thus the matrix element (Eq.\,\ref{eqn:mes-matrix-el}), averaged over the vector polarization and electron spins, factorizes as shown in~\cite{Koch:1992sk,Li:1996mi,Ko:1996is} to 
%
\begin{eqnarray}
\mid \mathcal{T}(V\rightarrow \ell^+\ell^-) \mid^2 & = &
 \mid \mathcal{W}(V \rightarrow\gamma^*)\mid^2 \cdot \frac{4}{M^4} \cdot \mid L(\gamma^*\rightarrow \ell^+ \ell^-) \mid^2 \\
& = &  \frac{16\pi^2\alpha^2}{3} \frac{M_V^4}{g_{V}^2} (M^2+2m_\ell^2)\frac{1}{M^4} \,.
\label{eqn:T-fact}
\end{eqnarray}
Note that $q^2 = M^2$ with $M$  the invariant mass of the dilepton. 
The matrix element 
\begin{equation*}
\mid \mathcal{W}(V\rightarrow \gamma^*)\mid^2=4\pi \frac{\alpha}{g^2_{V}} M_V^4
\end{equation*}
is given by (Eq.\,\ref{eqn:vdm}) at the meson pole $M_V$ . 
The conversion of the virtual photon to a lepton pair is written as  
\begin{equation}
\mid L(\gamma^*\rightarrow \ell^+\ell^-)\mid^2=\frac{4\pi\alpha}{3}(M^2+2m^2_\ell) \,.
\end{equation}
Finally, using  Eq.~(\ref{eqn:2body-gamma}), the dilepton decay width  $\Gamma(M)$  can be obtained as a function of the dilepton mass as
\begin{equation}
\Gamma(M)_{\ell^+\ell^-}  = 
\frac{4\pi\alpha^2}{3g^2_{V}}\frac{M_V^4}{M^3}\left(1-\frac{4m_\ell^2}{M^2}\right)^{1/2}\left(1+\frac{2m_\ell^2}{M^2}\right)\,.
\label{eqn:gamma-vdm}
\end{equation}
Table~\ref{tab:VM-decays} summarizes the total decay widths, the branching ratios to dileptons and the coupling constants $g_V$ of the light vector mesons defined at the pole positions ($M=M_V$).
\begin{table}
\begin{center}
\begin{tabular}{ |c|c|c|c|c| } 
\hline
 & Decay & BR & $\Gamma$ [\mev ] & $g_{\mathrm{V}}$ \\ 
\hline
\textrho\   &   \epem\ &  $(4.72 \pm 0.05)\cdot 10^{-5}$ & $149.1\pm 0.8$
& $5.03$  \\
$\rho $   &   \mumu\  &   $(4.55 \pm 0.28)\cdot 10^{-5}$ & &  \\
$\omega $  & \epem\ & $(7.28 \pm 0.14)\cdot 10^{-5}$ &$8.49\pm 0.08$ & $17.1$ \\
\textomega\   & \mumu\ & $(9.0 \pm 3.1)\cdot 10^{-5}$ & & \\
\textphi\ &  \epem\ &  $(2.954 \pm 0.03)\cdot 10^{-4}$ & $4.27 \pm 0.03$ & $-12.9$ \\
\textphi\ &  \mumu\ &  $(2.87 \pm 0.19)\cdot 10^{-4}$ & &  \\
\hline
\end{tabular}   
\end{center} 
\caption{Branching ratios for dilepton decays, total widths $\Gamma$ and coupling constants $g_\mathrm{V}$ of light vector mesons. \label{tab:VM-decays}}
\end{table}
One should note that the ratios of the coupling constants $g_V$, extracted from the measured branching ratios~\cite{Faessler:1999de}, are in good  agreement with the SU(3) flavor symmetry predictions: $g_{\rho}:g_{\omega}:g_{\phi}= 1:3: -3/\sqrt(2)$. 
Hence, in the context of VDM, they can be expressed in terms of a single universal vector meson-photon coupling constant $g_{\gamma}\simeq 5.6$. 

The shape of the dilepton invariant mass
distribution $dN/dM_{\ell^+\ell^-}$ of the two-body decay of the broad \textrho\ meson depends on the production process.
This is an important aspect as it will become evident later in the discussion of the dilepton spectra obtained in hadronic collisions.  
In general, the mass distribution $dN/dM$ of a resonance can be approximated by the relativistic Breit--Wigner formula, multiplied with $dN_p(M)/dM$ quantifying the number of initial states populating the resonance at mass $M$ as
\begin{equation}
\label{eqn:VM-2decay}
\frac{dN}{dM} = 
\frac{dN_p(M)}{dM} 
\frac{\Gamma_{in}\Gamma_{\ell^+\ell^-}M^2}{(M^2-M_\mathrm{V}^2)^2+M^2 \Gamma_\mathrm{tot}^2}\,.
\end{equation}
Here, $\Gamma_{in}$ and $\Gamma_{\ell^+\ell^-}$ account for the partial widths related to the production and the decay process, respectively.  
The number of states $dN_p/dM$ depends on the production process and available phase space.
In a baryon dominated hadron resonance gas \textrho\ mesons are likely produced via excitation and decay of baryonic resonances.
For such a two-step production process 
NN$\rightarrow$N$^\star$(\textDelta)$\rightarrow$NN\textrho$\rightarrow$NN$\ell^+\ell^-$ 
the mass distribution of the baryon resonance will strongly affect the dilepton mass distribution. 
Indeed, low mass resonances like N$^*(1520)$ or \textDelta(1720) have a significant branching to the N\textrho\ channel.
The production of \textrho\ mesons in such two-step processes appears as a special type of (baryon) Dalitz decay as will be discussed in Sec.\,\ref{sec:dalitz-decays}. 
In an equilibrated hadron resonance gas such baryon resonances will be excited with preference of the lower-side of their mass distribution.  
This limits also the phase space for the decay and in turn favors decays populating the low-mass tail of the \textrho\ meson, an effect, which is further enhanced by the $1/M^3$ dependence of decay width $\Gamma_{\ell^+\ell^-}$ (\mycf Eq.\,\ref{eqn:gamma-vdm}).

One should note, however, that the latter only holds for strict VDM, which assumes that the coupling $g_V$ is constant.
For narrow resonances like the \textomega\ and the \textphi\ meson such an assumption is justified but the situation is more complex in case of the broad \textrho\ meson state.
For example, the extended VDM model of~\cite{OConnell:1995nse} uses two coupling schemes for the hadron-photon transition: a direct coupling to photons and a coupling via intermediate vector mesons which vanishes at $q^2=0$. 
Consequently, a linear dependence on the dilepton mass is predicted~\cite{Effenberger:1999nn}.

An important channel for dilepton production in heavy-ion reactions at ultra-relativistic energies is the annihilation of two pions in the fireball. 
The respective elementary cross section has been calculated using a Breit--Wigner resonance description  in~\cite{Koch:1992sk} and \cite{Ko:1996is}: 
%
\begin{eqnarray}
\label{eqn:vdm-M3}
\sigma_{\pi\pi\rightarrow\rho\rightarrow \ell^+\ell^-} (M)
&=& \frac{4\pi}{p_{\pi}^2}\frac{M\,\Gamma_{\pi\pi}\,M\,\Gamma_{\ell^+\ell^-}}{(M^2-M^2_{\rho})+M^2\Gamma_{\mathrm{tot}}^2}\\
&=&\frac{8\pi\alpha^2 p_{\pi}}{3M^3}\frac{M_{\rho}^4}{(M^2-M_{\rho}^2)^2+M^2\Gamma^2_{\mathrm{\mathrm{tot}}}(M)} 
\left(1-\frac{4m_\ell^2}{M^2}\right)^{1/2}\left(1+\frac{2m_\ell^2}{M^2}\right) \\
&=&\frac{4\pi}{3}\frac{\alpha^2}{M^2}\left(1-\frac{4m_{\pi}^2}{M^2}\right)^{1/2}\left(1-\frac{4m_\ell^2}{M^2}\right)^{1/2}\left(1+\frac{2m_\ell^2}{M^2}\right)\mid F_{\pi} \mid ^2\,,
\end{eqnarray}
%
It demonstrates the dependence of the cross section on the  \textpi\textpi center of mass collision energy $\sqrt{s}=M$ with $F_{\pi}$ denoting the eTFF of charged pions and 
\begin{equation*}
\Gamma_{\pi\pi}=\frac{g^2_{\rho}}{2\pi}\frac{p^3_{\pi}}{M^2}
\end{equation*}
defining the two pion decay width of the \textrho\ meson ($\Gamma_{\ell+\ell^-}$ is given by Eq.\,\ref{eqn:gamma-vdm}). 
The proper threshold behavior is guaranteed by $\Gamma_{\pi\pi}$ as it cuts off the dilepton mass distribution from pion annihilation at $2m_{\pi}$.  

On contrary, for dileptons from the previously discussed production mechanism involving baryon resonances such a cut-off does not exist since the mass of the \textrho\ meson is generated from the excitation energy of the baryonic resonance.
These production mechanisms have been applied in the transport codes GiBUU~\cite{Weil:2012ji}, UrQMD~\cite{Endres:2015fna}, (P)HSD~\cite{Bratkovskaya:2007jk,Bratkovskaya:2013vx}, RQMD~\cite{Cozma:2006vp,Santini:2008pk} and SMASH~\cite{Weil:2016fxr}.
The threshold behavior is particularly visible in calculations carried out for SIS18/BEVALAC energies. 
We will return to this aspect in the discussion of the data (\mycf Sec.\,\ref{sec:status}).
%
%

For dilepton decays of vector mesons lepton angular distributions were calculated in~\cite{Gottfried:1964nx,Falciano:1986wk,Faccioli:2011pn} and expressed, in agreement with Eq.\,\ref{eqn:mes-matrix-el}, by
\begin{equation}
\label{eqn:vector-mesons-helicity}
\frac{d\sigma}{d\Omega}\propto 1 + \lambda_\vartheta\, \cos^2\vartheta_l + \lambda_{\vartheta\phi}\, \sin2\vartheta_l\, \cos\varphi_l + \lambda_\varphi\, \sin^2\vartheta_l\, cos 2\varphi_e\,.
\end{equation} 
The anisotropy coefficients $\lambda_\vartheta,\lambda_{\vartheta,\varphi}$ and $\lambda_{\varphi}$ are related to the spin density matrix elements (\mycf Eq.\,\ref{eqn:ang-distr-hel}).
Compared to Eq.\,\ref{eqn:ang-distr-full}, the coefficients related to the angle relative to the scattering plane vanish. 
For example, in the case of pion annihilation and the quantization axis oriented in the direction along the pion momentum, the angular distribution integrated over azimuthal angle further simplifies to \cite{Speranza:2016tcg}:
\begin{equation}
\label{helicty_a_distr}
\frac{d\sigma}{d\Omega}\propto (1+cos^2\vartheta_e)(\rho_{-1,-1}+\rho_{1,1})+2(1-cos^2\vartheta_e)\rho_{0,0}\propto \rho_{0,0} sin^2\vartheta_e \,.
\end{equation}
Note that the only non-vanishing element of the spin density matrix is $\rho_{00}$. 
Hence, the only relevant asymmetry coefficient is $\lambda_{\vartheta}=-1$ (longitudinal polarization). 
For the Drell-Yan process similar considerations lead to  $\lambda_{\vartheta}=+1$, \myie the virtual photon is fully transversely polarized.  
%
\subsubsection{Dalitz decays}
\label{sec:dalitz-decays}
The decay of a particle with mass $M$ into a three particle final state with
momenta $\vec{p}_i$ and masses $m_i$
can be expressed by its partial decay width $d\Gamma$, which is determined by the Lorentz invariant matrix element $\mathcal{T}$ \cite{Olive:2016xmw}
\begin{equation}
d \Gamma = \frac{1}{(2\pi)^5}\frac{1}{16M^2} \mid \mathcal{T} \mid^2\, \mid \Vec{p}_1^{\,*}\mid \; \mid \Vec{p}_3\mid dm_{12}\, d\Omega_1^{\,*} d\Omega_3\,.
\label{eqn:dalitz-decays}
\end{equation}
In this expression $\Vec{p}_1^{\,*}$ and $d\Omega_1^{\,*}$ define the momentum and solid angle element of particle~1 in the rest frame of the two-body subsystem (1-2), respectively. 
The invariant mass of this subsystem is $m_{12}$ and the momentum $p_1^{\,*}$ can be calculated with Eq.\,\ref{eqn:mom} by setting $M=m_{12}$. 
The solid angle element $d\Omega_3$ is defined in the rest frame of the decaying mother particle with mass $M$ into particle 3 and the two-particle subsystem (1-2), while the momentum $p_3^{\,*}$ can also be obtained from Eq.\,\ref{eqn:mom}.  
All three momenta of the daughter particles lie in a plane.  
For dilepton decays ($P \rightarrow P' \ell^+ \ell^-$) it is convenient to relate the system $1-2$ to the dilepton (equivalently to the virtual photon $\gamma^*$) with the invariant mass $M_{\ell^+\ell^-}$. 
Next, assuming that the decay is independent of the azimuthal emission angle of $\gamma^*$ \mywrt the decay plane, one can introduce three angles to characterize the decay process: the polar angle related to the emission of the virtual photon and the polar and azimuthal angles describing the lepton decay axis in the $\gamma^*$ rest frame (\mycf Eq.\,\ref{eqn:vector-mesons-helicity}). 
The latter two are related to the polarization of the virtual photon and are in particular important for the characterization of the decay source.

\vspace{2ex}
\noindent
\textit{Meson decays}\\[8pt]
%
Dileptons from Dalitz decays of scalar mesons make up a significant fraction of the LMR.
Electron pairs with masses below 140~\mevcc\ come dominantly from \piztodal.
Strong contributions to dileptons with masses up to the 500~\mevcc\ originate from  \etatodal.
At even higher masses pairs from \etatodal\ as well as from the vector mesons \omegatodal\ and \phitodal\ contribute, yet with less strength. 
In the calculations of the respective transition amplitude (Eq.\,\ref{eqn:mes-matrix-el}) the $P\to P'$\textgamma$^\star$ vertex for the decays of interest is identical for pseudoscalar-vector,vector and vector-pseudoscalar, vector transitions. 
In both cases the respective hadronic current (\mycf Eq.\,\ref{eqn:hadronic-tensor}) can be written as~\cite{Landsberg:1986fd}
\begin{equation}
\label{eqn:meson-delitz-tranistion}
\mathcal{M^\mu(P\rightarrow P'\gamma^*)}=f_{PP'}(q^2)\varepsilon^{\alpha\beta\gamma\mu} p_\alpha q_\beta \epsilon_\gamma
\end{equation}
where $p_\alpha$ and $q_\beta$ are the four-momenta of the daughter particle $P'$ and the virtual gamma \textgamma$^\star$, respectively.  
The polarization of the vector particle is denoted with  $\epsilon_\gamma$ and $\varepsilon^{\alpha\beta\gamma\nu}$ is the totally antisymmetric tensor. 
The function $f_{PP'}(q^2)$ is the 
eTFF, which describes the virtual photon-hadron vertex and encodes the electromagnetic structures of the involved hadrons. 
For a point-like particle it is just a constant while for extended objects it generally depends on the four momentum transfer which, in the time-like region, is equivalent to the dilepton invariant mass $M_{\ell^+\ell^-} = \sqrt{q^\beta q_\beta}$. 
%
\begin{figure}[tbh]
\begin{minipage}{\linewidth}
\includegraphics[width=0.32\linewidth,height=0.35\linewidth]{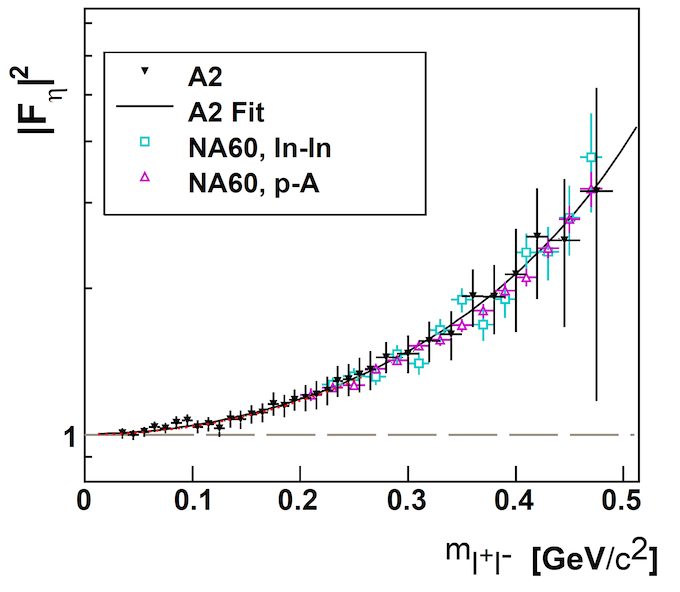}
\includegraphics[width=0.32\linewidth,height=0.35\linewidth]{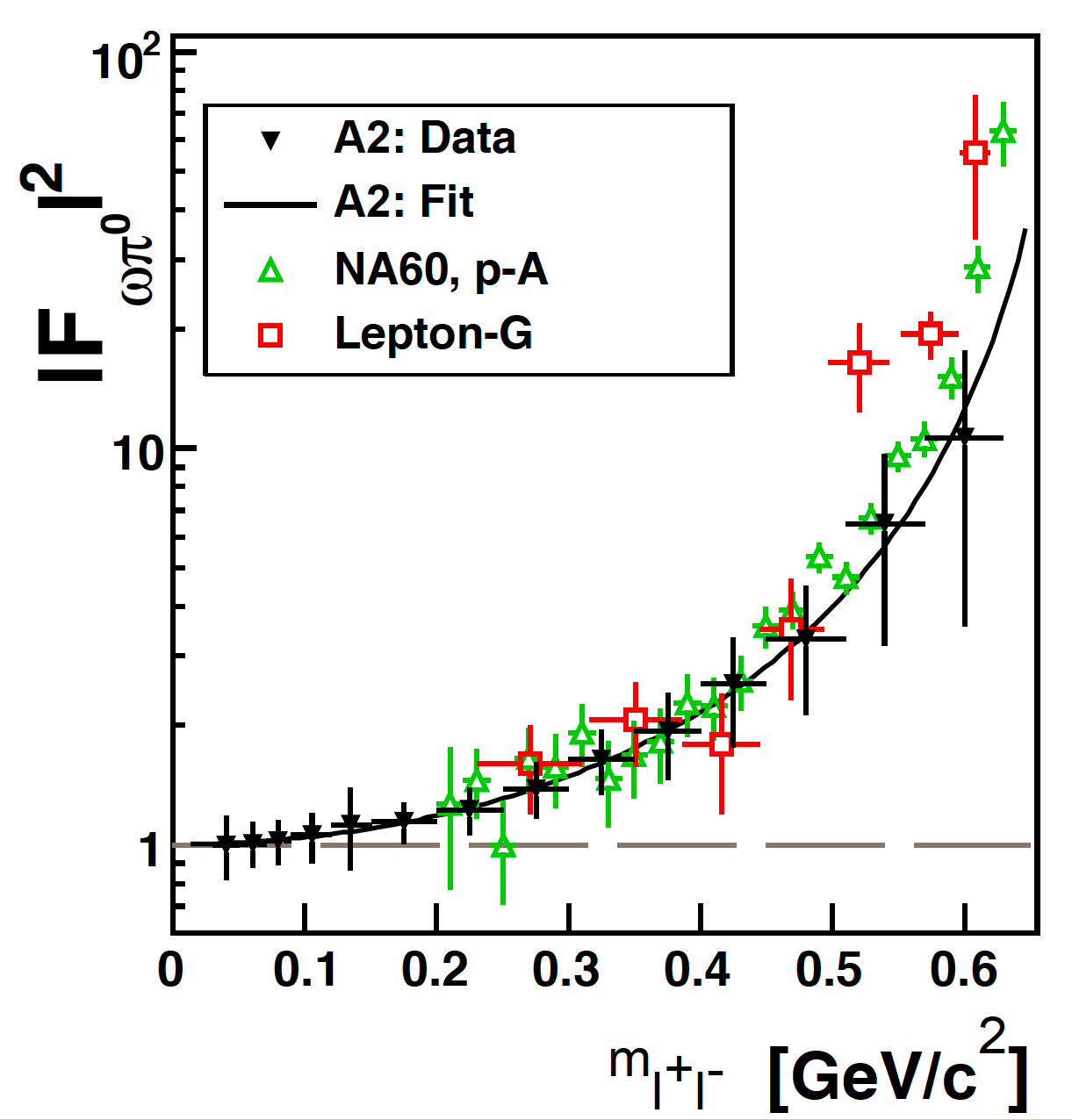}
\includegraphics[width=0.3\linewidth,height=0.33\linewidth]{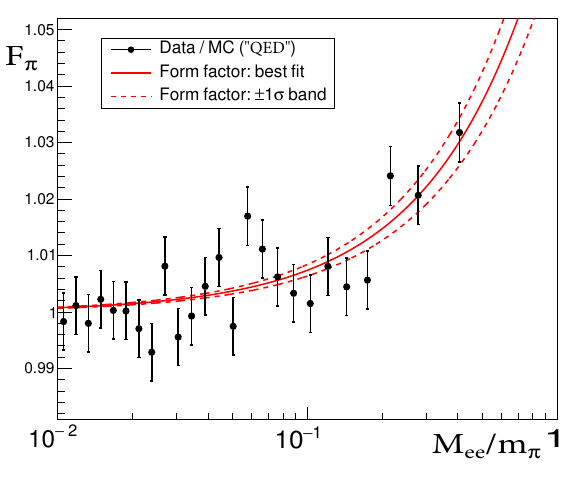}
\caption{\label{fig:dalitz-meson} Mass dependence of eTFF's derived from measurements of the dilepton Dalitz decays of \texteta\ (left), \textomega\ (middle) ~\cite{Adlarson:2016hpp,Arnaldi:2016pzu} and \piz\ (right) ~ \cite{TheNA62:2016fhr} mesons. Lines are results of fits using Eq.\,\ref{eqn:VDM-FF-a}.
 performed at MAMI (A2 collaboration)~\cite{Adlarson:2016ykr, Adlarson:2016hpp}, SPS/NA60~\cite{Arnaldi:2016pzu} and NA62~\cite{TheNA62:2016fhr} for \texteta\, \textomega\ and \piz\ mesons together with older data from Lepton-G.
}
\end{minipage}\hspace{2pc}%
\end{figure}
Therefore, measurements of the invariant mass of dileptons from Dalitz decays enable extracting this important quantity from data. 
The differential decay widths (Eq.\,\ref{eqn:dalitz-decays}) can be calculated using a similar factorization scheme as given by~Eq.\,\ref{eqn:T-fact} (see~\cite{Landsberg:1986fd,Koch:1992sk,Faessler:1999de} for details):
%
\begin{equation}
d\Gamma(P \rightarrow P' \ell^+ \ell^-)  =  
\frac{\alpha}{3\pi} d\Gamma\left(P \rightarrow P' \gamma^*\right) 
\left[1-\frac{4m^2_l}{M^2_{\ell^+\ell^-}}\right]^{1/2}\left[1+\frac{2m^2_l}{M^2_{\ell^+\ell^-}}\right]\frac{dM^2_{\ell^+\ell^-}}{M^2_{\ell^+\ell^-}}\,.
\label{eqn:dgamma-dm}
\end{equation}
%
Using the invariant transition matrix element $\mathcal{T}$ given by Eq.\,\ref{eqn:mes-matrix-el}, and after normalization to the partial width for a decay with a real photon in the final state, one arrives to
%
\begin{align}
\label{eqn:meson-dalitz-transition}
\frac{d\Gamma(P\rightarrow P' \ell^+\ell^-)}{d M^2_{\ell^+\ell^-} \Gamma(P \rightarrow P' \gamma)} & =
\frac{\alpha}{3\pi} 
\left[1- \frac{4m^2_l}{M^2_{\ell^+\ell^-}}\right]^{1/2}
\left[1+2\frac{m^2_l}{M^2_{\ell^+\ell^-}}\right] \,
\frac{1}{M^2_{\ell^+\ell^-}} \; \\ \nonumber
& \times 
\left[\left( 1+\frac{M^2_{\ell^+\ell^-}}{M^2-M^2_{p'}}\right)^2 
 - \frac{4M^2M^2_{\ell^+\ell^-}}{(M^2-M^2_{p'})^2}\right]^{3/2}
\left\vert \frac{f_{PP'}(M^2_{\ell^+\ell^-})}{f_{PP'}(0)}\right\vert^2
\end{align}
%
which can be written in a compact way as
\begin{equation}
\frac{d\Gamma(P\rightarrow P' \ell^+\ell^-)}{d M^2_{\ell^+\ell^-} \Gamma(P \rightarrow P' \gamma)} = [QED]\times F^2_{pp'}(M^2_{\ell^+\ell^-})\,.
\end{equation}
The function $F_{PP'}(M^2_{\ell^+\ell^-})$ is the eTFF normalized to its value at $M^2_{\ell^+\ell^-}=0$ corresponding to the emission of a real photon (\myie at the photon point).  
With the help of the VDM the eTFF can be expressed as weighted sum of the vector meson's Breit--Wigner distributions over all low-mass vector mesons with mass $M_V$ 
\begin{equation}
F_{PP'}(M^2_{\ell^+\ell^-}) =
\left[\sum_V \frac{g_{PP'V}}{g_{V}} \frac{M^2_V}{M^2_V-M^2_{\ell^+\ell^-}-i\Gamma_V M_V}\right] \,
\left[\sum_V \frac{g_{PP'V}}{g_{V}}\right]^{-1}\, ,
\label{eqn:VDM-FF}
\end{equation} 
where the weights are the ratio of the coupling strength' $g_{PP'V}$ and $g_{V}$ characterizing the $P\rightarrow P'V$ and $V\rightarrow \gamma^*$ vertices, respectively. 
They can be fixed from quark models, as described in~\cite{Landsberg:1986fd} (note that Eq. 3.10 in \cite{Landsberg:1986fd} uses constants with extra factor 2) or experimental data on meson decays. 
In the limit of small width' and using the pole approximation, the expression  equation\ref{eqn:VDM-FF} can be expanded in terms of $q^2$ to arrive at the known monopole form  
\begin{equation}
F_{PP'}(M^2_{\ell^+\ell^-}) \simeq \frac{1}{1-q^2/\Lambda^2} \simeq 1+\frac{q^2}{\Lambda^2}= 1 + \frac{1}{6} q^2\expval{r^2_{PP'}}^{1/2}\,.
\label{eqn:VDM-FF-a}
\end{equation} 
\begin{table}[tb]
\label{tab:meson-dalitz}
\begin{center}
\begin{tabular}{ |c|c|c|c|c| } 
\hline
& Decay & BR & $\Lambda^{-2}$ (G$e$V$^-2$) & Reference \\ 
\hline
\piz\   &   \epem\textgamma &  $(1.174 \pm 0.035)\cdot 10^{-2}$
&  $2.020\pm 0.031$ & \cite{TheNA62:2016fhr} \\
\texteta   &   \epem\textgamma  &   $(6.9 \pm 0.4)\cdot 10^{-3}$ & $1.97\pm 0.13$ & \cite{Adlarson:2016hpp} \\
\texteta   & \mumu\textgamma & $(3.1 \pm 0.04)\cdot 10^{-4}$  &  $1.934\pm0.084$ & \cite{Arnaldi:2016pzu}\\
\texteta' & \mumu\textgamma & $(1.08 \pm 0.27)\cdot 10^{-4}$ &  $1.7\pm 0.4$ & \cite{Landsberg:1986fd}\\
\texteta' & \epem\textgamma & $(4.73 \pm 0.3)\cdot 10^{-4}$ &  $1.6\pm 0.19$   & \cite{Ablikim:2015wnx}\\
\textomega\ &  \epem\piz &  $(7.7 \pm 0.6)\cdot 10^{-4}$ & $1.99 \pm 0.22$ & \cite{Adlarson:2016hpp}\\
\textomega\ &  \mumu\piz &  $(1.3 \pm 0.4)\cdot 10^{-4}$ &    $2.223\pm 0.045$ & \cite{Arnaldi:2016pzu}\\
\straightphi\   &  \epem\piz    & $(1.35 \pm 0.11)\cdot 10^{-5}$ &    $2.02\pm0.11$ & \cite{Anastasi:2016qga} \\
\hline 
\end{tabular}
\end{center} 
\caption{Summary decay parameters extracted from Light Meson Dalitz decays. For a definition see text. \label{tab:dalitz-table}}
\end{table}
The root mean square radius $\expval{r^2_{PP'}}^{1/2}$ defines a characteristic size of the transition which reflects the hadron size and $\Lambda^{-2}$ is the derivative of the eTFF \mywrt $q^2$ at the photon point.
The above approximation can be applied in the time-like and space-like regime.
Fig.\,\ref{fig:dalitz-meson} shows results of recent high precision measurements of eTFF of mesons  and fits to the data using Eq.\,\ref{eqn:VDM-FF-a}.  
The extracted values for~$\Lambda^{-2}$ are summarized in Tab.\,\ref{tab:meson-dalitz}. 
They generally confirm a growth of the eTFF as a function of the invariant mass and agree well with the prediction of VDM (given in Tab.~4 of \cite{Landsberg:1986fd}), except for the \textomega\ and \textphi\ mesons. 
A particularly significant disagreement can be seen for the \textomega\ meson.

State of the art calculations of eTFF's is performed using various approaches: effective Lagrangians ~\cite{Terschlusen:2012xw,Terschluesen:2010ik}, Dyson-Schwinger formalism ~\cite{Roberts:2000aa,Weil:2017knt}, quark models~\cite{Ramalho:2012ng,Ramalho:2016zgc} and dispersion relations~\cite{Hoferichter:2014vra,Schneider:2012ez,Hanhart:2013vba} (for a recent review see also \cite{Czerwinski:2012ry}). 
Recently, substantial progress has been made in extracting eTFF's from lattice-QCD calculations, both on the mesonic (pionic)~\cite{Meyer:2011um,Erben:2017hvr} and baryonic sector~\cite{Stokes:2017otq,Capitani:2015sba}. 
In particular the \textomega\ eTFF has triggered a lot of interest due to apparent violation of strict VDM. 
The description has been significantly improved by leading order chiral Lagrangian calculations~\cite{Terschluesen:2010ik} and dispersive theories~\cite{Schneider:2012ez}, all of which still failing to account for its strong rise at very large $q^2$.

A different motivation to study eTFF's is related to physics beyond the standard model.
The results from measurements of the muon $(g-2)$ differ \mywrt the Standard Model predictions~\cite{Jegerlehner:2009ry,Nyffeler:2016gnb}. 
It appears that the eTFF's of light pseudo-scalar mesons enter in the hadronic light-by-light scattering calculation and, together with the vacuum polarization, are currently the most prominent part of systematic uncertainties of Standard Model predictions (\cite{Czerwinski:2012ry}). 
Furthermore, Dalitz decays constitute an irreducible background in searches for a possibly existing dark photon, an extra U(1) gauge boson which mediates the interaction between Standard Model photons and dark matter. 
Several measurements were performed searching for signals from two-body dilepton decays of the dark photon in meson and baryon Dalitz-decays and in elementary and heavy-ion collisions, with negative outcome but providing important limits on its coupling strength to the SM photon (see~\cite{Denig:2016dgi} for further information).

The angular distributions of the leptons in light neutral meson Dalitz-days can be calculated using Eq.\,\ref{helicty_a_distr}. 
For such transitions the spin density matrix $\rho^{had}_{\lambda,\lambda'}$ contains only the elements $\rho_{-1-1}$ and $\rho_{11}$ and consequently only contributions from transversely polarized photons are selected.
This leads to the general expression for the differential cross section defined in the helicity frame  
\begin{equation}
\label{eqn:dalitz-helicity}
\frac{d\sigma}{d\Omega}\propto p_\pi^2 \left[2m_\ell^2 + \mid \mathbf{k} \mid^2 (1+cos^2\vartheta_l)\right]\,.
\end{equation}
\begin{figure}[tbh]
\includegraphics[width=0.45\textwidth,height=0.45\textwidth]{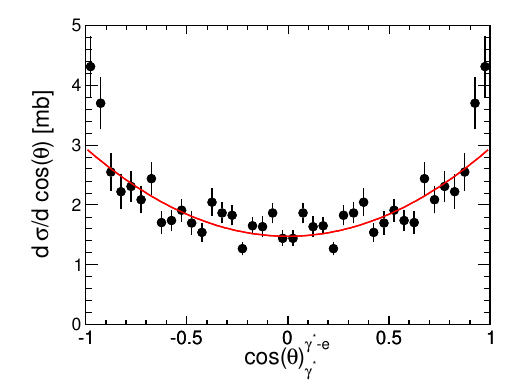}
\hfill
\includegraphics[width=0.45\textwidth,height=0.45\textwidth]{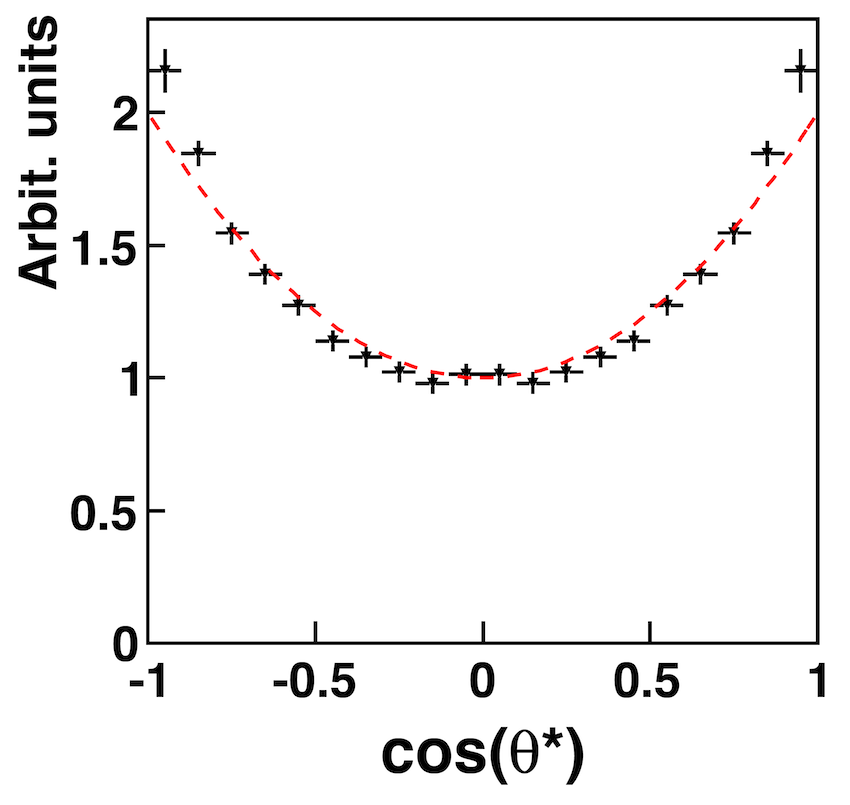}
\caption{Electron angular distributions in the helicity frame for the $\pi^0\rightarrow e+e-\gamma$ (left) \cite{Adamczewski-Musch:2017hmp} and the $\eta\rightarrow e^+e^-\gamma$~ (right)~\cite{Adlarson:2016hpp} Dalitz decays. Red lines denote the  $(1+B*cos^2\vartheta_e)$ anisotropy. 
}
\label{fig:helicity-mm}
\end{figure}
Experimental results for the decays \piztodal\ \cite{Adlarson:2016ykr,Adamczewski-Musch:2017hmp} and \etatodal\ \cite{Adlarson:2016hpp,HADES:2012aa}, shown in Fig.\,\ref{fig:helicity-mm}, corroborate the expected $1+cos^2\vartheta_l$ distribution (please note that for electrons one can neglect the electron mass term in Eq.\,\ref{eqn:dalitz-helicity}).

\vspace{2ex}
\noindent
\textit{Baryon decays}\\[8pt]
%
The matrix element for a radiative transition of an excited baryon to its ground state $\mathcal{M}$(R$\rightarrow$N\textgamma$^\star$) can be expressed as coherent sum of helicity amplitudes $A_{3/2}(q^2)$, $A_{1/2}(q^2)$ and $S_{1/2}(q^2)$, defined in the resonance decay frame~\cite{Krivoruchenko:2001jk,Aznauryan:2011qj}. 
The amplitudes describe transitions for a resonance of given spin $J$ and helicity $\lambda^*$ to a nucleon with the helicity $\lambda=\pm 1/2$ and a photon with  $\lambda_\gamma=1,0$. 
The projection axis is defined to be along the photon momentum and hence $\lambda^*=-\lambda+\lambda_\gamma>$ (please note that one can consider only $\lambda_\gamma \ge 0$ due to conserved parity). 
The first two amplitudes are related to the transverse photon polarization ($\lambda_{\gamma}= 1$) while the last one is related to the longitudinal photon polarization ($\lambda_\gamma=0$). 
Transverse helicity amplitudes have been determined in the space-like region for several resonances in electro- and photo-production experiments. 
However, they are unknown in the time-like region which is only accessible by their Dalitz decays R$\rightarrow$N\textgamma$^\star$ (for a review see~\cite{Aznauryan:2011qj}). 
The respective differential decay width $d\Gamma(\mathrm{R}\rightarrow \mathrm{N}\mree)/dM_{\ell^+\ell^-}$ can be expressed similar to Eq.\,\ref{eqn:dgamma-dm} but here with the R$\rightarrow$N\textgamma$^\star$ given as a function of three covariant or Magnetic ($G_M(q^2)$), Electric ($G_E(q^2)$) and Coulomb ($G_C(q^2)$) eTFF's.  
The form factors are related to the helicity amplitudes discussed earlier (respective relations can be found in~\cite{Krivoruchenko:2001jk}). 
The relations for the $\Delta(1232)$ resonance were introduced by Jones and Scadron~\cite{jones:1973} and for higher mass resonances by Devenish, Eisenschitz and K\"{o}rner~\cite{devenish}. 
It should be noted that the Coulomb form factor couples only to the $S_{1/2}$ while the electric and the magnetic form factors couple to all three helicity amplitudes.

The partial decay widths $d\Gamma$(R$\rightarrow$N\textgamma$^\star$)  for a resonance with mass $M_R$ and a spin $J \ge 3/2$ or $J =1/2$, as well as  with normal ($J^P=1/2^-,3/2^+,..$) and abnormal parities ($J^P=1/2^+,3/2^-,..$), are given by Krivoruchenko and F\"{a}ssler in~\cite{Krivoruchenko:2001jk}.
Equivalent expressions based on covariant form factors were obtained by Zetenyi and Wolf~\cite{Zetenyi:2002jy}).
The expressions read:
%
\begin{align}
\label{eqn:dalitz-baryons}
  d\Gamma (&\mathrm{R}_{J \ge 3/2} \rightarrow \mathrm{N}\mathrm{\gamma}^*) 
   = F(M_R,M_{\ell^+\ell^-})^{\pm} \\ \nonumber
  \times & \left( \frac{l+1}{l}  \, |G_{M/E}^{\pm}(M_{\ell^+\ell^-})|^2  +  (l+1)(l+2)\,|G_{E/M}^{\pm}(M_{\ell^+\ell^-})|^2+\frac{M^2_{\ell^+\ell^-}}{M_R^2}|G_C(M_{\ell^+\ell^-})|^2 \right) \,,
\end{align}
\begin{align}
 F(M_R,M_{\ell^+,\ell^-})^{\pm} & =  \frac{9\alpha}{16} \, \frac{(l!)^2}{2^l\,(2l+1)!}\; \frac{(M_R \pm M_{N})^2}{M_R^{2l+1} M^2_N}\; \Big[(M_R \pm M_N)^2 - M^2_{\ell^+\ell^-}\Big]^{l-1/2} \\ \nonumber
& \times \Big[(M_R \mp M_N)^2 - M^2_{\ell^+\ell^-}\Big]^{l+1/2} 
\end{align}
and
\begin{align}
d\Gamma(\mathrm{R}_{J = 1/2} \rightarrow \mathrm{N}\gamma^*) & =   H(M_R,M_{\ell^+\ell^-})^{\pm} \, (2\,|G_{E/M}^{\pm}(M_{\ell^+\ell^-})|^2 +\frac{M_{\ell^+\ell^-}^2}{M_R^2}|G_C(M_{\ell^+\ell^-})|^2) \,, 
 \\
H(M_R,M_{\ell^+\ell^-})^{\pm} & = \frac{\alpha}{8\,M_R}\; \Big[(M_R \mp M_N)^2 - M^{2}_{\ell^+\ell^-}\Big]^{1/2} \times \Big[(M_R \pm M_N)^2 - M^{2}_{\ell^+\ell^-}\Big]^{1/2}\,.
\end{align}
%
Note that for $J=1/2$ only two eTFF are defined and that there is no contribution from the Coulomb form factor at $M_{\ell^+\ell^-}^2=0$. 
The $+$ and $-$ signs stand for resonances with normal and abnormal parity, respectively ($P=P_{intr}(-1)^l$, $l=0$ for $J=1/2$ and  $l=J-1/2$ for $J\ge/3/2$). 
The functions $F$ and $H$ depend on the resonance mass ($M_R$) and the virtual photon mass $M_{\ell^+\ell^-}$ \cite{Krivoruchenko:2001jk}). 
For a point like particle the form factors $G_{M/E}$ are constant and can be fixed at the $M_{\ell^+\ell^-}^2=0$ (photon point) from the existing data. %
\begin{figure}[tbh]
\begin{center}
\includegraphics[width=0.5\textwidth,height=0.5\textwidth]{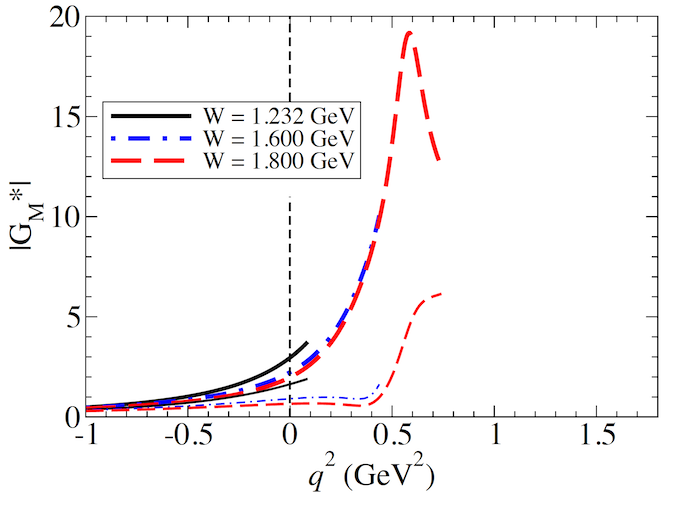}
\includegraphics[width=0.46\textwidth,height=0.46\textwidth]{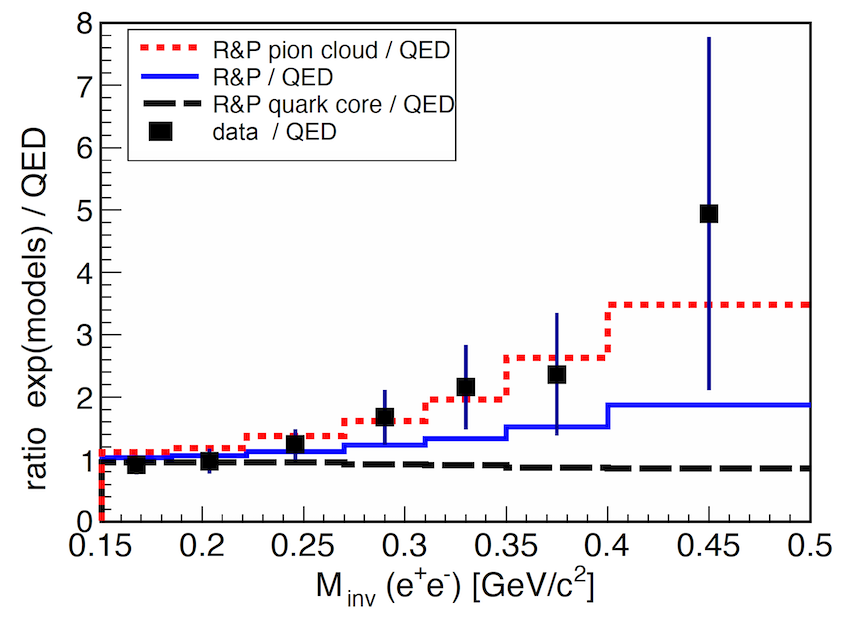}
\end{center}
\caption{\label{fig:GM-Delta} (Left panel) Dependency of $G_M$ on $M^2_{\ell^+\ell^-}$ for the  \textDelta(1232)$\rightarrow$R\textgamma$^\star$ transition for various \textDelta masses obtained from the covariant quark model of \cite{Ramalho:2015qna}.
(Right panel) Ratio of the measured dielectron yield to model predictions  as a  function of the invariant mass for \textDelta(1232)$\rightarrow$p\epem\ (see text and \cite{Adamczewski-Musch:2017hmp} for details).}
\end{figure}
%

In such cases, the mass dependence of the decay width is given by the functions $F$ and $H$ and can be regarded as "QED" reference, as in case of the meson decays. 
One should however note, as pointed out in \cite{Krivoruchenko:2001hs}, that there were earlier also other expressions used in model calculations of dilepton production which provided inconsistent results.
The influence of various expressions on $d\Gamma(\mathrm{R}\rightarrow$N\epem) and the parametrization of resonance mass distributions for dielectron spectra have been investigated in detail for the $\Delta(1232)$ in~\cite{Bratkovskaya:2013vx}. 

Actually, the $q^2$ dependence of the eTFFs is still an open issue as well from the theory as from the experiment viewpoint. 
For example, calculations presented in~\cite{Schafer:1994tz} assume strict VDM for the \textDelta(1232)$\rightarrow$N\textgamma$^\star$ transition and a dominance of the \textrho\ meson in the baryon--photon coupling, which strongly affects the mass dependence of the eTFF. 
However, this approach overestimates the radiative decay widths of baryonic resonances. 
The extended VDM (eVDM) developed in~\cite{Krivoruchenko:2001jk,Faessler:2000md} includes the coherent contribution of several excited vector meson states (like \textrho$^{'}$, \textrho$^{''}$ and \textomega$^{'}$) to calculate the decay transition rates.   
The model has several parameters which have been fixed by the quark counting rules, by available data on helicity amplitudes in the space-like region and based on known branching rations for R$\rightarrow$N\textrho\ decays. 
This approach allows to reduce the otherwise too large branching ratios of the R$\rightarrow$N\textgamma\ transitions computed with strict VDM models. 

The covariant constituent quark model by Ramalho and Pe\~na~\cite{Ramalho:2012ng,Ramalho:2015qna} provides a description with a dominant $G_M$ form factor for the \textDelta(1232)$\rightarrow$N\textgamma$^\star$ transition while the other two form factors come out to be small, in accordance with data obtained in the space like region. 
The eTFF is build from two contributions originating form the quark core and the pion cloud. 
The quark core component describes the resonance as a quark-diquark system in $S$-wave with a small admixture ($1\%$) of a $D$ state. 
This contribution is determined from lattice QCD and also in agreement with the data in the space-like region at high $q^2$. 
The comparison with time-like data visualizes the role of the meson cloud component which is dominantly of pionic type.
The contribution of the pion cloud in the time-like region is parameterized using two terms: a photon, either directly coupling to a pion or to an intermediate baryon state. 
The eTFF of the pion is fixed from available data.
In their most recent work~\cite{Ramalho:2015qna}, in contrast to previous implementations \cite{Frohlich:2009eu,Ramalho:2012ng} based on the Iachello--Wann model developed for the nucleon eTFF~\cite{Iachello:2004aq, Bijker:2004yu}, the correct position of the meson pole and the width has been included. 
A similar model has been applied for the N$^\star$(1520) resonance \cite{Ramalho:2016zgc}. 
The striking feature of both calculations is an important and increasing role of the pion cloud contribution in the time-like region. 

For the \textDelta(1232) a rapid enhancement of the $|G_M|^2$ as a function of the dielectron invariant mass is predicted, particularly significant for higher resonance masses (see Fig.\,\ref{fig:GM-Delta} left panel). 
The pion cloud plays a dominant role, as can be seen from the comparison to the separately plotted (thin lines) quark-core contribution. 
The latter provides an almost constant strength below the \textrho\ meson pole which rapidly increases when the invariant mass approaches the pole region. 
In the same figure (right panel) the first experimental data from HADES on the \textDelta(1232)$\rightarrow$p\epem\ transition are shown \cite{Adamczewski-Musch:2017hmp}.
Plotted is the ratio of the measured dielectron yield to the one expected for a decay of a point-like $\Delta(1232)$ (denoted as "QED") as a function of the invariant mass. 
An increase of the ratio with increasing mass is clearly visible, similarly to the observation for Dalitz decays of mesons discussed in the previous section. 
The data are also compared to predictions of the covariant quark model of Ramalho-Pe\~na \cite{Ramalho:2015qna} (blue solid line). 
The increase in the measured ratio can be interpreted, within this model, as predominately due to pion a cloud effect (red dotted line) since the core contribution (black dashed line) shows an almost flat behavior.
%
\subsection{Nucleon-nucleon bremsstrahlung}
\label{sec:N-N-brems}
%
In the most general case \nnnn\ (nucleon-nucleon) bremsstrahlung can be defined as a process associated with production of a photon (real or virtual) in a \nnnn\ interaction, regardless of the number of particles in the final state. 
Ultra-relativistic \nnnn\ collisions are governed by partonic interactions and can result in very hard photons through $q\bar{q}$ annihilation (Drell-Yan) or to very complex states with charged particle multiplicities reaching rapidity densities in excess of ten and even giving rise to thermal radiation. 
Such contributions from hard processes are usually not subsumed under the term bremsstrahlung. 

At much lower collision energies, in the range of a few \gev\ beam energy per nucleon, the dominant contribution to dielectron production from \nnnn~collisions originate from the $\Delta(1232)$ isobar Dalitz decay as discussed in the previous section.
However, also other amplitudes contribute to the scattering process NN$\rightarrow$ NN\textgamma$^\star$.  
In the non-resonant ("quasi-elastic") process one or both nucleons go off-shell for some time after or before the photon is emitted. 
Their virtuality is lifted by strong interaction between the nucleons, conveniently calculated in the one-boson exchange approximation. 
Such bremsstrahlung is suppressed in \pp\ over \np\ collisions due to the absence of a dipole moment in the time-dependent electromagnetic field generated by the colliding protons. 
In microscopic transport model calculations ``quasi-elastic'' bremsstrahlung is often calculated using the soft photon approximation~\cite{Gale:1987ki, Gale:1988yk}. 
It assumes photon emission following elastic nucleon-nucleon interactions with an appropriate phase space modification induced by the produced virtual photon; any interference processes and emission from a meson exchange line are neglected.
Contributions from the $\Delta$ isobar and higher resonances are treated as separate processes and added incoherently as an extra source of lepton pairs.

Various calculations of \nnnn\ bremsstrahlung were performed in the framework of the One Boson Exchange model in the nineties~\cite{Schafer:1994tz,Schafer:1994vr,deJong:1996ej}, all treating the resonant and non-resonant channel coherently.
The results show a dominance of $\Delta$ contributions and the importance of interference terms originating from various meson exchange graphs. 
Moreover, the calculations demonstrate a strong sensitivity of the dielectron yield on off-shell electromagnetic form factors of nucleons, and on the eTFF of the $\Delta(1232)$. 
Also the modeling of the nucleon-nucleon interaction requires form factors to dress the respective vertices and also various nucleon-meson coupling constants, which all need to be fixed from experimental data.
While the coupling constants and hadronic form factors can be extracted from meson production data reasonably well, the electromagnetic form factors can only be inferred from dilepton data which are very scarce. 

The comparison of model calculations to dielectron data from \pp\ collisions generally confirms the dominance of $\Delta$ Dalitz decays for dilepton invariant masses above the pion mass. 
In \pp\ collisions only a tiny contribution originates from non-resonant radiation~\cite{Adamczewski-Musch:2017hmp}. 
This situation changes dramatically in case of \np\ collisions where the non-resonant \nn\ bremsstrahlung contribution plays a significant role~\cite{Agakishiev:2009yf}.
There are two reasons for that. 
First, the non-vanishing dipole moment and, 
second, the additional channel of charged pion exchange not possible in \pp\ collisions. 
The internal charged current gives rise to substantial extra radiation.
Moreover, the respective eTFF at the photon (internal) pion vertex can introduce significant changes in the dielectron invariant mass distribution towards the \textrho\ meson pole mass.
The respective one-boson-exchange (OBE) calculations differ in how gauge invariance is implemented, in particular in graphs related to the production of lepton pairs from the internal (charged) pion exchange line. 
This particular mechanism of dielectron production has been  realized already in the $90s$ \cite{deJong:1996ej}. 
The role of charged pion exchange was emphasized by more recent calculations of Kaptari and K\"{a}mpfer \cite{Kaptari:2005qz} who predicted cross sections by a factor 2-4 larger for the ``quasi-elastic'' contribution as compared to former results. 
After implementing these newly obtained cross sections for bremsstrahlung in transport calculation performed by the HSD group \cite{Bratkovskaya:2007jk} it became possible to explain the long standing``DLS puzzle'' \cite{Porter:1997rc}, related to unexplained dielectron yield measured in low energy light ion collisions.
However, the subsequent calculations of Shyam and Mosel, using a similar effective Lagrangian model but a different recipe to conserve gauge invariance, did not confirm such a large increase of the "quasi-elastic" contribution but stressed the importance of off-shell \textrho\ contributions in eTFF of the internal charged pion ~\cite{Shyam:2003cn,Shyam:2008rx}. 

In order to differentiate between various calculations and mechanisms of dilepton production in \nnnn\ collisions information on the lepton angular distributions are very valuable. 
As already discussed for the Dalitz decays of neutral meson, the distribution of the lepton decay angle \mywrt the \textgamma$^\star$ axis (in the helicity frame) is expected to show a $1+B\,cos^2(\vartheta_l)$ distribution with $B=1$ (see Eq.\,\ref{eqn:dalitz-helicity}).
OBE calculations of~\cite{Bratkovskaya:1995my} also predict significant anisotropies of the electron distributions from \nnnn\ collisions.
For the $\Delta$ Dalitz decay a similar anisotropy as for the \piz\ Dalitz with $B \sim 1$, with some slight dependence of $B$ on the dielectron mass, was obtained.  
This can be understood as being due to a dominance of transverse photon polarization. 
On the other hand, for the non-resonant neutron--proton bremsstrahlung a smaller anisotropy of $B\sim 0.4$, decreasing with increasing invariant mass, was predicted. 
The only experimental results on electron angular distributions from \nnnn\ bremsstrahlung were recently obtained by the HADES experiment and will be discussed in (\mycf Sec.\,\ref{sec:sis-bevalac}).
\subsection{Dileptons from hard processes}
\label{sec:hard-processes}
With increasing collision energy partonic processes start to dominate the initial \nnnn\ reactions. 
As a result, the collision partners can fragment into quark diquark states and the final states are characterized by an increasing number of produced particles.
A particular process of relevance for dilepton production is the annihilation of a quark with an antiquark, each from one of the two collision partners.
The cross section for this Drell-Yan process is readily written down in case of very high collision energies, where the parton-parton annihilation can be calculated perturbatively. 
The cross section for a purely electromagnetic transition of two quarks (partons) can be inferred from the respective Feynman diagram of an annihilation process into a dilepton (one-photon exchange):
\begin{equation}
\frac{d\hat{\sigma}}{dQ^2} =
\frac{4 \pi \alpha^2}{3 Q^2} e^2
\delta\left( Q^2 - \hat{s}\right) \,.
\end{equation}
Here $\hat{\sigma}$ is the ``off-shell'' cross section to produce a dilepton with squared invariant mass $Q^2 = \hat{s}^2$ and $\alpha$ is the fine structure constant.
To arrive at a cross section for production in a \pp\ collision one uses the factorization concept of QCD.
Along this concept, the annihilation is treated as hard process while the probability to find a parton with given momentum fraction $x$, determined by non-perturbative QCD, is given by the experimentally accessible parton distribution (structure) functions~\cite{Collins:1989gx}.
In that way the parton-parton (off-shell) cross section has to be weighted with the probabilities to find partons with the proper momenta $p_{q}$ and $p_{\bar{q}}$ given by the respective structure functions $f_q$ and $f_{\bar{q}}$  
\begin{equation}
\frac{d \sigma}{dQ^2} =
\beta_c
\sum_{q}
\int dx \int dy \:
f_q \left( x \right)
f_{\bar{q}} \left( y \right)  
\frac{d\hat{\sigma}}{dQ^2}\,.
\end{equation}
Here $x$ and $y$ are denoting the momentum fractions carried by the quark and antiquark, respectively. 
Then, the $q\bar{q}$ collision energy can be expressed in terms of the \pp\ center of mass energy $\sqrt{s}$ as $\hat{s} = (xp_1 +yp_2)^2 \simeq xys$. 
The fact that the Drell-Yan cross section at high energies is determined solely by a QED cross section times the structure functions makes this process a formidable tool to investigate the internal structure of nucleons.
The comparison of such leading order results with experiment have revealed that higher order effects do play a role.
This has led to the introduction of so-called K-factors, included in the calculation to force agreement with data. 
More care has to be taken in the evaluation of Drell-Yan cross sections in $p+A$ and $A+A$.
Not only the structure functions are modified when nucleons are embedded in a medium, but also additional effects occur like gluon shadowing and a $p_\perp$ enhancement caused by parton multiple-scattering (Cronin effect).
The precise knowledge of in-medium partition functions, in particular those of the gluons, is of great importance to determine the initial conditions of an ultra-relativistic heavy-ion collision (\mycf Sec.\,\ref{sec:initial} (see \myeg \cite{Peng:2014hta} for a recent review on this topic).
 
Both hard processes important for dilepton continuum physics in the IMR, open-charm production and Drell--Yan, can be calculated using PYTHIA \cite{Sjostrand:2000wi} tuned in \pp\ collisions and scaled by the number of binary collisions. 
The correlated charm production can also be  determined experimentally by measurements of $D$-meson production.
Unfortunately, such measurements of correlated $D$ meson production are not easy because of their short decay path and have so far been obtained at SPS only by the NA60 experiment. 
The production of open heavy-flavor hadrons has been measured via their hadronic and semi-leptonic decays at mid- and forward-rapidity in p--p collisions at the RHIC and LHC. 
The results on $D$ meson production are in agreement (within factor 2) with pQCD calculations performed FNOLL, PYTHIA , POWHEG.  
Nuclear modification factors for charm mesons, identified by their hadronic decay channels, have been recently also measured at LHC in \pbpb\ collisions by ALICE \cite{Acharya:2018hre} (for recent overview of LHC results on heavy flavor see \cite{Wilkinson:2018cpq}, ).
An analysis of dilepton emission in \auau\ collisions at \sqsnn{200} based on STAR data obtained with the Heavy Flavor Tracker (HFT) has been ongoing at the time this review was written. 
The HFT should help to reduce the contribution of correlated open charm decay while at the same time it produced additional conversion background due to its material budget.
Excellent tracking performance of the TPC can help to identify electrons and positrons emerging from conversion in the HFT. 

At beam energies below around \sqsnn{10} perturbative approaches are difficult as ever higher-order terms have to be taken into account.
As of today, the calculation of prompt dilepton radiation from initial semi-hard\footnote{We distinguish semi-hard process from hard processes where perturbative calculations are applicable and soft processes, which are driven by baryon resonance production and meson exchange.} scattering processes has not been investigated from first principles.
An attempt to provide a consistent picture of particle production going from soft to hard processes is realized with the NEXUS project, introduced in Sec.\,\ref{sec:initial} and  described in \cite{Drescher:2000ha,Drescher:2000ec}.
However, radiative (electromagnetic) processes have not been treated yet. 

The TAMU group has investigated possible contributions of semi-hard processes to the dilepton yield by studying \textrho\ meson production in a schematic jet-quenching model~\cite{vanHees:2007th}.
The initial transverse momentum spectrum of \textrho\ mesons is assumed to follow a power law with
\begin{equation}
\frac{1}{p_\perp}
\frac{d N_{\mathrm{prim.}}}{d p_{\perp}} =
A ( 1 + B p_\perp )^{-a}\,.
\end{equation}
Here $A$, $B$ and $a$ are parameters adjusted to \nnnn~ scattering data.
Second, the transverse momentum distribution is modified to implement the Cronin effect for which the parametrization is adjusted such as to reproduce measurements of direct photon production in $p+A$ collisions. 
Next, the fraction of \textrho\ mesons which are destroyed by re-scattering before they can leave the surrounding medium is evaluated.
For this, hadronic as well as pre-hadronic cross sections and the local particle density are used. 
Finally, the remaining \textrho\ mesons are decayed into dileptons according to vacuum branching ratios. 
Such a model can account for the contribution of hard dileptons around the \textrho\ pole mass as observed \myeg by the NA60 experiment~\mycf Sec.\,\ref{sec:sps}.

Non-equilibrium models with partonic degrees of freedom can trace the dilepton emission from initial state dynamics consistently by following the collision history of the partons explicitly. 
In the pHSD model, the initial state parton collisions are obtained from the event generator PYTHIA~\cite{Sjostrand:2006za}.
Leading partons can re-scatter and strings either hadronize according to the rules of PYTHIA or melt into partons if the energy density of the surrounding medium exceeds 0.5~\gev ~fm$^3$.
Partons are treated as dynamical quasi-particles with spectral functions depending on the medium properties.
Dileptons are mainly produced in bremsstrahlung-like $qg$ or $\bar{q}g$ quark-gluon scatterings or in $q\bar{q}$ annihilation processes. 
For details see~\cite{Linnyk:2010vb}.
Since the partons are dressed in this approach, the respective dilepton spectrum from $\bar{q}q$ annihilation is cut off towards small invariant masses.
\subsection{Thermal radiation}
\label{sec:thermal-radiation}
In this section we turn to a different concept of calculating the dilepton emission from strongly interacting processes.
The goal is to describe radiation not based on single collision or decay process but to take thermal averages of such processes.
As a result one obtains so-called emissivities, that is an expression for the number of dileptons emitted with given four-momentum ($d^4q$) and per unit of four-volume ($d^4x$) out of strongly interacting matter with fixed microscopic composition (or equation of state).

Dilepton radiation rates from QCD matter in thermal equilibrium for a perturbative gas of quarks and gluons have been first derived in ~\cite{McLerran:1984ay} and later also in \cite{Strickland:1994rf}. 
The rates per unit of the four-volume can be calculated in accordance with the general expression for coupling leptons to the electromagnetic current of hadrons (\mycf Eq.\,\ref{eqn:mes-matrix-el}) as:
\begin{equation}
\frac{dN^{\ell\ell}}{d^4 x}=-4e^4\int \frac{d^3 k_1}{(2\pi)^32E_1}\frac{d^3 k_2}{(2\pi)^32E_2}L_{\mu\nu}(q)\frac{1}{q^4}W^{\mu\nu}(q)\,. 
\label{eqn:rad-rate}
\end{equation}
The leptonic tensor is integrated over the momenta of the final state leptons with $q=k_1+k_2$ and $M^2 = q^2 =q_\nu q^\nu$, which, for momentum transfers substantially larger than the lepton's rest mass, can then be written in a compact form as~\cite{Rapp:1999ej}
\begin{equation}
L_{\mu\nu}(q) \frac{1}{q^4} = -\frac{\alpha^2}{3\pi^3 M^2}\left(g_{\mu\nu}-\frac{q_{\mu}q_{\nu}}{M^2}\right)\,.
\end{equation}
Note that the virtual photon propagator has been combined with the lepton tensor in this expression. 
The hadronic tensor $W^{\mu\nu}$ contains all information about the partonic medium and is obtained as thermal average (indicated by the double brackets) of the electromagnetic current-currents correlator~\cite{Shuryak:1993kg} assuming essentially $q\bar{q}$ annihilation in an ideal gas of quarks and gluons:
\begin{equation}   
W^{\mu\nu} = \int d^4 x\,e^{-iqx}  \left<\left< j^\text{em}_{\mu}(x)\,j^\text{em}_{\nu}(0) \right>\right>\,.
\label{eqn:current-current}
\end{equation}
The thermal average is obtained by integrating over the momenta of all possible initial state partons and applying the proper Boltzmann weights for quarks and gluons in a medium with given temperature and chemical potential. 
For $q^2\gg m_\text{e}$, the spectral rate can be expressed as
\begin{equation}   
\frac{dN^{\ell\ell}}{d^4x\,dq^4} =
\frac{\alpha^2}{12\pi^3\,M^2}
W^\mu_{\:\:\mu}(q) \,
\label{eqn:parton-rates}
\end{equation}
where $W^\mu_\mu(q)$ is the reduced scalar correlator~\cite{Klingl:1996ps}.

For all practical purposes in the description of thermal radiation emitted from heavy-ion collisions it is sufficient to take into account the three lightest quarks only.
The (vector) current then reads
\begin{equation}   
j^{em}_{\mu} = 
\frac{2}{3} \bar{u} \gamma_{\mu} u -
\frac{1}{3} \bar{d} \gamma_{\mu} d -
\frac{1}{3} \bar{s} \gamma_{\mu} s
\,,
\label{eqn:partonic-current}
\end{equation}
where the factors in front of the bi-linear forms are the electric charges of the respective quarks.
\begin{figure}[tb]
\begin{minipage}[c]{0.5\textwidth}
\includegraphics[width=\textwidth]{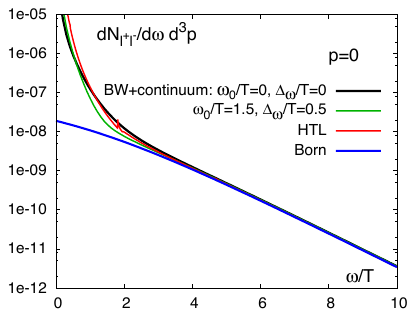}
\end{minipage}
\hspace{2pt}
\begin{minipage}[c]{0.4\textwidth}
\includegraphics[width=\textwidth]{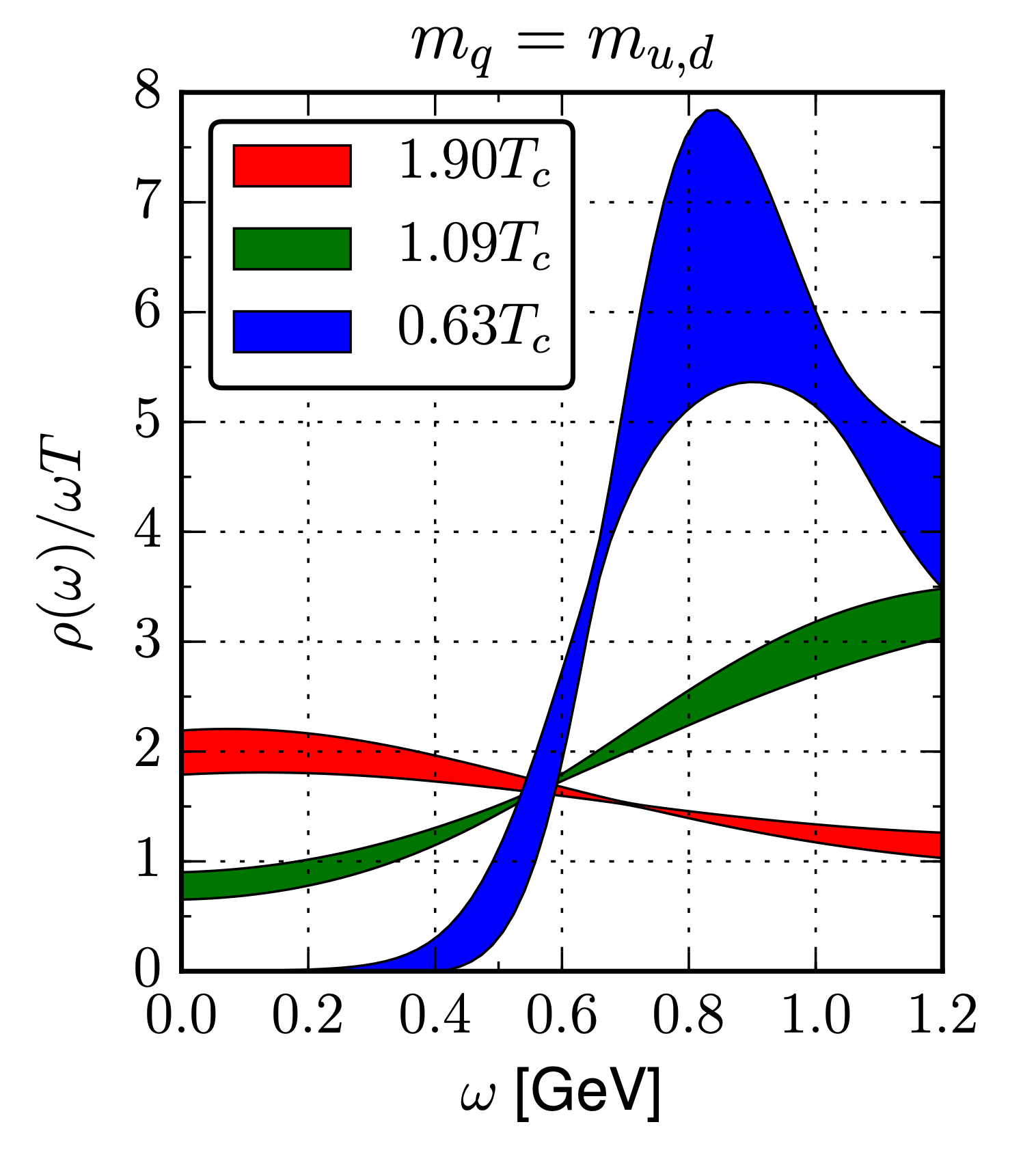}
\end{minipage}
\caption{
Left: thermal dilepton rates as a function of the (dimensionless) dilepton energy $\omega / T$ (with $\omega^2 = M^2 + \vec{p}^2$) calculated on the lattice and compared to rates obtained from effective theories.
The lattice results are derived from fitting the spectral distributions with a Breit--Wigner plus continuum contribution. 
The legend, from top to down, refers to (a) lattice calculation without a cutoff on the continuum contribution and (b) with a large cut-off providing an upper limit acceptable. 
(c) is a calculation in the hard-thermal-loop approximation (with thermal quark mass $m_T=T$) and (d) uses the Born approximation (free vector spectral function) \cite{Ding:2010ga}.
Right panel: Spectral functions $\rho(\omega)/\omega T$, obtained from lattice QCD for the light quarks for three temperatures indicated in the legend~\cite{Aarts:2014nba}.
}
\label{fig:thermal-lattice}
\end{figure}
Evaluating the thermal average for an ideal plasma of quarks and gluons at vanishing chemical potential (\myie matter antimatter symmetric) and finite temperature 
yields 
\begin{equation}
\dfrac{dN^{\ell\ell}}{dM^2} = R_{\mathrm{part.}} \frac{\alpha^2}{6\pi^2} 
M^2 T K_1 \left( M/T \right)\,,    
\label{eqn:plasma-emissivity}
\end{equation}
where the Bessel function $K_1$ results from the integration over the 3-momentum of the dilepton.
Furthermore, $M$ is the dilepton invariant mass, $T$ the temperature of the emitting thermal source and 
\begin{equation}
R_{\mathrm{part.}} = N_c \sum_{f=u,d,s} e_f^2 = 3 \left(\frac{4}{9}+\frac{1}{9}+\frac{1}{9} \right)
\label{eqn:partonic-ratio}
\end{equation}
is representing the squared charges of the thermalized quark-gluon phase (shown as reference in Fig.\,\ref{fig:thermal-lattice} as ``Born'' approximation). 
The thermal average has been taken by working out the respective phase-space integrals of the initial and final states of all partons in the fireball, and by taking into account Bose-Einstein and Fermi--Dirac distribution functions as well as Pauli-blocking factors for the parton final states~\cite{McLerran:1984ay}. 
The microscopic properties of the QGP enter the emissivity of the QGP just through the number $R_{\mathrm{part.}}$. 
For $M/T \gtrsim 5$, corresponding to the IMR, the functional form given by Eq.\,\ref{eqn:plasma-emissivity} agrees within $5$\% with the description of black body radiation.
Thus, the detection of thermal radiation from the QGP provides a direct measurement of the mean temperature of the emitting medium.
The mean is taken over the 4-volume, \myie the respective volume $V(T)$ in which the matter is of partonic character. 
Due to the comparatively high temperature of the QGP phase this radiation of partonic origin dominates the thermal radiation in the IMR.

Fireballs formed in heavy-ion collision rapidly expand and by the time the system is thermalized the temperatures do likely not exceed a few hundred M$e$V.
It is understood that at such temperatures the plasma is not close to an ideal plasma. 
To investigate the impact of a (strong) coupling of the plasma on the emissivity, higher order effects have been studied (see \myeg \cite{Gelis:2002yw} for a concise overview ). 
A step forward in deriving the vector correlator in a partonic medium, beyond the perturbative approach, has been achieved by introducing so-called hard-thermal-loops (HTL) \cite{Braaten:1989mz,Braaten:1990wp}.
This concept overcomes the detailed book-keeping of Feynman diagrams,   otherwise needed to avoid singularities occurring as the invariant mass approaches zero. 
The vector correlator can also be evaluated on the lattice.
For that, the spectral properties of the correlator have to be extracted from the correlator calculated in Euclidean space--time.
The spectral function is then obtained from a Fourier transformation of the Euclidean correlator. 
Technically, this is achieved by employing Maximum Entropy Methods (MEM) as the correlator is only known on the intersections of the space--time grid.
Such calculations have been carried out \myeg by the Hot QCD collaboration within the quenched approximation, \myie using static quarks only~\cite{Ding:2010ga}.  
The resulting dilepton rates as a function of the dimensionless ratio of dilepton energy to temperature for virtual photon momentum $p=0$ are shown on the left panel of Fig.\,\ref{fig:thermal-lattice}.
The vector correlator has recently been calculated also with dynamical quarks for 2+1 flavors using anisotropic lattices~\cite{Aarts:2014nba}.
This calculation addresses a range of temperatures crossing the pseudo-critical temperature.
For temperatures below $T_{\mathrm{pc}}$, a clear resonance peak is obtained at around $800$~\mevcc for the vector correlator using $m_q = m_{u,d}$ (see Fig.\,\ref{fig:thermal-lattice} (right)).
Once the temperature of the radiating source drops below the deconfinement temperature $T_c$ (\mycf Sec.\,\ref{sec:freezeout}), the spectral distribution of the thermal radiation is modulated due to the presence of hadronic resonances  as depicted in the right panel of Fig.\,\ref{fig:thermal-lattice}. 

For a gas of hadrons, the hadronic current $J^{em}$ can be composed such as to reflect the dominant vector meson states (VDM) as
\begin{equation}
J^{\mu}_{em} = \frac{2}{3} \bar{u} \gamma^\mu u - \frac{1}{3} \bar{d} \gamma^\mu d - \frac{1}{3} \bar{s} \gamma^\mu s =
J^\rho + 
J^\omega + 
J^\phi\,.
\end{equation}  
The vector meson currents read
\begin{eqnarray}
J_{\rho}^{em} &=&
\frac{1}{2} \left(
\bar{u} \gamma^\mu u
 - \bar{d} \gamma^\mu d
\right)\\
J_{\omega}^{em} &=&
\frac{1}{6} \left(
\bar{u} \gamma^\mu u
 + \bar{d} \gamma^\mu d
\right)\\
J_{\phi}^{em} &=&
\frac{1}{2} \left(
\bar{s} \gamma^\mu s
\right)
\,.
\end{eqnarray}  
\begin{figure}[tbh]
\includegraphics[width=0.95\textwidth]{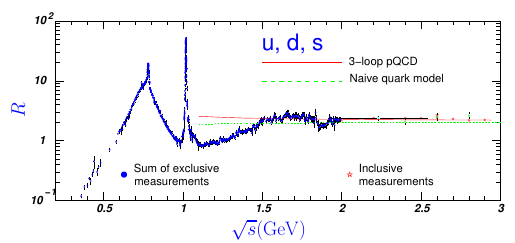}
\caption{\label{fig:muon-hadron} Cross section ratio of hadron to dimuon production in positron-electron annihilation~\cite{Rapp:2016xzw}. }
\end{figure}
For addressing the LMR it again suffices to take only the low-mass vector mesons into account. 
In an analogous procedure as above, the differential rate of virtual photons emitted from a gas of hadrons  is then given by:  
\begin{equation}
\dfrac{dN^{\ell\ell}}{d M^2} = R_{\mathrm{had.}} \frac{\alpha^2}{6\pi^2} 
M T K_1 \left( M/T \right)\,,    
\label{eqn:hadron-emissivity}
\end{equation}
where the factor 
\begin{equation}
R_{\mathrm{had.}} = 
\frac{
\sigma \left(e^+ e^- \rightarrow \mathrm{hadrons}\right)
}
{
\sigma \left(e^+ e^- \rightarrow \mu^+ \mu^-\right)
}
\propto 
\frac{1}{M^2} \Im\Pi^{em}
\end{equation}
in the region($M \lesssim 1.5$ \gev ) is reflecting the resonant structure of the hadronic vector current.
The latter is evident from the results of annihilation experiments as shown in Fig.\,\ref{fig:muon-hadron}.
The prominent resonant hadron production for \sqrts $\leq 1.5$~\gev~ is due to intermediary light vector mesons.  
Note that for the vector mesons composed of the light quarks the \textrho\ is dominating over the \textomega\.
Indeed, since the dilepton radiation from a fireball is proportional to $\Gamma_{l+l-}/\gamma \tau_{FB}$ the most dominant role plays the short-lived \textrho\ meson with the largest leptonic decay width (see Tab.\,\ref{tab:VM-decays}). 
For $M > 1.5$~\gevcc~the ratio approaches $R_{part.}$, consistent with the result of perturbative QCD (\mycf Eq.\,\ref{eqn:partonic-ratio}).

The most general expression for the emissivity of QCD matter is given by a 8-fold differential rate per four momentum and four volume unit
\begin{equation}
\frac{dR}{d^4qd^4x}=-\frac{\alpha^2}{\pi^3}\frac{L(M^2)}{M^2}\, \Im\Pi_{em} (M,q,T, \mu_b)\,f^{B}(q_0,T)
\label{eqn:rad-rate-bol}
\end{equation}
where now the spectral function, defined by the imaginary part of the current-current correlator $\Im\;\Pi_{em}$, is used.  
The latter is defined in analogy to the hadronic tensor (see Eq.\,\ref{eqn:current-current}) as time-ordered correlation function of the vector currents defined above
\begin{equation}
\Pi^{\mu\nu}(q)=i\int d^4xe^{iqx}\theta(x_0)<\Omega|[j^{\mu}(x),j^{\nu}(0)]|\Omega>    
\end{equation}
with $|\Omega>$ characterizing the medium.

Using strict VDM the spectral function in the hadronic phase can be expressed in terms of the vector meson retarded propagators $D_V$ as
\begin{equation}
\Im \Pi_{em}^{\mathrm{had.}} =
\sum_{V=\rho,\omega,\phi}
\left(
\frac{m_V^2}{g_V}
\right)^2
\Im D_V(M) \,.
\label{eqn:ratio-hadron}
\end{equation}  
The spectral function is weighted by the Bose factor $f^B(q_0,T)$ which takes care of the thermal ``bath'' from which the dileptons are radiated.
The term $1/M^2$ is the free (virtual) photon propagator and the quantity 
\begin{equation}
L(M^2)=\left(1+\frac{2m^2}{M^2}\right)\sqrt{1-\frac{4m_\ell^2}{M^2}}
\label{eqn:rho_propagator}
\end{equation}  
is the dilepton phase space factor .
This is an important result. 
Keeping in mind that the dominant vector current is $J_{\rho}$, the thermal dilepton rate in the LMR is essentially determined by the \textrho\ propagator.
In vacuum, the latter is dominated by the pion loop which gives rise to the dominant \textrho\ decay channel into a pair of charged pions. 
The time-reversed process is the formation of a \textrho\ meson through $\pi^+ \:\pi^-$ fusion. 
It is hence conceivable that the dilepton radiation from a hot hadron gas, in which pionic states dominate (\mycf Fig.\,\ref{fig:freezout_densities}), is essentially propelled by pion annihilation.
On the other hand, pions directly link to the Spontaneous Chiral Symmetry Breaking ( S$\chi$BS)  as they constitute the respective Goldstone modes. 
Possible consequences for the spectral distribution of dileptons emitted from a hot hadron gas are discussed in Sec.\,\ref{sec:VDM} further below. 

The vector correlator for a partonic phase using can be calculated with perturbative QCD as  
\begin{equation}
\Im \Pi_{em}^{\mathrm{part.}} =
N_c \sum_{u,d,s,c}
\frac{M^2}{12\pi}(e_q^2)
\left(
1 + \frac{\alpha_s(M)}{\pi} + ...
\right)
\label{eqn:ratio-parton}
\end{equation}  
which collapses into $R_{\mathrm{part.}}$ for large enough invariant masses $M$ such that asymptotic freedom is reached. 
The perturbative expression can be used to estimate the lowest order $q\bar{q}$ annihilation rates in IMR (see~\cite{Cleymans:1986na}). 
However, as already discussed above, the calculations using hard-thermal-loop corrections provide rates larger by one order of magnitude than the estimates obtained with one-loop approximation~\cite{Braaten:1990wp}. Similar results for dilepton radiation rates have also been obtained in recent coarse grained~\cite{Endres:2015fna, Endres:2015egk} and by fireball models~\cite{ Rapp:2014hha,vanHees:2007th} using lattice QCD results (for $\mu_b=0$) for the vector spectral functions above $T_{\mathrm{pc}}$. 
One should note that effects of in-medium hadronic spectral function, as we will discuss them below, are suppressed in this mass region by $T^2/M^2$ \cite{Rapp:2014hha}.

The yields determined by the quantities $\Pi_{em}^\mathrm{had.}(M)$ and $\Pi_{em}^\mathrm{part.}(M)$, representing the contributions to the dilepton spectrum from a hadronic and a partonic phase, respectively, dominate distinct regions in the invariant mass spectrum.
The LMR is governed by the \textrho\ propagator and is as such sensitive to possible modification of the \textrho\ spectral function in the medium. 
The IMR, in contrast, shows a smooth exponential decay solely determined by the temperature of the partonic medium. 
The reason is that the constituents are elementary particles.
A particular role is played by the transition region between LMR and IMR, \myie the mass range between 1 and 1.5 \gevcc . 
This region is indicated in Fig.\,\ref{fig:muon-hadron}
by a continuous distribution rising from a ratio of about unity to about 2 (given by pQCD -- Eq.\,\ref{eqn:partonic-ratio}).
In the hadronic picture it is characterized by multi-pion processes, but it can also be understood as the transition regime between partonic and hadronic degrees of freedom, a phenomenon known as quark-hadron duality~\cite{Rapp:1999ej}.
An important multi-pion process occurring in a dense fireball is the mixing of \textrho\ and a$_1$ states through pion dressing \textpi\textrho$\rightleftarrows$a$_1$
Hence, this region is the ideal place to search for signs of a partial restoration of chiral symmetry in a hot and dense hadronic medium, where spectral functions of both mesons are expected to merge (see next chapter). 

Yet another aspect of thermal emission is the degree of polarization of the virtual photons. 
It has been pointed out that leptons emitted from a thermalized QGP show polarization effects as a result of the non-uniform momentum distributions like Bjorken-flow~\cite{Hoyer:1986pp}.  
Recently, detailed calculations have been carried out to quantify the polarization signal assuming different production mechanisms.
Generalized spectral functions for dilepton emission out of an locally anisotropic plasma have been derived in~\cite{Baym:2017qxy}.  
The emission through both processes, $q\bar{q}$ annihilation in a QGP and \textpi\textpi\ annihilation in a hadron gas has been studied in~\cite{Speranza:2018osi} including collective expansion and quantum statistics. 
The calculations show that indeed effects of a few percent modulation of the lepton angular distribution are to be expected, surprisingly already for emission out of a static source depending on the virtual photon transverse momentum, the invariant mass and the flow. 
The reason is that the momenta of the mother particles, ``picked'' from the thermal distributions and ``combined'' to form a dilepton of given four-momentum, are constrained by energy-momentum conservation.
Hence, only a sub-region of the two-particle phase space is contributing thus breaking isotropy.  
It was found that isotropic distributions are obtained only close to the threshold (for example two-pion mass for the pion annihilation) and that the anisotropy parameters approach zero for high pair momentum and/or invariant mass.
Yet, the effects are much smaller than for elementary processes, \myie for production of dileptons annihilation processes in vacuum (\mycf Sec.\,\ref{sec:cocktail-phenom}).
%
\subsection{Vector mesons and chiral symmetry restoration}
\label{sec:chiral}
%
Chiral symmetry is a fundamental symmetry of the strong interaction in the limit of massless quarks. 
In nature, it is an approximate symmetry and is best justified for SU(2) flavor symmetry because current masses of u and d quarks amount to about $m_\text{u} \simeq m_\text{d} \simeq 5$~\mevcc~ and are hence small against the hadronic mass scale. 
The symmetry refers to the invariance of the QCD Lagrangian under global vector and axial-vector transformations of quark wave functions in the flavor space. 
It is equivalent to the conservation of the chirality components (left and right) of quark wave functions under the influence of the strong force.  
In the QCD vacuum, the chiral symmetry is spontaneously (dynamically) broken ($\chi$SB) due to gluon self interactions leading to the appearance of antiquark-quark and gluon-gluon condensates (for recent reviews of this subject we refer the reader to~\cite{Leupold:2009kz,Hayano:2008vn}).
Important consequences of $\chi$SB are: the appearance of a mass splitting 
of parity partners
(for example the mass difference between the \textrho\ and $a_1$ is almost 500 \mevcc ) and the appearance of Goldstone bosons (in SU(2) pions) with properties linked to the expectation value of quark condensates and to the quark mass. 
The latter is given by the Gell-Mann, Oakes, Renner (GOR) relation~\cite{gor}:
\begin{equation*}
 m_{\pi}^2f_{\pi}^2=-\frac{m_\text{u} + m_\text{d}}{2}\:\expval{\bar{q}q}\,,    
\end{equation*}
with the pion decay constant $f_{\pi}\simeq 92.4$~\mev\ and the expectation value of the scalar two-quark condensate $\expval{\bar{q}q} \simeq (-240$\mev $)^3$. 
Lattice results show that $\expval{\bar{q}q}$ departs from its vacuum value at finite temperatures and finally approaches zero when the pseudo-critical temperature $T_{\mathrm{pc}} \simeq 158$ is surpassed~\cite{Borsanyi:2010bp}. 
The evolution of $\expval{\bar{q}q}$ at non-vanishing baryon densities is not directly accessible by lattice calculations due to the sign problem. 
Various techniques have been applied to extend lattice calculations into the region of finite $\mu_B$ like Taylor expansion or using imaginary chemical potentials and respective results will certainly appear soon.

An early result on the in-medium expectation value of the chiral condensate was obtained from calculations based on the Nambu and Jona-Lasino model~\cite{Lutz:1992dv}.
A feature of this result is a linear depletion as a function of increasing net-baryon density at low temperature, while a more sudden drop is observed along the temperature axis at around $kT\simeq M_{\pi}$ (\myie for zero net-baryon density).
The approximately linear depletion as a function of density can be understood as being due to the reduced chiral condensate in the interior of baryons, a fact which is encoded in the nucleon sigma term and was first obtained in low density approximation \cite{Drukarev:1991fs}. 
More recent calculations in the context of chiral perturbation theory do not support a vanishing condensate with increasing baryon density, essentially because of the hard-core repulsion in the nucleon-nucleon interaction, which enters as a parameter into the calculation~\cite{Fiorilla:2012bc}.
Unfortunately, a direct measurement of the condensate value is not possible but it is conceivable that changes in the vacuum properties driven by temperature and/or density are reflected by changes in the hadron spectrum. 
In the context of dilepton production it is important to discuss the connection between vector mesons and $\chi$SB.  
The light vector mesons can be considered as the lowest order $\expval{\bar{q}q}$ vector excitations of the QCD vacuum and are consequently connected to its properties.
For example, a relation linking the \textrho\ meson mass with the decay constant of the pion, the Goldstone boson of $\chi$SB, has been derived from current algebra~\cite{ksfr} as
\begin{equation*}
m_{\rho}=\sqrt{2}\,g_Vf_{\pi}\,,
\end{equation*}
where the universal coupling constant of VDM $g_V$ is given in Tab.\,\ref{tab:VM-decays}.
Hence, it is appealing to assume that changes of the order parameters of $\chi$SB are reflected in modification of the vector meson properties like their masses or/and widths. 
QCD Sum Rules (QCDSR) provide such non-trivial correlations resulting from dispersion relations connecting the energy weighted integral of the imaginary part of vector current-current correlator (\myie spectral function see eq.\,\ref{eqn:ratio-hadron}) with an Operator Product Expansion (OPE) (for review see \cite{Leupold:2009kz}).  
The latter is calculated in the space-like regime and contains terms obtained from pQCD and terms containing the expectation values of local operators, the condensates. 
In order to enhance the low energy part of the integrand in the dispersion relation, relevant for the low mass vector mesons, the weighting function is chosen as a Borel transformation with a Borel Mass ($M$) being a parameter of the expansion (see left side of Eq. \ref{eq:qcdsr} )
QCDSRs have been worked out most extensively for the \textrho\ meson because of its prime importance for dilepton spectroscopy.  
The respective relation reads as follows:
\begin{eqnarray}
\frac{1}{\pi M^2} \int ds \Im\Pi(s) e^{-s/M^2}
&=& 
\frac{1}{8\pi^2}
\left(1 + \frac{\alpha_s}{\pi}\right) +
\frac{1}{M^{4}}
\left(m_q \expval{\bar{q}q} +
A\, G^2 + B\, m_N a_2\rho_N\right) \\
&-& 
\frac{1}{M^6}
\left(C\, O^V_4 + D\, m_N^3 a_4 \rho_N\right)
+ \order{\frac{1}{M^{8}}} \, ,\nonumber
\label{eq:qcdsr}
\end{eqnarray}
where the $A,B,C,D$ are numerical constants, $G^2$ , $\expval{O^V_4}$ are the expectation values of the gluon and the four-quark condensates, respectively.
The terms containing $a_2$ and $a_4$ stem from a non-scalar higher twist-2 condensate and are related to moments of parton distribution functions (for details see \cite{Leupold:2009kz}).   
The numerical analysis of the terms on the right hand side of Eq.\,\ref{eq:qcdsr} has been performed for finite densities and shows that the most important contributions originate from the gluon, the four-quark and the $a_2$ term, with the explicit dependence on the density ($\rho_N$). 
One should emphasize, however, that not all terms break chiral symmetry. 
The term containing $a_2$ is significant as well for the vacuum as for the in-medium case but does not break the chiral symmetry. 
Similarly, the gluon condensate provides an important contribution to the meson vacuum spectral distribution but is also symmetric \mywrt chiral transformation and depends only weakly on the density (at least in the investigated low density limit ($\rho_n<1.5\, \rho_0$)). 
Properties of the four-quark condensate are known with much less precision but estimations show that its numerical value is significant. 
The $O_4$ contribution for the \textrho\ meson can be separated into two parts: a symmetric and odd one, with respect to the chiral transformation. 
It can be estimated for vacuum from the combined analysis of the spectral distributions of the \textrho\ and its chiral partner $a_1$, which are precisely known from $\tau$ decays  \cite{Leupold:2001hj,Hohler:2012xd}. 
On the other hand, the two-quark condensate, which is the order parameters of the $\chi$SB, does not play the paramount role because of the smallness of the light current quark mass appearing as a scaling parameter in the operator expansions. 
Nevertheless, it may contribute indirectly since for the evaluation of the four-quark condensate it is routinely assumed that a factorization scheme works for which the expectation value of the four-quark condensate is simply taken to be the product of the two-quark condensates (as for example applied in the above mentioned analysis of the \textrho/a$_1$ vacuum spectral functions).  

Despite the large uncertainties in the knowledge of the four-quark condensate, the analysis of QCDSR, presented in\cite{Leupold:2009kz}, predict significant in medium-modifications to both the mass and the width of \textrho\ meson. 
In more detail, it neither provides independent information on both nor a stringent relationship between the two properties but rather a broad corridor of allowed masses and widths of the meson and the density.
One should emphasize that in this analysis the spectral function of the \textrho\ was assumed to be a of a Breit--Wigner type.  
Furthermore,  different scenarios for the estimation of the four quark condensate contribution have been worked out: (i) using aforementioned factorization scheme and  modifications of the two-quark condensate according to the low density approximation, (ii) the factorization and the two-quark condensate modifications but only for the part with chirally odd terms  and (iii) no modification in both types of terms. 
All scenarios reveal significant modifications of the OPE value \mywrt vacuum with the largest effect obtained for the case (ii). 
In other words, QCDSR predict significant changes of mass and widths (or both at the same time). 
As already pointed-out, only part of the effect can be attributed to the modification of the terms breaking $\chi$S.  
This result provides a slightly different picture as compared to the earlier calculations of~\cite{Hatsuda:1991ez}. 
In the latter, a shift of the vector meson pole mass in medium was obtained along the suggestion made by Brown and Rho based on QCD scale invariance and predicting the scaling law ~\cite{Brown:1991kk} 
\begin{equation}
\label{eqn:Brown-Rho}
\frac{M_V^*}{M_V}
\sim
\frac{f_{\pi}^*}{f_{\pi}}
=
\left(\frac{\expval{\bar{q}q}^*}{\expval{\bar{q}q}}\right)^{1/3}
\,,
\end{equation}
where the asterisk denotes in-medium values. 
The predicted effect is now understood as a consequence of the applied factorization scheme for the four-quark condensate, mentioned above, and the assumption of $\delta$ function for the \textrho\ meson spectral distribution assumed in the calculations. 
It is worth to mention that this original conjecture triggered enormous experimental activities focused on investigations of in-medium properties of vector mesons, also in cold nuclear matter.

A more direct connection between the properties of hadronic spectral functions and the order parameters for the chiral symmetry breaking is provided by inspection of the Weinberg sum rules (WSR).
These sum rules relate weighted moments of the spectral distributions of the parity doublets, \textrho\ and a$_1$, directly with the respective order parameters of $\chi$SB:
\begin{eqnarray}
\int ds\,s^{-1}\,
\left(D^V(s) - D^{A}(s)\right) & = &  f^2_{\pi}\;,\\
\int ds \, \left(D^V(s) - D^{A}(s) \right) & = & f_{\pi}^2 m_q = -  2 m_q \expval{\bar{q}q}\;,\\
\int ds\,s \left(D^V(s) - D^{A}(s) \right) & = & - 2\pi\alpha_s O_4^{\mathrm{SB}}\;.
\label{eq:weinbergsr}
\end{eqnarray}
The first two sum rules were derived by Weinberg ~\cite{weinbergsr}, while the third one has been added by the Lastly, Kapusta and Shuryak \cite{Kapusta:1993hq}. 
In the latter sum rule $O_4^{\mathrm{SB}}$ stands for the chirally odd combination of four-quark condensates in the vector and the axial channels.
For example, using the first sum rule and assuming $\delta$ function for the \textrho\ meson, the relation between the $a_1$ and \textrho\ masses  $m_{a1}=\sqrt{2} m_{\rho}$ has been derived~ \cite{weinbergsr}.
As already pointed-out above, the spectral distributions of the vector and axial-vector states are very well known in vacuum from the $\tau$ decays into odd (even) number of pions $\tau\rightarrow \nu_{\tau}(2n+1)(2n) \pi$. 
They provide the reference for the study of the QCDSR and WSR in vacuum and were shown to be satisfied within a few percent if, in addition to the dominant \textrho\ and a$_1$, contributions from excited states (a$_1^{\prime}$, \textrho$^{\prime}$) and the continuum are included in the respective vector and axial-vector spectral distributions \cite{Hohler:2012xd}.
Similar studies were also performed in  \cite{Leupold:2001hj} and \cite{Kwon:2010fw}. 

\begin{figure}[th]
\center\includegraphics[width=0.95\textwidth,height=0.5\textwidth]{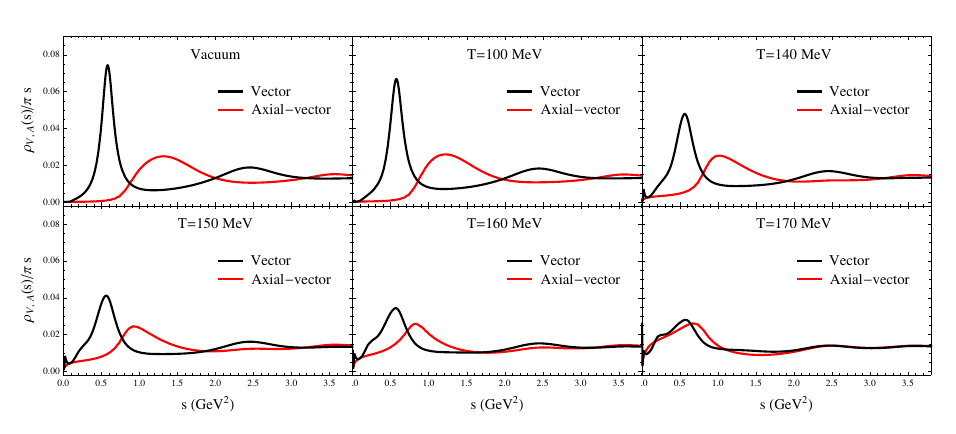}
\caption{\label{fig:sum_rules} Evolution of vector(black) and axial-vector (red) spectral functions with temperature towards critical temperature ~\cite{Hohler:2013eba} ) }
\end{figure}

A formulation of sum rules expected at finite temperatures was first obtained in \cite{Kapusta:1993hq} and more recently also in \cite{Kwon:2010fw,Hohler:2013eba}. 
Experimentally, the in-medium spectral function of the \textrho\ meson is presently accessible by dilepton spectroscopy. 
The results of various heavy ion experiments, that will be discussed in Sec.\,\ref{subsec:heavy_ion}, provide convincing evidence for its substantial modifications in dense and hot nuclear matter.  
Hence, low-mass dilepton data provide a valuable constraints for theoretical models. 
Up to now, no complementary experimental information on the in-medium $a_1$ spectral function is available. 
Nevertheless, the detailed understanding of the vector meson spectral function alone together with constraints imposed by the WSR and QCDSR sum rules provide already very valuable insight to  chiral restoration effects on hadronic spectra.
This has been demonstrated by Hohler and Rapp \cite{Hohler:2013eba} in an analysis using the spectral function of the \textrho\ meson calculated with hadronic models \cite{Rapp:1999ej,vanHees:2007th}, described in more details in the next section.
The calculation allows to establish the full evolution of the  vector spectral function as function of the temperature and the baryo-chemical potential.  
Using these spectral functions and a temperature dependence of the condensates (defined in the right hand side of Eq.\,\ref{eq:qcdsr}) obtained from lattice-QCD calculation for $\mu_b=0$, and assuming otherwise a hadron gas equation-of-state as the input, QCDSR were calculated for vanishing chemical potential. 
The result shows that the sum rules are satisfied within few percent,  which is already a non-trivial result. 
In the next step, the evolution of the axial vector spectral function has been calculated requesting conservation of the QCDSR and WSR.
The axial spectral function was modeled in the calculations by $a_1$ with a Breit--Wigner distribution with variable peak position and a width and a continuum contributions representing vector-axial vector mixing. 
The calculation demonstrates, see Fig.\,\ref{fig:sum_rules}, gradual reduction of the $a_1$ mass and an increasing width of the Breit--Wigner peak, while the \textrho\ mass stays nearly constant but the width is increasing as well with a shoulder appearing on the low-mass side of the distribution (see for details Fig.~3 in \cite{Hohler:2013eba}). 
The obtained behavior shows that the spectral functions of vector-axial chiral partners become degenerate at $T=T_{\mathrm{pc}}$ , as expected for the chiral symmetry restoration. 
%
\subsection{In-medium spectral function in microscopic models}
\label{sec:hadronic-model}
%
Most of the models addressing in-medium hadron properties focus on vector mesons and in particular the short-lived \textrho\, which is ideally suited for experiments with dileptons.
For comprehensive overview on this topic we refer to excellent dedicated works \cite{Rapp:1999ej}, \cite{Leupold:2009kz}). 
Below we want to shortly summarize the most important conclusions obtained from various hadronic models with emphasis on the model worked out by Rapp and Wambach, commonly used for comparisons with results from Heavy Ion collisions. 
The model  incorporates achievements worked-out by many other authors, which we briefly discuss below (for details see \cite{Rapp:1999ej,vanHees:2007th}).
The calculations in hadronic models are based on effective hadronic Lagrangians of \textrho\ interactions with mesons and baryons with vertices and coupling constants constrained by available experimental data.
Among them are: results on vector meson production in pion induced and photon induced reactions, photo-absorption and available decay widths of  resonances to vector meson channels. 
Medium effects on the mesons self energy are then  obtained using many body theory.
For example, the main effects on the in-medium spectral function of \textrho\ meson  are given via  modifications of the pion loop in the \textrho\ meson self-energy term and via the direct interactions of \textrho\ meson with  a resonance-nucleon hole (\textrho-RN$^{-1}$) and mesonic  (\textrho-M) states. 
The  \textrho-RN$^{-1}$  interactions are governed by baryonic resonances (including also hyperons) with significant couplings to (\textrho,N) final states.  
\begin{figure}[th]
\begin{center}
\includegraphics[width=0.45\textwidth,height=0.4\textwidth]{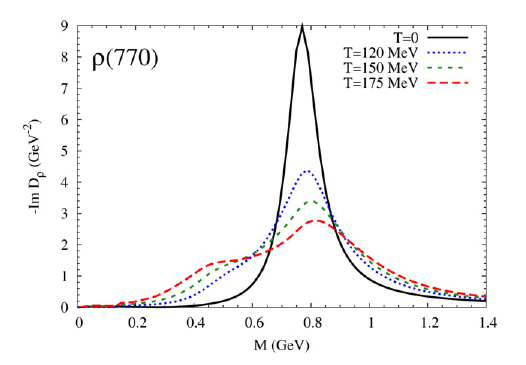}
\includegraphics[width=0.45\textwidth,height=0.4\textwidth]{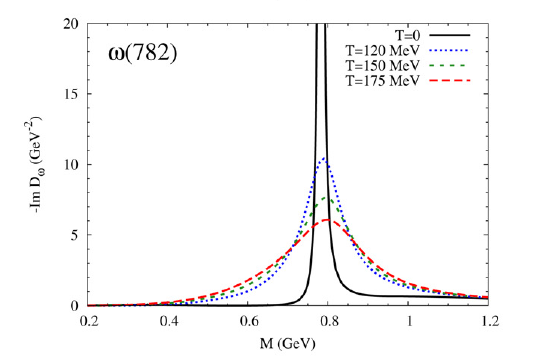}
\end{center}
\caption{ In-medium spectral functions of \textrho\ (left) and \textomega\ (right) for sets of $\mu_b, T$ representative for the evolution of the fireball in  heavy-ion collisions at around $160$\agev\ (SPS full energy)~\cite{vanHees:2007th} 
Four variants are shown ($T$/\mev , $\mu_B$/\mev): vacuum $(0,0)$, ($120,449$),  ($150,331$) and ($175,232$) (labeled by the respective temperature only).
Note that the case of highest temperature clearly reflects a condition where likely a partonic medium is
realized.
}
\label{fig:spectral-fun}
\end{figure}
The pion loop is modified by interactions of the off-shell pions with nucleons forming \myie \textDelta-hole states  
while interactions with mesons include resonances with significant branching to the \textrho--\textpi like \myie the \textomega\, the a$_1$(1260) and the h$_1$(1170) states.
With the help of Eq.\,\ref{eqn:ratio-hadron} the in-medium propagator of \textrho\ meson can thus be written as
\begin{equation}
D_{\rho}=\frac{1}{M^2-(m^0_{\rho})^2-\Sigma_{\rho\pi\pi}-\Sigma_{\rho M}-\Sigma_{\rho B}}\,,
\end{equation}
where the individual self energy terms in the denominator reflect the different coupling schemes used in the calculation (M = mesons, B = baryon). 
The modification of the \textrho\ meson via a dressing of the pion loop was calculated in \cite{Asakawa:1992ht}, \cite{Herrmann:1992kn},\cite{Rapp:1995zy}, \cite{Rapp:1997fs} within the framework of the \textDelta-hole model and generally leads to the increase of the meson width. 
The interactions with baryon resonances modify also the mass distribution and produce additional structures. 
The interactions with higher mass baryon resonances were first considered for cold nuclear matter in \cite{Friman:1997tc} and were limited to $P$-wave \textrho--N states. 
Later, the calculations were extended to $S$-wave states~\cite{Peters:1997va} thus elucidating the important role played by the low mass N$^\star$ ($D_{13}(1520)$) and  \textDelta$^\star$ ($S_{31}(1620),S_{33}(1700)$) resonances. 
In particular strong effects on the \textrho\ spectral function for small relative momentum of the vector meson \mywrt\ the medium ($p \simeq 0$) were observed. 
The momentum dependence of the \textrho--N coupling was further studied in \cite{Post:2000qi} by accounting also for relativistic effects. 
The work confirmed the important role of $S$-wave resonances and extended the study to the \textomega\ meson \cite{Post:2001am,Muehlich:2006nn}, where a significant broadening of its spectral function, associated with appearance of low mass tail, was found. 
The latter is due to strong off-shell coupling of the meson to S11(1535) resonance.

The extension of the hadronic models to account for the interactions of higher mass resonances and interactions with anti-baryons  produced in the fireball was also included in the model calculation of Rapp and Wambach. 
The calculated spectral function has been parameterized as a function of baryon densities (or baryo-chemical potential) and temperature and applied in the emissivity formula for dilepton calculations. 
The imaginary part of the in-medium \textrho\ propagator is depicted in Fig.\,\ref{fig:spectral-fun} (left panel) for the selected intensive quantities as characteristics for the evolution of the fireball in heavy-ion collision at full SPS energy.  
The \textrho--BN$^{-1}$ loops, especially including low-mass baryonic \textDelta$^\star$ and N$^\star$ resonances, are responsible for a substantial distortion of the \textrho\ meson spectral distribution towards lower masses.
A strong broadening of the spectral distribution occurs and features a characteristic emergence of a second bump on the low-mass side of the resonance pole as the baryon densities increases. 
This structure can be understood as ``bound'' \textrho\ mesons due to the direct coupling of the \textrho\ to a baryon resonance-hole state.
The respective in-medium modification of the \textomega\ meson is shown on the right panel which does not show a distinct second bump.  
 
As we discussed in previous section a scenario for the evolution of the in-medium a$_1$ spectral has been computed in~\cite{Hohler:2013eba}. 
This work demonstrates that the mass splitting of the two chiral partners can vanish close to the chiral pseudo-critical temperature $T_{\mathrm{pc}}$. 
Indeed, both spectral functions significantly broaden and finally overlap. 
The same calculations also show a reduction of the pion decay constant f$_{\pi}$, in line with the expected pattern of a restoration of spontaneously broken chiral symmetry. 
This important observation (although limited to the region of vanishing chemical potential) provides a strong hint that calculated mass modification of the \textrho\ meson are consistent with the scenario of chiral symmetry restoration around $T_{\mathrm{pc}}$. 
The vector axial-vector mixing (\textrho--a$_1$) is an interesting effect which can be studied in dilepton spectra above 1\,\gevcc.  
The corresponding rate is given by the vector correlator which can be expressed at finite temperature by mixing of the vector and axial vacuum correlators.
It has been first worked out in \cite{Dey:1990ba} in the chiral limit ($m_\pi=0$):    
\begin{equation}
\Pi^V_{\mu\nu}(q,t)=(1-\epsilon)\Pi^V_{\mu\nu}(q,0)+\epsilon\Pi^A_{\mu\nu}(q,0)
\label{eqn:va-mixing}
\end{equation}
with a mixing parameter $\epsilon$ determined via the loop integral $\epsilon=\frac{2}{f_\pi^2}\int \frac{d^3k}{2\pi^3\omega_k}f^{\pi}(\omega_k;T)$ (\textomega\ is on-shell pion energy). 
For $\epsilon=1/2$ ($T=T_{\mathrm{pc}}$) vector and axial-vector spectral distributions completely overlap signaling chiral-symmetry restoration (see Fig.\ref{fig:sum_rules}). 
For $\epsilon<1/2$ the axial admixture fills the dip in the the $1-1.5$ \gevcc mass region. 
More elaborate considerations including finite pion mass and momentum in the axial vector correlator and a non-vanishing pion chemical potential were worked out in \cite{Urban:2001uv,vanHees:2007th}.   
%
\subsection{Mass modifications in transport codes}
\label{sec:mass-transport}
%
A full quantum field theoretical treatment of the one-particle information of an interacting many-body system out of equilibrium is provided by the Kadanoff--Baym equations (\cite{KadanoffBaym:1962}). 
By regarding one-particle Greens functions the many body aspects of the problem are taken care of through proper self-energies of the states
and their virtuality is consistently treated in the time evolution.   
The concept of nonequilibrium Greens functions has been generalized to relativistic field theories in~\cite{Danielewicz:1982kk, Mrowczynski:1989bu, Mrowczynski:1992hq}. 

Under some conditions, the Kadanoff--Baym equations can be reduced to a semi-classical treatment describing the time-evolution of the Wigner transform of the system by the Boltzmann equations.  
The semi-classical transport theory emerges from the Kadanoff--Baym equations requiring smooth changes of the internal variables.
Three characteristic time scales can be inspected to assess the validity of the simplification which is considered justified if the separation between the time scales exists with $\tau_{\mathrm{int.}}<\tau_{\mathrm{coll.}}<\tau_{\mathrm{bulk}}$.
Here $\tau_{\mathrm{int.}}$ refers to the range of the interaction between two constituents and the time it takes the collision partners need to reach asymptotic states. 
The typical time between two subsequent collisions is denoted by $\tau_{\mathrm{coll.}}$ and defines the dilute gas limit if $\tau_{\mathrm{int.}}<<\tau_{\mathrm{coll.}}$ is fulfilled.
The time scale over which the bulk properties like density or shape of the many body system changes (hydro time scale) is given by $\tau_{bulk}$.
Indeed, the ratio $\tau_{\mathrm{int.}}/\tau_{\mathrm{coll.}}$ serves as an expansion parameter and the first order is treated to arrive at the Boltzmann equations. 
A consequence of this quasi-particle ansatz is that the virtuality finally collapses to a delta function and pushes the states on the mass shell. 

In the traditional approach to the microscopic transport is to average over the spectral distribution. 
This is reasonable if the distribution is sufficiently well localized around the pole mass that represents the on-shell energy-momentum relation.
For heavy-ion collisions, however, there are many studies that suggest that states with broad spectral distributions should be included as (propagating) degrees of freedom. 
States with a long life time in vacuum (\myie much longer than the average time between subsequent collisions) or stable particles can experience significant collisional broadening if the state is destroyed in a collision.
In a dynamical treatment of off-shell particles it is important to guarantee that particles get back their vacuum properties before they leave the dense region. 
Corresponding equations have been suggested by Effenberger et al.~\cite{Effenberger:1999ay}. 
Therefore, if one wants to propagate resonances or states including a dynamical width, one has to devise a scheme how to propagate the spectral information that is traditionally averaged. 
Conceptually this implies quite a change of the original ideas that lead from quantum field theory to the traditional transport equations sketched above.
Knoll et al.~\cite{Knoll:2001jx} formulated coarse grained transport equations for states with broad spectral distributions such that the original expressions for particle number and energy are still exactly conserved in the coarse grained theory. 

Unfortunately, the transport equations cannot be solved exactly by a test particle method. 
To describe states with a broad spectral distribution one needs test particles that can have arbitrary momenta and masses.  
The development of Leupold~\cite{Leupold:1999ga} and Cassing, Juchem~\cite{Cassing:1999mh} provided a justification for the equations for the test particles that were proposed by Effenberger~et~al.\ and formulated coarse grained transport equations that can be solved by test particle methods.
For the BUU code developed in Gie{\ss}en the off-shell transport of broad resonances has been realized absorbing the effect of the back-flow term in the equation of motion into the mean field. 
While it still is a major simplification relative to the full treatment, nevertheless a correct transformation of the in-medium state to the vacuum state is guaranteed once the particle travels into the region of vanishing local baryon density~\cite{Buss:2011mx}.

A non-relativistic case was treated by Leupold and scalar relativistic case was treated by Cassing and Juchem and implemented in the HSD transport (for a recent review \cite{Linnyk:2015rco}).  
As shown in the first of these references, the obtained transport equations do not fully conserve the original particle number and energy. 
However, it was argued \cite{Leupold:2000ma}, that one can find a modified particle number for the coarse grained theory that is exactly conserved. 

There are two phenomenological scenarios applied commonly in transport model calculation to account for the vector meson modifications. The first one assumes a mass shift according to the Brown-Rho scaling (see Eq.\,\ref{eqn:Brown-Rho}) usually parameterized as
\begin{equation}
M^*(\rho_N)=\frac{1}{1+\alpha\rho_N/\rho_0}\,,
\end{equation}
where $\rho_N$ is the local baryon density at the actual decay position ($\rho_0$ is the nuclear ground state density), $\alpha\simeq0.16$ 
and the second one takes into account collisional broadening (in natural units),
\begin{equation}
\Gamma_{coll}(M,q,\rho_N)=\gamma\
\rho_N<v\sigma_{VN}^{\mathrm{tot}}>\simeq\alpha_{coll}\frac{\rho_N}{\rho_0}\,, 
\end{equation}
where $\sigma_{VN}^{\mathrm{tot}}$ is the total vector meson-nucleon cross section and $\alpha_{\mathrm{coll.}} \propto \tau^{-1}_{\mathrm{coll.}}$. 
For example, in the HSD transport model $\alpha_{coll} \simeq 150$ \mev\ was fixed for the \textrho\ meson and $\alpha_{coll} \simeq 70$ \mev\ for the \textomega\ \cite{Linnyk:2011hz}.
A comparison of these two approaches will be presented in Sec.\,\ref{sec:status} by comparison of respective HSD calculations with experimental data.
%
%
\section{Survey on Experimental Results}
\label{sec:exp}
\subsection{Nucleon-nucleon collisions}
\label{sec:exp-elementary}
%
We start our survey on experimental results with the discussion of \nnnn\ collisions. 
Dilepton production in \nnnn\ collisions provides an important reference for the interpretation of results obtained from heavy-ion collisions since they help to constrain dilepton contributions from the initial and late stage of the reaction. 
For such small collision systems thermal radiation is  generally not expected to occur.
An exception is \nnnn\ collisions at collider energies.
It has been observed that in a sub-class of the collisions, featuring high final state charged particle multiplicities, the event topology clearly exhibits signatures of collective behavior. 
For all other cases \nnnn\ collisions allow to verify the understanding of ``elementary'' processes contributing to the dilepton cocktail which substantially changes in composition as a function of the collision energy. 
\subsubsection{RHIC and LHC}
\label{sec:elementary-rhic}
\begin{figure}[tbh]
\begin{center}
\includegraphics[width=0.36\textwidth,height=0.45\textwidth]{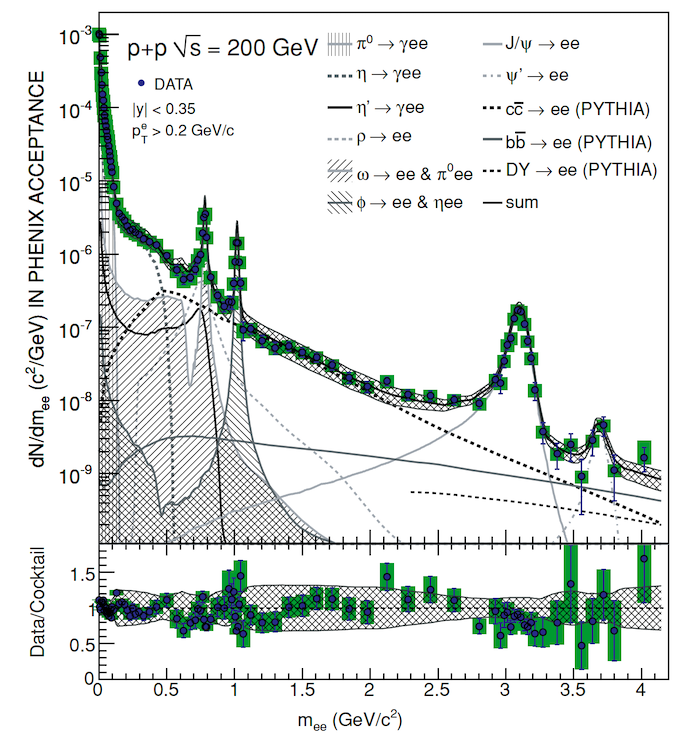}
\includegraphics[width=0.3\textwidth,height=0.4\textwidth]{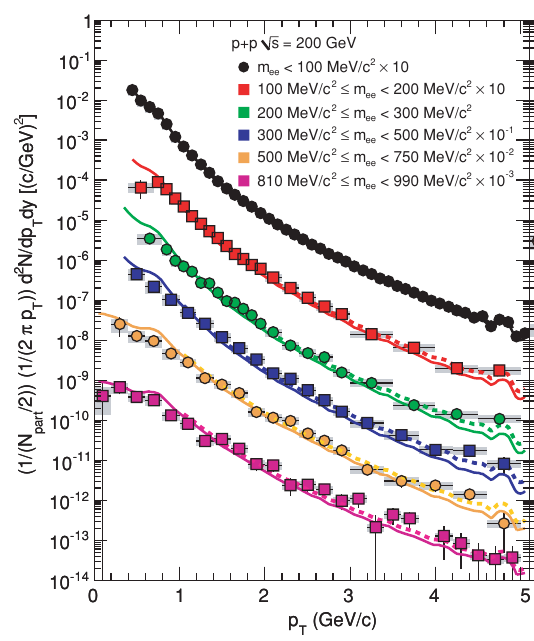}
\includegraphics[width=0.3\textwidth,height=0.4\textwidth]{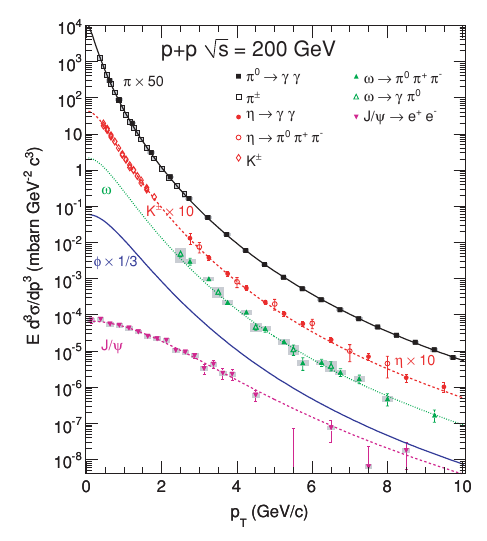}
\end{center}
\caption{Left panel: Inclusive invariant mass distribution of $\mree$ pairs measured by PHENIX in \pp\ collisions at \sqsnn{200} compared to a cocktail of sources originating from the decay of light hadrons (\piz, \texteta, \texteta'), vector meson decay (\textrho, \textomega, \straightphi, J/\textpsi, \textpsi') correlated weak charm ($c\bar{c}$) and bottom ($\bar{b}b$) decays, and Drell-Yan scattering (DY)~\cite{Adare:2009qk}. 
The ratio of the data to the cocktail is shown as inset. 
Respective systematic uncertainties of the data are depicted as boxes while the uncertainty on the cocktail is shown as band around unity. 
Middle panel: Transverse momentum distributions of dielectrons for various invariant mass bins compared to the expected yield adding all contributions to the hadronic cocktail.  
Right panel: Invariant transverse momentum distributions of neutral mesons used as basis for the cocktail calculation. 
Further explanations in the text. 
}
\label{fig:phenix-pp} 
\hspace{2pc}%
\end{figure}

We start our review with results from \pp\ collisions at the top RHIC energy of \sqsnn{200} \gev .
Fig.\,\ref{fig:phenix-pp} (left panel) shows dielectron invariant mass distributions measured with the PHENIX detector~\cite{Adare:2009qk}, covering the mid-rapidity region ($\mid \eta \mid <0.35$), in comparison to a hadronic cocktail composed of the expected sources and spanning over all three mass regions. 
Evidently Dalitz decays of \piz , \texteta and \textomega mesons and two body decays of light vector mesons are the dominant sources in the LMR. 
The integral yield originating from these cocktail sources fully saturates the data up to the \textphi meson  beyond which mesonic contributions break down. 
%
%
Meson multiplicities needed for the construction of the hadronic cocktail were obtained from fitting transverse momentum spectra of various mesons with a modified Tsallis function (right panel of Fig.\,\ref{fig:phenix-pp}).
The respective multiplicities vary little within systematic uncertainties compared to numbers extracted from the same data using the original Tsallis distribution~\cite{Adare:2010fe}.
Cross sections were derived by multiplying the fully corrected and integrated differential multiplicity distributions with the total inelastic cross section taken to be $\sigma^{\mrpp}_{\mathrm{inel.}} = 42$~mb. 
The differential cross section for \texteta\ and \textomega\ production were obtained from hadronic and photon decay channels in the similar way.
For unobserved phase space regions $m_t$ scaling was assumed to hold and all cross sections to be constant in the measured $\Delta \eta$ bin (boost invariance). 
The cross sections parameterized in that way were input to the PHENIX decay generator used to derive the respective dielectron yields (hadronic cocktail). 
The meson decay generator (EXODUS) uses mass dependent eTFF's for the meson Dalitz decays and samples angular distributions of electrons \mywrt\ the virtual-photon axis as explained in Sec.\,\ref{sec:dalitz-decays}.
For the two-body decay no polarization of vector mesons and partial decay widths $\Gamma_{\mree}$ given by VDM with finite width corrections calculated by Gounaris~\cite{gounaris} are assumed. 
The \textrho\ production is assumed to proceed via pion--pion-annihilation and hence a cut-off in the meson mass distribution towards $2 m_{\pi}$ appears.
The measured transverse momentum distributions of dielectrons, shown in the middle panel of Fig.\,\ref{fig:phenix-pp} for various invariant mass bins, are well described by the cocktail calculation based on the meson production cross sections derived as described above and shown as full lines.

Correlated semi-leptonic D/$\bar{\text{D}}$ meson decays largely dominate the IMR.
The PHENIX collaboration extracted the cross section for $\bar{c}c$ production by saturating the experimental dielectron invariant mass distribution in the IMR and through extrapolation of the theoretical invariant mass distribution down to zero mass in the LMR. 
The cross section obtained this way agrees with calculations based on perturbative QCD.
It is interesting to note that correlated charm decay makes up a significant contribution to the dilepton spectrum in the LMR already at this energy. 
The contributions from Drell--Yan and bottom decays are at least one order of magnitude smaller than the correlated D/$\bar{\text{D}}$ decays and only becomes relevant in the HMR.
Their invariant mass distributions are taken from  PYTHIA calculations with settings reproducing the measured open charm yield.  
The prominent feature in the HMR is the peaks from J/\textpsi\ and \textpsi' decay. 
The authors observe that the dielectron transverse momentum distribution simulated on the basis of PYTHIA appears softer than the measured ones.  
In summary, the data are very well described in the full mass range by the calculation assuming free hadron decays only.
The results of PHENIX have been confirmed by independent measurements and analyses of dielectron emission from \pp\ collisions by the STAR collaboration~\cite{Adamczyk:2012yb}. 

At LHC, continuum dielectron production in \pp\ collisions has recently been measured by the ALICE collaboration at $\sqrt{s}=7$ and $14$~\tev\ \cite{Acharya:2018ohw,Acharya:2018kkj}. 
The hadronic cocktail describing the LMR and IMR is qualitatively similar to the one extracted at RHIC. 
The LMR region is properly described by a hadronic cocktail based on the known cross sections of mesons measured at LHC. 
In the IMR, the dielectron yield is given by correlated charm and, to a lesser extent, by bottom contributions. 
The relative contributions of charm and bottom decay can be assessed by inspection of the dielectron transverse momentum distribution as dielectrons from bottom decay feature a harder spectrum.  
However, the extracted cross sections show a model-dependence (factor two difference) depending on which  event generator, PYTHIA or FONLL, POWHEG is used to calculate the phase space distribution of heavy-flavor production.      
%
\subsubsection{SPS}
\label{sec:elementary_sps}
\begin{figure}[tbh]
\begin{center}
\includegraphics[width=0.45\textwidth]{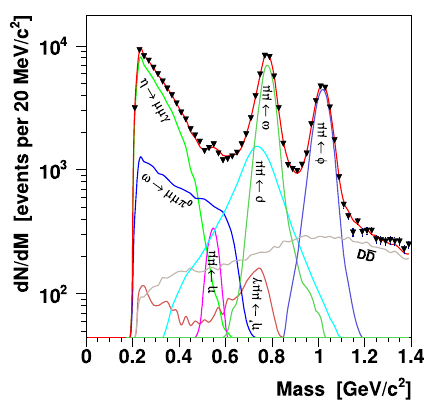}
\includegraphics[width=0.45\textwidth]{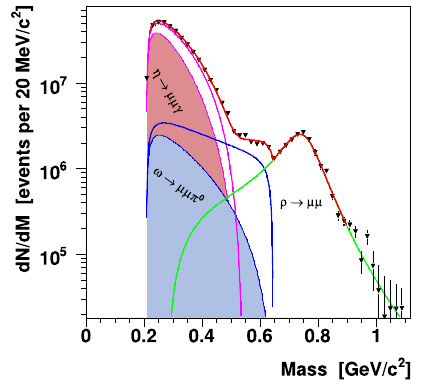}
\end{center}
\caption{Di-muon invariant mass distribution measured by NA60 for $400$\gevc\ p--A collisions in comparison to a cocktail of various sources: (left panel) without acceptance corrections and (right panel) with acceptance/efficiency corrections and after subtraction of the two body decays of narrow vector mesons~\cite{Arnaldi:2016pzu,Uras:2011qs}. 
}
\label{fig:na60-pA}
\hspace{2pc}%
\end{figure}

%
Investigation of lepton pair production in \pp\ and \pA\ collisions at the SPS started already in seventies (see \cite{Specht:2007ez} for a detailed account). 
At the beginning most of the experiments claimed to observe enhancements in the LMR, above the contributions expected from hadronic sources. 
It was, however, disproved one decade later by more detailed evaluations of the \texteta\ Dalitz contribution which had been underestimated before. 
The HELIOS-1 experiment~\cite{Akesson:1994mb} measured electron and muon pairs as well as photons in p--A collisions at $450$\gevc\ (\sqsnn{29.5}). 
With the detection of a photon in coincidence with  lepton pairs, the reconstruction of the Dalitz decay \texteta$\rightarrow \ell^+\ell^-$\textgamma\ became possible and a model independent determination of the respective contributions to the dilepton spectra was obtained. This was an important breakthrough since upper limits for a possible existence of "unconventional" sources in the \texteta\ mass region had been reported.  
The upper limit of a possible unconventional contribution to the LMR dielectron yield was further reduced to at maximum of $23\%$ in an experiment by the CERES collaboration investigating p-Be at $450$\gevc\ \cite{Agakishiev:1998mw}. 
The main source of systematic uncertainty which remained was the \textomega\ meson eTFF.
Recently this uncertainty has been significantly reduced by the dimuon measurements of the NA60 collaboration using a $400$\gev\ proton beam on nuclear targets (Be, Cu, In, W, Pb and U )~\cite{Arnaldi:2016pzu,Uras:2011qs}.
Fig.\,\ref{fig:na60-pA} shows the respective dimuon invariant mass distribution after combinatorial background subtraction in comparison to a cocktail of free (vacuum) hadron decays. 
The important outcome of the NA60 analysis is the determination of the individual components, based on an iterative procedure making use of the measured inclusive data (mass and transverse momenta) only. 
This became possible due to the eminent statistics of the collected data and the high precision of the spectrometer. 
In the first step of the analysis the yields due to two-body dilepton decays of the \textomega\ and \straightphi\ mesons were determined by fitting the respective peaks using line shapes known from simulation of the detector response. 
The correlated charm contribution, dominating the mass spectrum above the \straightphi\ mass, was calculated using the PYTHIA string fragmentation model.
The narrow vector mesons were then subtracted and the remaining yield was corrected for acceptance (right panel).
In order to account for the acceptance effects, the detector response was simulated using extracted meson transverse mass distributions~\cite{Uras:2011qs}. 
The resulting spectrum was then fit a second time allowing variation of the three overall normalizations for \texteta, \textrho, \textomega\ and the two shape parameters ($\Lambda_{\eta/\omega}$) controlling the eTTF's for the \textomega\ and \texteta\ Dalitz decays ( see Eq.\,\ref{eqn:VDM-FF-a} and Tab.\,\ref{tab:dalitz-table} in \mycf Sec.\,\ref{sec:dalitz-decays} ). 
The angular distributions of electrons emitted in the Dalitz decays of the pseudo-scalar and vector mesons (\mycf Sec.\,\ref{sec:dalitz-decays}) were also included in the calculations.
The sum of these sources provides a good description of the total yield and leaves little ambiguity in the determination of the vector meson contributions.  
In that way the eTFF's could be extracted without exclusive reconstruction of the respective Dalitz decays. 
The effect of the mass dependent eTFF's can be seen in Fig.\,\ref{fig:na60-pA} (right) by comparing the shapes given by the hatched area (representing decays of point like particles) and the solid line (including the effect of eTFF's) plotted separately for both mesons.

Another important outcome of the analysis is the conclusion that cold matter effects on the spectral distribution of the short-lived \textrho ~meson could be neglected in case of NA60.
This can be understood as there is a large rapidity gap between the projectile and target regions at such high collision energies.
Indeed, the acceptance of the NA60 spectrometer covers the more forward rapidity region\footnote{For \inin\ at 158\agev\ $y_{CM} = 2.95$}, \myie $3.3<y<4.3$ for low-$p_{\perp}$ \textrho ~mesons.
On the other hand, strong medium effects occur if the propagating mesons couple to the baryons in the target.
Hence, the region of interest for medium modifications in cold matter is outside the acceptance of the NA60 spectrometer at upper SPS energies and consequently the \pA\ data can be regarded as a good approximation for \nnnn\ collision. 
This is further supported by a detailed investigation of the accurately measured \textrho ~meson mass distribution shown in the right panel of Fig.\,\ref{fig:na60-pA} . 
It was found that the data is best described by a ``vacuum line-shape'', given by Eq.\,\ref{eqn:VM-2decay}, with a total decay width $\Gamma_{tot} = 146\pm 6$~\mev\ and a density of initial states $dN_p(M)/dM$ according to thermal pion gas with $T=161\pm 5(stat)\pm 7(syst)$.  
%
\subsubsection{BEVALAC/SIS18} 
\label{sec:sis-bevalac}
%
Dielectron production in 1--5\gev\ \nnnn\ collisions was investigated in the nineties by the DLS experiment at the BEVALAC~\cite{Wilson:1997sr} and more recently by HADES at SIS18~\cite{Agakishiev:2009yf,Agakishiev:2012tc,HADES:2011ab}. 
In contrast to the situation at high energies, where the dilepton cocktail in the LMR is dominated by neutral meson decays, an important role is played by baryonic sources. 
Most important are the Dalitz decays of baryonic resonances N$^\star$(\textDelta) $\rightarrow$N $\mree$ and nucleon-nucleon bremsstrahlung. 
In particular, contributions from \textDelta\ Dalitz decays dominate at beam energies below the \texteta\ meson production threshold ($E_{beam} < 1.25$\gev).
\begin{figure}[tbh]
\begin{center}
\includegraphics[width=0.45\linewidth,height=13.4pc]{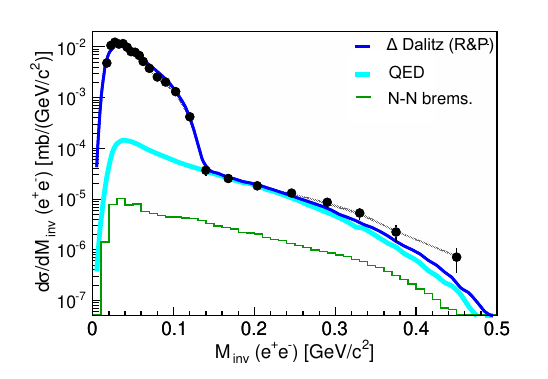}
\includegraphics[width=0.45\linewidth,height=13pc]{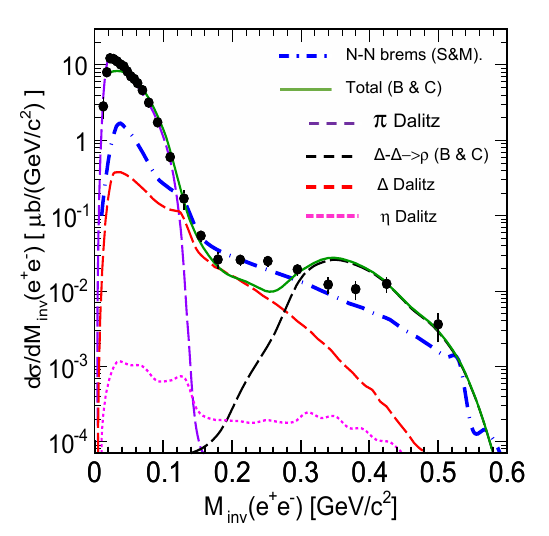}
\end{center}
\caption{
Dielectron Invariant mass distribution from exclusive $\mrpp\mree$(left panel) and $np\mree$ (right panel) at $1.25$~\gev\ in comparison to model calculations~\cite{Adamczewski-Musch:2017hmp, Adamczewski-Musch:2017oij}.
The cocktail contributions are labeled in the following way: full \pp\ cocktail including the eTFF description of Ramalho and P\~ena (\textDelta\ Dalitz (R\&P)) \cite{Ramalho:2015qna}, same calculation but removing \piz\ Dalitz and w/o using an eTFF (QED), yield due to non-resonant bremsstrahlung (N-N brems.).  
Likewise for \np\ collision displayed in the right panel: full OBE calculation of Shyam and Mosel \cite{Shyam:2008rx} w/o \piz\ Dalitz (N-N brems (S\&M)), full cocktail according to Bashkanov and Clement~\cite{Bashkanov:2013dua} assuming double \textDelta\ excitation (Total (B\&C)), respective contribution of the double excitation ($\Delta-\Delta>\rho$~(B\&C)), and various Dalitz contributions separately shown. }
\label{fig:hades-nn125} 
\end{figure}
The first experimental data on inclusive dielectron production in \pp\ and \xdp\ collisions in this energy range were provided by the DLS~\cite{Wilson:1997sr} and several years later also by HADES~\cite{Agakishiev:2009yf}. 
A recent comparison of HADES dielectron results to hadronic cocktails obtained from GiBUU, UrQMD, IQMD and HSD models (for details see~\cite{Weil:2012ji,Bratkovskaya:2013vx}) shows a good agreement for \pp\ collisions. 
The dominant contributions stem from $\Delta(1232)$ into $\mrN(\pi^0\rightarrow \mree\gamma)$, \myie \piz\ Dalitz decay, and from the resonance Dalitz decays into $\mrN(\gamma^*\rightarrow \mree)$. 
On the other hand, the hadronic cocktails show deficits for the \np\ (and for the \xdp\ collisions), which can be considered as a superposition of quasi-free \pp\ and \pn ) in the description of the invariant mass region towards the kinematic limit. 

Various explanations were put forward focusing on specific channels contributing in \pn\ collisions but being absent in \pp . 
In particular a stronger bremsstrahlung in case of  \pn\ was considered as a natural candidate.
As already discussed in Sec.\,\ref{sec:N-N-brems} in \pn\ collisions charged pion exchange contributes significantly to the scattering process, a channel, not available in $\mrp\,\mrp\rightarrow \mrp\,\mrp\,\mree$.  
In the OBE calculations of \cite{Kaptari:2005qz} and \cite{Shyam:2008rx} a coherent sum of amplitudes from non-resonant \nnnn\ bremsstrahlung, with graphs including emission from the (internal) meson exchange line, and a resonant $\Delta(1232)$ contribution were considered.
The results of the two calculations, however, differ by about a factor $2-4$, with \cite{Shyam:2008rx} coming closer to the inclusive dielectron data~\cite{Agakishiev:2009yf}, in particular towards higher invariant masses. 
A different phenomenological approach, based on the successful description of two-pion production in \pp\ and \np\ collisions via simultaneous excitation of two \textDelta\ isobars $\Delta\Delta\rightarrow \mrN\,\mrN\,\rho\rightarrow \mrN\,\mrN\,\mree$, was proposed in~\cite{Bashkanov:2013dua}.   

Fig.\,\ref{fig:hades-nn125} shows a comparison of the HADES dielectron data~\cite{Adamczewski-Musch:2017hmp,Adamczewski-Musch:2017oij} with the results of the calculations mentioned above. 
We focus here on the exclusive channels $\mrp\,\mrp\rightarrow \mrp\,\mrp\,\mree$ (left panel) and the quasi-free scattering $\mrn\,\mrp\rightarrow \mrn\,\mrp\,\mree$ (right panel) that characterize more directly the relevant channels. 
The latter has been obtained from \xdp\ collisions by tagging the forward-going spectator proton. 
The dielectron distribution from \pp\ is very well described considering a cocktail including \piz\ and $\Delta(1232)$ Dalitz decays using the \textDelta\ eTFFs from the covariant quark model~\cite{Ramalho:2015qna} in addition to bremsstrahlung derived in the OBE frame work.
The effect of the eTFF is moderate as can be seen by comparing the full calculation (cocktail) to the \textDelta\ Dalitz invariant mass distribution obtained assuming a point-like \textDelta N\textgamma$^*$ coupling 
(\mycf Fig.\,\ref{fig:GM-Delta} and the related discussion about the $\Delta(1232)$ transition form factor). 
Also shown in the left panel is the contribution of non-resonant bremsstrahlung used for the cocktail. 

Strikingly, the \np\ data exhibits a very different shape above the \piz\ pole mass as compared to \pp\ data. 
This can be recognized as an apparent excess above the  \textDelta\ Dalitz contribution, which has here a similar shape like in the \pp\ case.
The excess is almost completely accounted for by the OBE  model of~\cite{Shyam:2008rx} (N--N brems (S\&M)). 
The additional yield can be attributed to virtual photon emission from the (internal) charged pion line strongly affected by the pion eTFF. 
On the other hand, the model calculation assuming  \textDelta-\textDelta\ fusion of~\cite{Bashkanov:2013dua} (\textDelta\textDelta$\rightarrow$\textrho\ ($B\&C$)) also produces a strong excess and even slightly overestimates the data. 
The dominance of the \textDelta\ Dalitz contribution in the invariant mass range $0.14 < M_{\mree} < 0.28$~\gevcc\ is corroborated by the electron angular distributions in the helicity frame revealing a characteristic $~(1+B\,cos^2(\vartheta_l))$ distribution with $B=1.58\pm0.52$  \cite{Adamczewski-Musch:2017oij}, as $B\sim 1$ is expected for $\Delta(1232)$ Dalitz decay (~\mycf Sec.\,\ref{sec:N-N-brems}). 
The coefficient changes to $B=0.25\pm 0.35$ for the higher mass bin $M_{\mree}>0.28$~\gevcc.   
This can be interpreted as a dominance of the emission from the meson line via an off-shell \textrho ~meson. 
Indeed, the anisotropy parameter $B$ extracted in the reference frame with the $z$ axis, fixed to the direction of the charged pion exchange, becomes slightly negative $B=-0.4\pm 0.25$ which is more in line with the expected $B=-1$ for the pion annihilation process (\mycf Sec.\,\ref{sec:dalitz-decays}).

With increasing beam energy, \myie above 2\gev,  Dalitz decays of the \texteta, higher mass baryon resonances and the two-body decays of vector mesons start to contribute to the invariant masses above the \piz\ mass~\cite{Agakishiev:2009yf,Agakishiev:2012tc,HADES:2011ab}. 
The contribution of baryon resonances is particularly interesting because it is known that \textrho ~mesons couple to several N$^\star$ and \textDelta$^\star$ resonances~\cite{manley:92}. 
Calculations based on resonance models describe the \textrho ~meson production as a two-step process with the excitation of baryon resonance which subsequently decays into a \textrho ~meson.
The short-lived vector meson further decays into a pair of leptons (\epem\ or \mumu). 
Such a picture is equivalent to strict VDM where the resonance-virtual photon coupling is saturated by the \textrho.
However, as discussed in Sec.\,\ref{sec:dalitz-decays}, there is an alternative and more consistent approach to describe such a production via a Dalitz decay of a baryonic resonance, which coherently adds intermediate meson states. 

\begin{figure}[tb]
\begin{center}
\includegraphics[width=0.44\linewidth,height=0.45\linewidth]{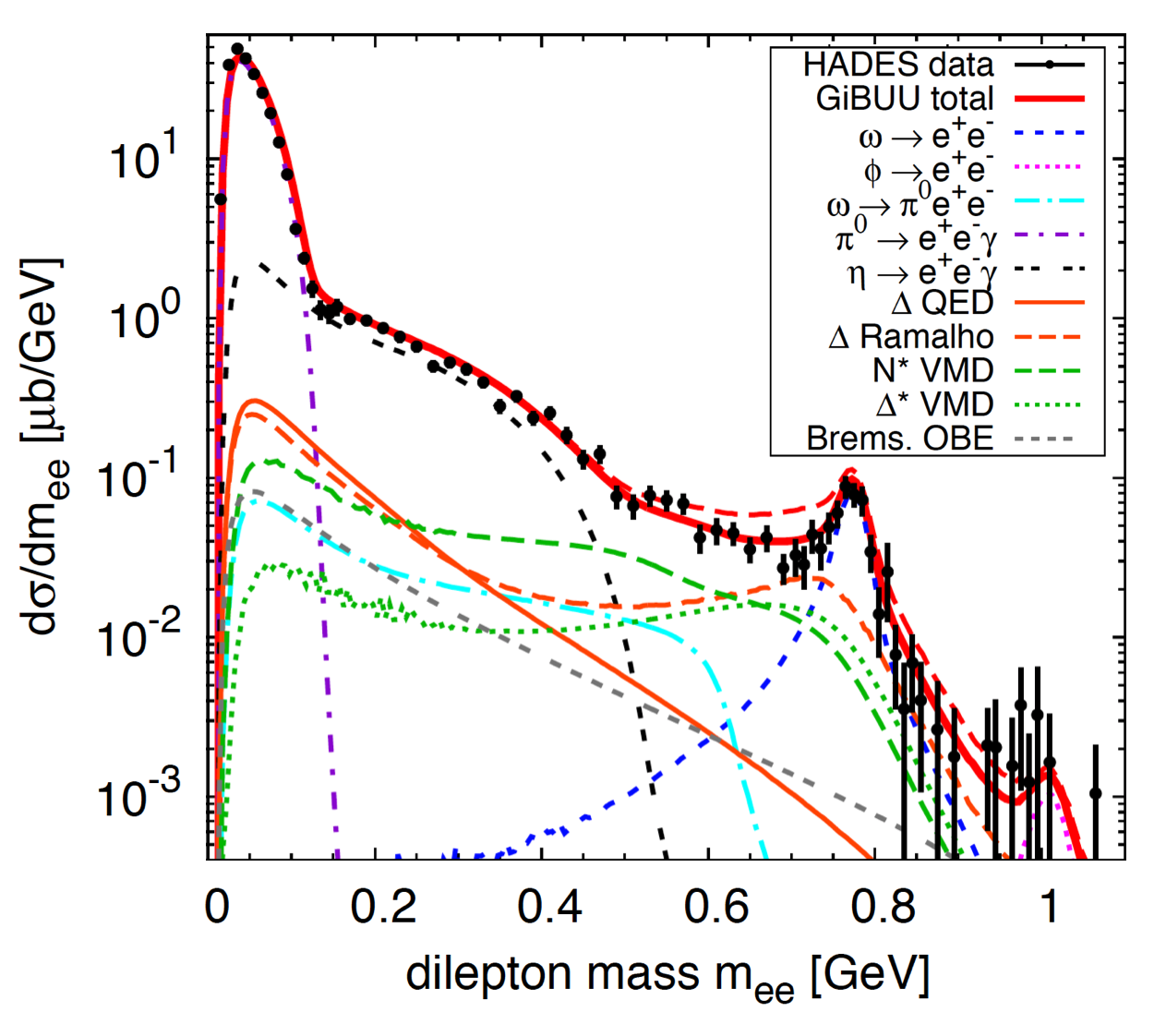}
\includegraphics[width=0.42\linewidth,height=0.45\linewidth]{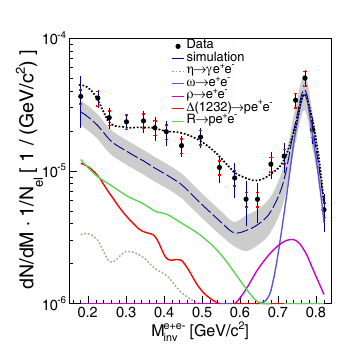}
\end{center}
\caption{Left: Invariant mass distribution for inclusive $\mree$ production measured in \pp\ collisions at $3.5$\gev~ by HADES \cite{HADES:2011ab} in comparison of hadronic cocktail calculated with GiBUU with ("\textDelta\ Ramalho") and without ("\textDelta\ QED") an $\Delta\rightarrow N\gamma^*$ eTFF \cite{Ramalho:2015qna}. 
For higher mass baryonic resonances strict VDM  was applied (see text for details). 
Right: Dielectron Invariant mass distribution for the exclusive  $\mrpp\mree$ at $3.5$~\gev\ final state~\cite{Agakishiev:2014wqa} in comparison to expected contributions from (i) Baryon Dalitz decays assuming constant eTFFs (dashed line with band) and two-body $\rho/\omega$ decays (ii) incoherent sum of $R\rightarrow p\rho\rightarrow p\mree$ (dotted-line) and two body \textomega\ decays (see text for further explanations).
}
\label{fig:hades-pp35}
\end{figure}

In the left panel of Fig.\,\ref{fig:hades-pp35} the inclusive dielectron invariant mass distribution for \pp\ collisions at $3.5$\gev\ taken by HADES~\cite{HADES:2011ab} is presented in comparison to a hadronic cocktail calculated with the microscopic transport code GiBUU~\cite{Ramalho:2015qna}. 
In the region $M_{\mreen} > 0.5$~\gevcc\ the calculation can saturate the yield  by contributions from decays of various higher-mass baryonic resonances and the two body decay of the \textomega\ meson. 
The decays of these resonances were modeled using the already mentioned two-step production mechanism involving intermediate \textrho ~meson (\mycf Sec.\,\ref{sec:dalitz-decays}). 
The resulting contributions are shown separately for the N$^* (I=1/2)$ and \textDelta$^*(I=3.2)$ resonances (VMD).
The cross sections for the different production channels were determined from fits to available data on meson production in pion and proton induced reactions (for details see~\cite{Weil:2012ji} and references therein).
It appears that the obtained parametrization of the production cross sections can be applied, with slight modifications (for details see \cite{Agakishiev:2014wqa}), to model also the exclusive one pion and dielectron production reconstructed from the same data . 

The right panel of Fig.\,\ref{fig:hades-pp35} shows the respective dielectron invariant mass distribution for the exclusive $\mrpp\,\mree$ final state. 
The data are compared to two different scenarios assumed in the modeling of R$\rightarrow$N\epem\ transition.  
While the dashed line represents the expected contribution for the transition with constant eTFFs (\myie no mass dependent form factors -- "QED" scenario), the dotted line shows the result of GiBUU calculations accounting for the two-step process with intermediate \textrho's. 
In the latter approach the branching ratios for the R$\rightarrow$N\textrho\ decays were taken from recent results obtained from a partial wave analysis of electron and pion scattering experiments (for details see~\cite{Agakishiev:2014wqa}). 
The shaded area spanning around the dashed line ("QED") uncovers the systematic error accounting for uncertainties in the baryon resonance production cross sections.
Furthermore, the contributions from the \textDelta(1232) and higher mass resonances (denoted by R) are plotted separately for the QED scenario.  
This is an interesting study demonstrating how the spectral distributions of contributions from \textrho\ decay are altered  depending on the  mechanism chosen for the baryon resonance decays. 
As one can see, abundant dielectron production through resonances with pole masses $M_R < M_N + M_{\rho}$ enhances the \textrho$\rightarrow$ \epem\ contribution right below the \textrho ~meson pole mass in accordance to the experimental data.  

An alternative approach to explain the dielectron spectral distribution below the \textrho ~meson pole masses is to apply dedicated electromagnetic baryon transition form factors at the resonance photon vertex.  
An attempt in this direction has been made including eTFFs only for the $\Delta\rightarrow \mrN\,\gamma^*$ transition based on the covariant quark model of Ramalho and Pe\~na. 
In contrast to the \pp\ collisions at $1.25$\gev~ discussed above, a prominent enhancement at the vector meson pole is visible (left panel of Fig.\,\ref{fig:hades-pp35}, (\textDelta\ Ramalho)). 
However, the inclusion of the predicted eTTF seems to deteriorate the agreement with data (dashed line). 
It is also questionable if such an approach is justified for the $\Delta(1232)$ resonance for masses far above the pole ($1232$\mevcc). 
Moreover, there are various parameterizations used to model the high-mass tail of the $\Delta(1232)$ and it is not clear which one has to be chosen. 
Its exact shape has a substantial influence on the calculated dielectron production rates (see for discussion \cite{Bratkovskaya:2013vx}).  
A more consistent calculation including eTFF's for all higher mass resonances is not available yet from this model.  It should be also noted that in the exclusive dielectron spectrum contributions from three body decays, like \texteta\ Dalitz, are almost completely removed by the kinematics, and the invariant mass spectrum is sensitive to the dilepton mass distribution from $\mathrm{R}\,\rightarrow \mrp\,\rho (\rightarrow \mree)$ decay. 
Indeed, the mass distribution shown in the right panel of Fig.\,\ref{fig:hades-pp35} can be considered, after subtraction of the \textomega\ line, as the line shape of the \textrho ~meson produced in this reaction. 
It shows a broad distribution located below the meson pole and hence cannot be described by a simple Breit-Wigner form. 
The reason is the strong coupling of the \textrho ~meson to baryons.
Taking a more microscopic view one can argue that the intermediary \textrho ~meson is modified by the vicinity of the baryon. 
Hence, ``medium effects'' are already apparent in \pp\ collisions.
This phenomenon is of particular importance in the description of thermal radiation of a baryon dominated resonance gas (\mycf Sec.\,\ref{sec:thermal-radiation}). 

The line shape of \textrho\ measured by HADES in \pp\ collisions is very much different from the line shape observed by CLAS \cite{Wood:2008ee} at similar energy but in photon induced reactions. 
The CLAS data is very well described by a relativistic Breit-Wigner distribution with $\Gamma_{\mree} \sim 1/M^3$. 
Although CLAS had a much smaller acceptance at low masses and low momenta ($p_{\mree}<1$ GeV/c$^2$), the acceptances of both spectrometers are comparable in the $0.5<M_{\mree}<0.8$\gevcc\ invariant mass range where the contribution from off-shell \textrho\ decay appears to be strong in the HADES data. 
This seems to indicate that contributions from baryonic resonances are much stronger in \pp\ than in $\gamma+p$ experiments. 
Such a conclusion is corroborated also by model calculations.
For photo-induced \textrho ~meson production they predict only a small contribution originating from baryonic resonance decay~\cite{Effenberger:1999ay}.
\subsection{Proton and Photon Induced Reactions on Nuclei}
\label{sec:coldmatter}
As discussed in Sec.\,\ref{sec:chiral}, the modification of vector mesons properties in cold nuclear has been proposed as a  precursor signal of chiral symmetry restoration. 
Two main approaches are presently pursued to conclude on in-medium mass/width modifications: (i) direct reconstruction of the invariant mass from detected decay products (line shape measurements) and/or (ii) measurements of the meson yields in reactions off targets of increasing size to infer on meson absorption in nuclear matter.
\begin{figure}[tbh]
  \begin{center}
    \includegraphics[width=0.45\textwidth,height=0.45\textwidth]{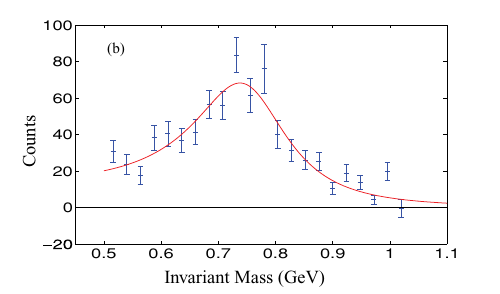}
    \includegraphics[width=0.45\textwidth,height=0.45\textwidth]{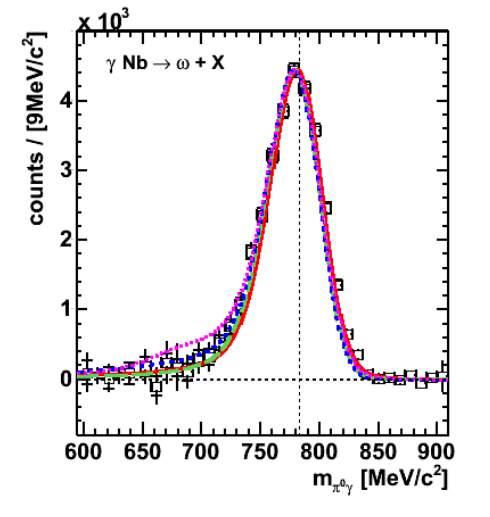}
  \end{center}
\caption{Left: \textrho ~meson line shape measured by CLAS in $\gamma+Fe,Ti$ reactions with a fit using a Breit-Wigner function~\cite{Wood:2008ee} (figure from aps). Right: \textomega\ meson line shape measured by CB/TAPS@MAMI in $\gamma+Nb$ compared to GiBUU calculations for various scenarios: assuming no in-medium modifications (red solid curve), only collisional broadening (green dashed curve), collisional broadening and mass shift by $-16\%$ at normal nuclear matter density (short dashed, blue curve) and mass shift without broadening (dotted, magenta curve) ~\cite{Thiel:2013cea}.
}
\label{fig:line-shape}
\end{figure}
%
\subsubsection{Vector Meson Line Shapes.}
Dilepton decays of \textrho ~mesons are best suited for line shape studies due to the absence of strong final-state interactions (FSI) and short live time of the meson ($c\tau=1.3$ fm/c). 
Dilepton signals from the longer-lived \textomega\ and \straightphi mesons achieve better signal to background but have stronger contributions from decays outside the nucleus.  
Furthermore, strong absorption of the mesons in nuclear matter leads to an increase of the total width (collisional broadening) and reduces the in-medium contribution to the total yield. 
Large statistics is needed to be sensitive to subtle effects on the meson tails arising from a small fraction of in-medium decays for the $\omega/\phi$~\cite{Weil:2012qh}. 
The sensitivity can be enhanced by selecting only mesons propagating slowly through the nucleus.
 
Pioneering experiments studying vector meson production off nuclei in the dilepton decay channels were performed at DESY in the seventies using few\gev\ photon beams~\cite{Alvensleben:1970zu}. 
They found a constructive interference between the \textrho\ and \textomega\ amplitudes in the coherent diffractive production with a phase difference $\Delta\phi_{\rho/\omega}=41\pm20^O$.
The interference pattern was later also investigated in the high statistics NA60 measurement in \pA\ collisions, described in Sec.\,\ref{sec:elementary_sps}.
Two possible solutions for the phase difference were derived, only one consistent with the DESY result. 

The first dedicated dilepton experiment searching for in-medium modifications in the dilepton channel was E325 at KEK~\cite{Ozawa:2000iw,Naruki:2005kd}.  
It reported a $\sim 9\%$ mass drop of the \textrho ~meson at nuclear ground state density in $12$\gev~ p--A collisions (the aforementioned interference effect was excluded as possible explanation).  
This finding was connected to a mass shifts predicted in the context of QCD sum rules~\cite{Hatsuda:1991ez} or due to ``Brown-Rho scaling'' (\mycf \ref{sec:chiral}). 
Also a mass modification of the \straightphi\ meson has been reported. 
The E325 collaboration observed a $\sim 3.4\%$ downward shift accompanied by $\sim 4$ larger $\Gamma^{\mree}$ width for ``slow'' ($\beta\gamma<1.2$) mesons~\cite{Muto:2005za}.

The follow-up experiment CLAS at JLAB investigated photo-production at energies of $E_{\gamma}=0.6-3.85$\gev~ off nuclear targets (H2, C, Fe, Ti, Pb).
However, they could not confirm the \textrho\ mass shift reported by the KEK experiment ~\cite{Wood:2008ee,Nasseripour:2007aa}.
Instead, some modest broadening of the \textrho ~meson mass distribution ($\sim 40$\mev~ ), most visible for the heaviest targets (Fe-Ti), was concluded (see Fig.\ref{fig:line-shape} --- left panel). 
It was also suggested that the \textrho\ mass shift reported by E325 could be due to the applied background subtraction method based on a fit, instead of exploiting the like-sign pair technique.
The latter could not be applied in E325 because of the trigger setting used in this two-arm experiment requiring oppositely charged leptons each in one of the arms. 
As a result, possible contributions from correlated background pairs modifying the shape and the absolute normalization of the combinatorial background could not be evaluated~(\mycf Sec.\,\ref{sec:CBdeterm}.) 
This could have lead to an overestimation of the background in the mass region above the \textrho\ peak, resulting in an apparent downward shift of the meson mass distribution.
The CLAS collaboration found no modification of the \textomega\ lines shape. 
It should be noted, however, that the specific detector acceptance permitted only the detection of large momentum mesons ($p \ge 1 $\gevc) only.

Another approach was pursued by the CBTAPS experiment using the MAMI accelerator in Mainz.
They measured the process \omegatodalg\ in photo-production off nuclei with different mass number.
The photon energy was selected close to the meson production threshold ($E_\gamma=0.9-1.3$\gev~\cite{Trnka:2005ey,Thiel:2013cea}). 
Reconstruction of the Dalitz-decay has the advantage of strongly reduced (by a factor $\simeq 10^{4}$) background due to \textrho ~meson production but the disadvantage of strong final state interactions of the daughter pion. 
The latter was suppressed requiring a minimum energy of the escaping pion of $E_{\pi^0}>150$\mev .
Unfortunately, this selection leads at the same time to a suppression of slow \textomega\ mesons and consequently to a reduced sensitivity on in-medium decays. 
While the first measurement indicated a downward shift of the \textomega\ spectral distribution~\cite{Trnka:2005ey}, subsequent measurements with improved background subtraction showed no effect.
Fig.\,\ref{fig:line-shape} (right panel) shows the respective $\pi^0\gamma$ invariant mass distribution in comparison to the expectation of the GiBUU transport model calculation~\cite{Weil:2012qh} for various scenarios: 
no in medium modifications (red solid curve), collisional broadening only, with a total width of $\Gamma(\rho) =\Gamma_0(=150 MeV)\rho/\rho_0$) (green dashed curve), in-medium mass scaling like $m = m_0(1-0.16\,\rho/\rho_0)$ (magenta dotted) and with combined mass shift and  collisional broadening (blue dotted). 
Only little sensitivity to the various assumptions is observed although the statistics of the measurement is very good, likely for the reasons discussed above.
Within the error bars, the scenario assuming a mass shift but no broadening seems to be less likely.        

\begin{figure}[tb]
  \begin{center}
    \includegraphics[width=0.9\linewidth]{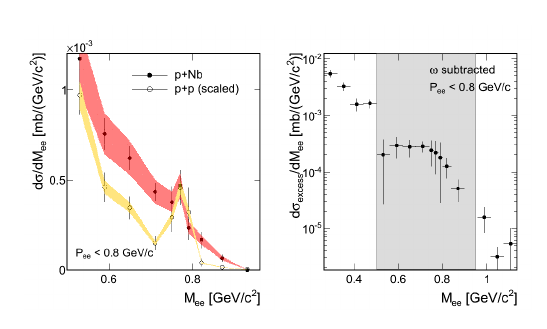}
  \end{center}
\caption{Left: Comparison of dielectron cross sections as a function of the invariant mass measured in \pp\ (shown in Fig.\,\ref{fig:hades-pp35}) and $\mrp+\mathrm{Nb}$ collisions at beam energy of $3.5$\gev\ zoomed in the vector meson region. 
The shaded bands represent the systematic uncertainties due to the normalization. 
Right: Excess yield in the \pnb\ data after subtraction of the scaled \pp\ reference data (the \textomega\ contribution has been subtracted in both data samples). 
The gray region corresponds to the invariant mass range plotted in the left picture \cite{Agakishiev:2012vj}. 
}
\label{fig:pNbexc}
\end{figure}

New results from proton-induced reactions and with dielectron final state were obtained by the HADES collaboration investigating \pnb\ collisions at 3.5~\gev~\cite{Agakishiev:2012vj}. 
The large acceptance of the detector and the moderate beam energy enabled the detection of electron pairs from \piz , \texteta\ (via Dalitz decays) and the \textrho ,\textomega\ with low laboratory momenta of $p_{\mree} < 1.0$~\gevc . 
Moreover, the measurement had also sensitivity in the low-mass continuum down to the \piz\ Dalitz region, not covered before by the CLAS and E325 experiments. 
The direct comparison of distributions measured in \pp\ and \pnb\ (see Fig.\,\ref{fig:pNbexc}) reveals additional strength below the vector meson pole mass in the \pnb\ case, most  pronounced for low momentum dielectrons $p_{\mree}<0.8$\gevc .
Such a selection enhances the decay probability in the medium and probes the in-medium spectral function in a region were the strongest effects due to meson-baryon coupling are expected (see \mycf Sec. \ref{sec:hadronic-model}).
For better comparison, the dielectron cross section for \pp\ is scaled 
by the average number of participants calculated with a Glauber model according to $\sigma_{pNb}/\sigma_{pp} \times <A_{part}^{pNb}>/<A_{part}^{pp}>$. 
With such a scaling the \piz\ production measured in the \pp\ describes the pion Dalitz yield in \pnb , \myie the \textpi\ production scales like $A_{\mathrm{part}}$. 
The comparison of the two spectra reveals a significant reduction of the \textomega\ signal in the \pA\ case which can be interpreted as strong absorption of the \textomega\ meson in nuclear matter (see next section).
The \textomega\ peak was removed in both data samples and the properly scaled continuum dielectron yield from \pp\ was subtracted from the \pnb\ yield.
The remaining yield is shown in Fig.\,\ref{fig:pNbexc} (right panel) and can be interpreted as the in-medium (excess) contribution to the total yield. 
It decreases exponentially revealing a bump structure directly below the \textomega\ pole mass. 
This excess might  be interpreted as a sign for contributions from secondary processes involving the formation and decay of baryon resonance states (R) like p$\rightarrow$\textpi ~X and  \textpi N$ \rightarrow$R$\rightarrow$N\epem\ and possibly in-medium meson modifications.
In terms of strict VDM the first contribution can as well be understood as due to decays of in-medium (or far off-shell) \textrho ~mesons. 
Indeed, as discussed in Sec.\,\ref{sec:sis-bevalac}, such decays show a strong enhancement in exactly this mass region due to strong baryon resonance-\textrho\ couplings. 
Such an interpretation is supported by calculations with the GiBUU transport model~\cite{Weil:2012ji} where also "mass dropping" and collisional broadening are taken into account explicitly. 
The effect of secondary collisions involving higher mass resonances was found to be very important. 
However, as already discussed in previous sections, the results of such calculations have to be taken with caution because of large uncertainties related to the description of baryon resonance dielectron decays.
A different modeling of dielectron production has been applied in the HSD transport calculations~\cite{Bratkovskaya:2013vx}.
In this model, the measurement is explained using string fragmentation rather than baryon resonance production. 
Only $\Delta(1232)$ resonances are explicitly produced and propagated, though with a production cross section about a factor two larger than the one used in the GiBUU transport code. 
Collisional broadening of the $\rho/\omega$ mesons is used as well. 

Hence, no clear conclusion about in-medium modifications of the \textrho ~meson could be achieved so far from this data, which reveals a strong cold matter effects in dilepton radiation.
The proper modeling of the baryon-resonance decays and treatment of in-medium effects remains a challenge for microscopic transport codes in this energy region.
A simultaneous description of multi-differential observables in different collision systems will help to foster the theoretical description of dilepton radiation from cold matter. 
It is expected that the on-going HADES program with pion beams will provide more constraints on the role of resonance decays herein.
\subsubsection{Production Experiments}
An observable to address meson absorption in cold matter is the so-called transparency ratio ($T_A$).
It is defined as the ratio of the meson production cross section in a given photon (or proton)-nucleus reaction to the respective cross section on a single nucleon, scaled with the nuclear mass number $A$ of the target.  
The relation between the transparency ratio and the in-medium width can be obtained from the eikonal approximation~\cite{Cabrera:2003wb}, which general holds better if the relative momentum between meson and medium is large.
It relates the transparency ratio to the imaginary part of the meson self-energy $\Im \Pi(p,\rho(r))$. 
The latter in turn is related to the meson in-medium width in the nucleus rest frame $\Gamma_{coll}\, (p,\rho(r))$ and the imaginary part of in-medium potential. 
Both quantities, $\Pi\,(p,\rho(r))$ and $\Gamma_{coll}\,(p,\rho(r))$, depend on the meson momentum $p$ and the local baryon density $\rho(r)$.  
The expression reads
\begin{equation}
T_A \equiv \frac{\sigma_A}{A\sigma}=\frac{1}{A}\int d^3 r\, \rho(\vec{r}) \,
\exp \left[\frac{1}{p}\int_{0}^{\infty} dl \, \Im \Pi \left(p,\rho (\vec{r'})\right)\right]
\label{eqn:transparency}
\end{equation}
where
$-\Im \Pi(p) = \Gamma_{coll(p)}\,\omega$,      
%
and
$\vec{r'} = \vec{r} + l \vec{p}/p\,$. 
The production cross section for dileptons off a nucleus with mass number $A$ is $\sigma_A$. 
The first integral is taken over the volume of the nucleus and the second integral along the (straight) path of a produced meson. 
In analogy to the optical model the energy is here labeled as \textomega.

When discussing experimentally obtained transparency ratios one should keep in mind the importance of any kind of nuclear effect, which have to be considered carefully in calculations.
The first is the initial production probability, which is large (surface dominated) for \myeg pion-induced reactions but small (``illuminating'' the whole nucleus) in photo-production experiments.
Moreover, at threshold energies the momentum distribution of the target nucleons has to be modeled precisely as details in high-momentum tails,  \myeg through short-range correlations, can significantly enhance the production probability. 
A further complication arises from two-step meson production processes, which involve intermediary hadrons produced in a first-chance collision (\myeg a pion) which then, in second collisions, produce the meson finally observed. 
Such processes are evidently not present in reactions on single nucleons.  
In order to reduce such effects, the transparency ratio is usually defined \mywrt a reference measurement on light nuclei (for example carbon). 
But even this does not completely guarantee full elimination of the effects related to two-step production, \myeg because of dependence on size and clustering phenomena. 
Last but not least, in cases where a daughter hadron is emitted along with the photon or dilepton, its FSI have to be considered carefully. 
One should also emphasize that meson production can be affected by the real part of the in-medium potential. 

The TAPS experiment obtained results for the transparency ratio for \textomega\, \straightphi, \texteta\ and \texteta'. 
The data demonstrate significant absorption in cold nuclear matter for all addressed mesons.
For more details on this topic we refer the interesting reader to recent reviews \cite{Leupold:2009kz,Metag:2017ixh}. 
%
\begin{figure}[tb]
  \begin{center}
    \includegraphics[width=0.45\textwidth,height=0.45\textwidth]{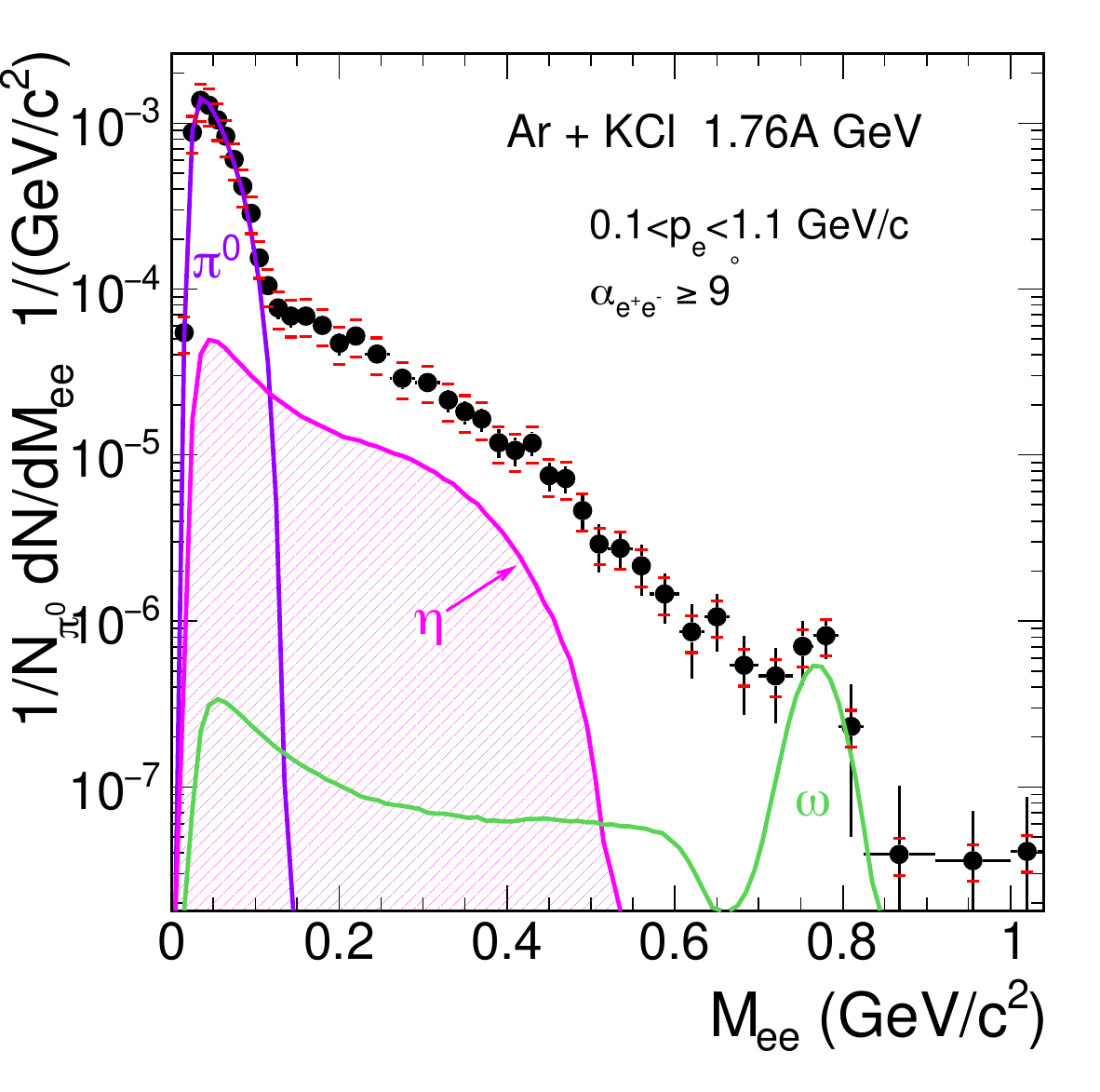}
    \includegraphics[width=0.45\textwidth,height=0.45\textwidth]{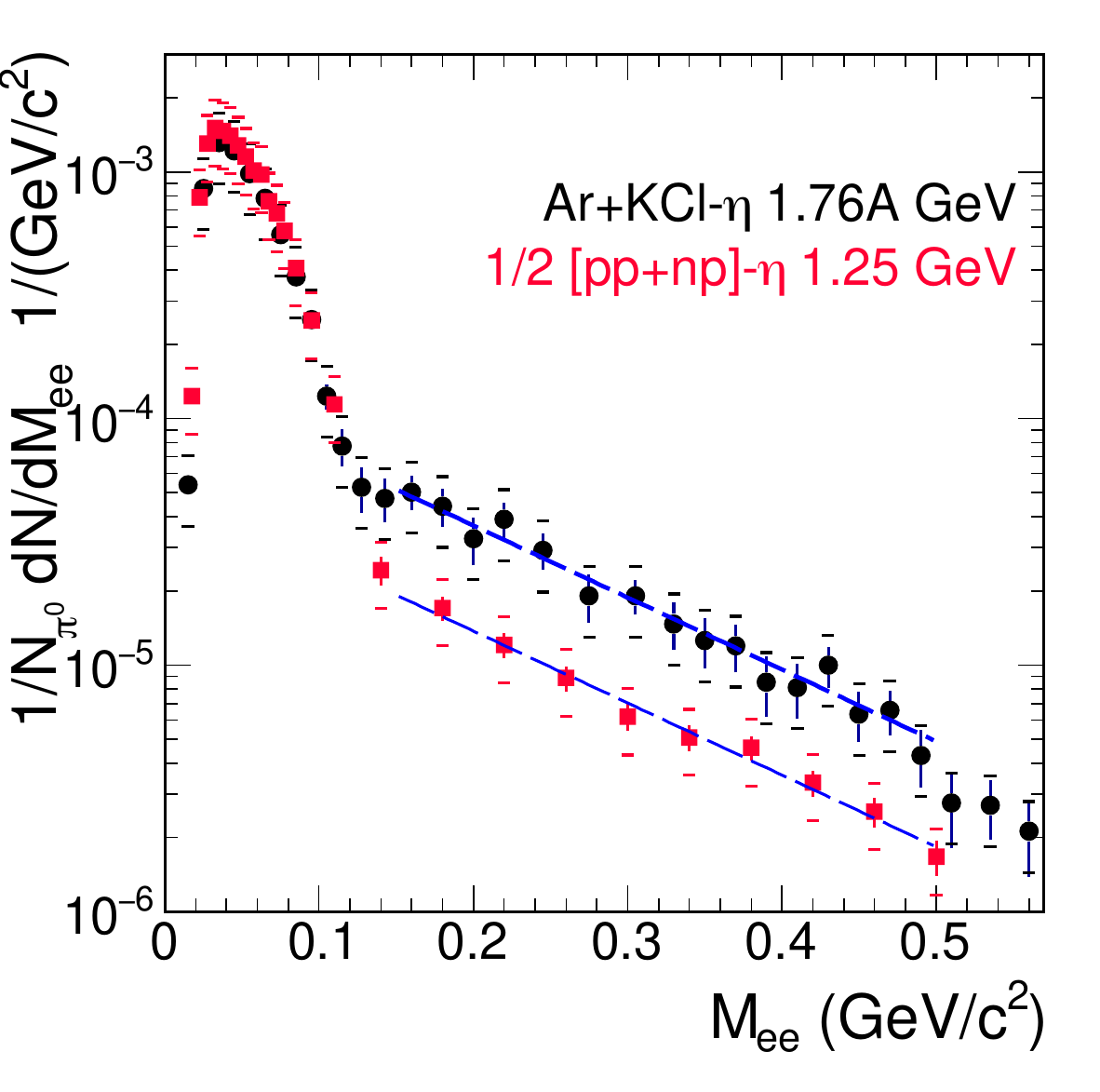}
  \end{center}
\caption{Left panel: Differential $\mree$ multiplicities normalized to the $1/2(M_{\pi^+}+M_{\pi^-})$ as function of the dilepton invariant mass. The data has been obtained from \arkcl\ collisions at 1.756\agev .
Also shown is the cocktail of dilepton from mesonic sources (\piz, \texteta, \textomega) accounting for contributions from decays after freeze-out~\cite{Agakishiev:2007ts}.  
The excess radiation clearly outshines the contribution from \texteta\ Dalitz decays (shaded). 
Note the conditions on the single lepton laboratory momentum and the opening angle. 
Right panel: Same data but after subtraction of the mesonic  cocktail, except \piz\ (black points, see inset).
Also shown is the reference spectrum scaled by to the mean number of participant nucleons assigned to the ion data (red dots).
The blue lines are shown to guide the eye and indicate the ``medium radiation''.
}
\label{fig:arkcl}
\end{figure}
%
 \subsection{Heavy-Ion Collisions}
\label{subsec:heavy_ion}
\subsubsection{BEVALAC/SIS18}
In heavy-ion collisions at 1-2\agev\ pions are the only abundantly produced mesons.
They originate mainly from \textDelta(1232) resonance decays.
The production of \texteta\ mesons is already suppressed by orders of magnitude but yet constitutes an important contribution to the hadronic dilepton cocktail.
Respective \piz\ and \texteta\ meson multiplicities were measured by the TAPS collaboration at GSI investigating the two-photon decay channel~\cite{Averbeck:1997ma, Holzmann:1997mu, Averbeck:2000sn}. 
It was found that the multiplicities follow $m_t$-scaling \cite{Bratkovskaya:1997dh}.
However, pioneering dilepton experiments performed by the DLS collaboration at the BEVALAC showed substantial dilepton yield above contributions from \texteta\ Dalitz decays~\cite{Porter:1997rc} already for comparatively light collision systems. 
This additional yield could not be explained by any transport calculation, even after inclusion of in-medium modifications of the \textrho ~meson (for details see~\cite{Rapp:1999ej}). 
This situation motivated next generation experiments conducted with the HADES detector at GSI.

HADES measured inclusive dielectron production in the light $\mrC+\mrC$ (at $1$ and $2$\agev)~\cite{Agakichiev:2006tg, Agakishiev:2007ts}, the medium--heavy \arkcl\ (at $1.756$\agev)~\cite{Agakishiev:2010rs} and the heavy $\mrAu+\mrAu$ (at $1.23 $\agev) collision system~\cite{Adamczewski-Musch:2019byl}. 
Fig.\,\ref{fig:arkcl} shows as example the result obtained for the \arkcl\ collision system. 
The invariant mass distribution of dielectron pairs is normalized to the mean charged pion multiplicity $\bar{M} = 0.5\, (M_{\pi^+}+M_{\pi^-})$.
Charged pions were measured with HADES as well and their respective yields extrapolated to full solid angle.
The mean charged pion multiplicity served as substitute for the neutral pion multiplicity $M_{\pi^0}$ and agreed very well with the \piz\ systematics measured by TAPS in the two-photon channel. 
The dielectron invariant mass spectrum for the 40\% most central collisions is compared to the hadronic cocktail in the left panel of Fig.\,\ref{fig:arkcl}.
%
\begin{figure}[tb]
   \begin{center}
\includegraphics[width=0.5\textwidth]{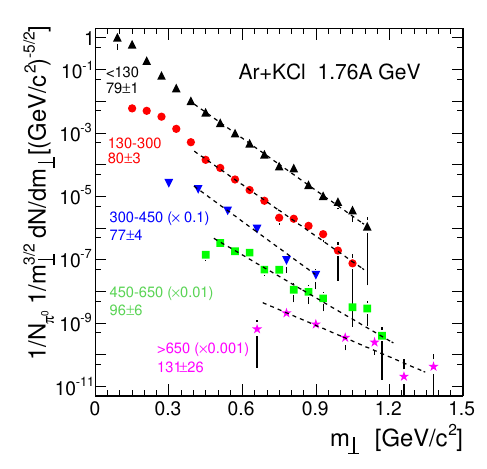}
   \end{center}
\caption{Transverse mass distribution of virtual photons corrected for the HADES detector acceptance and efficiency  for various mass bins and with exponential (Boltzmann) fits. 
Inverse slope parameters are given in the figure~\cite{Agakishiev:2011vf}.}
\label{fig:arkcl-mt}
\end{figure}  
The bump observed around 800\mevcc\ agrees with the cocktail \textomega\ two-body decay very well.
It is the first measurement of an \textomega\ signal in heavy-ion collisions at a collision energy below its nominal $\mrN+\mrN$ production threshold.
For the cocktail generation the \textomega\ yield was estimated using $m_T$ scaling (for details see~\cite{Agakishiev:2010rs}). 
As one can clearly see, the dielectron cocktail composed of meson decays after freeze-out does not explain the measured yield but does not yet contain expected contributions originating from baryonic sources, as they were found already in \nnnn\ collisions (\mycf Sec.\,\ref{sec:sis-bevalac}). 
The most relevant baryonic contributions to the dilepton yield are Dalitz decays of baryonic resonances (mainly $\Delta(1232)$) and \nnnn~bremsstrahlung.
Since the modeling of dileptons yields from baryonic sources is subject to large uncertainties, these contributions were approximated by reference measurements using \pp\ and \np\ collisions at $1.25$\agev. 
The inclusive invariant mass distributions obtained in these experiments were also normalized to the respective neutral pion multiplicities and averaged over all types of \nnnn\ collisions using proper isospin factors $(M_{pp}^{\mree}+M_{pn}^{\mree})/(2 M_{\pi^0})$ (for that it was assumed that dilepton production in \pp\ and \nn\ is the same).
The reference distribution can be regarded as contribution of baryonic sources from \nnnn\ first chance collisions and is normalized to $M_{\pi^0}$. 
Note that the small \texteta\ Dalitz yield measured in these \nnnn\ collisions has been subtracted to obtain the reference spectrum.
The subtraction of the \texteta\ contribution is important as the \texteta\ meson production mechanisms close to threshold differ substantially in nucleus--nucleus and nucleon--nucleon reactions at this beam energy and therefore has to be estimated separately. 
One should also note that the normalization to $M_{\pi^0}$ takes the excitation function of baryon resonance production with beam energy into account. 
Moreover, also the system size dependence is taken care of via implicit $A_{part}$ scaling since the pion multiplicity is known to be proportional to $A_{part}$ in heavy-ion collisions. 

Fig.\,\ref{fig:arkcl} (right) shows a comparison of dilepton yields, with subtracted \texteta\ meson contribution, recorded for \arkcl\ to the respective expected baryonic contributions from first chance collisions measured in the \nnnn\ reference reactions (the reference spectrum defined above).
The contribution from \piz\ Dalitz-decay is included in the measured distribution to confirm the normalization procedure but the contributions from the other meson Dalitz decay are consistently excluded to remain with baryonic sources only in the mass region above 150~\mevcc . 
A significant excess yield is, however, visible above the pion mass indicating the presence of medium radiation. 
To scrutinize this conclusion the same analysis was carried out for the \cc\ collision system at 1 and 2\agev.
No significant excess radiation above the reference spectrum could be observed for this light collision system
~\cite{Agakishiev:2009yf}.
This observation explained the long standing "DLS puzzle" connected to the dilepton yield measured in \cc\ collisions as being due to unexpected baryonic contributions not properly accounted for in the theoretical description of the data. 

The high statistics of the measurement enabled multi-differential studies. 
As an example, Fig.\,\ref{fig:arkcl-mt} displays transverse mass distributions of the virtual photons $m_t=\sqrt{M+p_t}$, integrated over rapidity, for the five indicated invariant mass bins. 
The distributions were fitted with the exponential function $1/m_t^{3/2} dN/dm_T\sim exp(-m_t/T)$ and revealed inverse slope parameters increasing for the high-mass bins.
Visible is a significant change in the inverse slope parameters for masses higher than the 450~\mevcc .   
Indeed, the thermal population of (in-medium) \textrho ~mesons states with significant densities requires high temperatures.
However, particles with higher mass also gain more transverse momentum if the emission system features collective radial expansion.
Hence, a unique interpretation of the underlying physics being responsible for the observed inverse slope parameters is not available yet. 
It is interesting to note in this context that the freeze-out temperature, obtained for this collision system from fitting the particle abundance in the context of a strangeness-canonical statistical hadronization model, was found to be $73\pm 5$\mev~\cite{Agakishiev:2010rs,Agakishiev:2015bwu}).
The fit converges at this high temperature only by driving the volume parameter to a rather small value.

\begin{figure}[tb]
  \begin{minipage}{\linewidth}
    \begin{center}
    \includegraphics[width=0.32\linewidth,height=10pc]{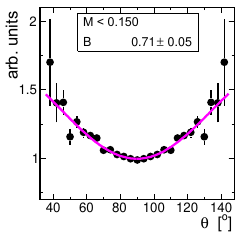}
    \includegraphics[width=0.32\linewidth,height=10pc]{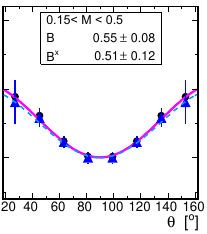}
    \includegraphics[width=0.32\linewidth,height=10pc]{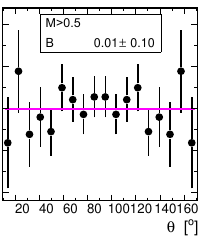} 
    \end{center}
  \end{minipage}
    \caption{Electron angular distributions in the helicity frame for three invariant mass bins measured in \arkcl\ at $1.756$\agev \cite{Agakishiev:2011vf}. 
    The asymmetry is quantified by adjusting the function $ dN/d\theta\sim(1+B\,\cos^2\theta)$.
    The result is shown as full (purple) line.}   
\label{fig:arkcl-hel}
\end{figure}
Another interesting information, addressed for the first time at such low collision energies, is electron angular distributions in the helicity frame which, as discussed in Sec.\,\ref{sec:theory} carry information about the virtual photon polarization.      
Fig.\,\ref{fig:arkcl-hel} shows electron helicity angular distributions, corrected for the detector acceptance and efficiency, for the three indicated mass bins.
The distributions are fitted with the function $dN/d(\theta)\sim 1+B\,\cos^2\vartheta$ to extract the asymmetry quantified by $B$ (the results are quoted in the figure). 
For the lowest mass bins, which are dominated by lepton pairs from  \piz\ Dalitz decay, the outcome is as expected (\mycf Sec.\,\ref{sec:dalitz-decays}).
One should, however, note that the explicit reconstruction of the $\pi^0\rightarrow\gamma^*\gamma$ decay is not possible in such an inclusive measurement as the real photon is not measured. 
Consequently, also the daughter $\gamma^*$ rest frame cannot be correctly reconstructed in the \piz\ decay frame.
Therefore, angular distributions of electrons in the $\gamma^*$ rest-frame reconstructed from the two daughter electrons only are slightly affected by this effect. 
Simulations accounting for this effect show~\cite{Agakishiev:2009yf} that for the \piz\ and \texteta\ mesons values of $B = 0.6-0.7$ are expected in agreement with the results obtained for the two mass bins, addressed in the left and center panels referring to \piz\ and \texteta-Dalitz, respectively. 
Similar anisotropies are also expected for \textDelta(1232) Dalitz decays which contribute mainly to the spectrum shown in the center panel and depicting the range $0.15<M=M_{\mreen}<0.5$.
But in this mass range contributions from \texteta\ Dalitz can hamper the distribution and complicate the interpretation.
In the middle panel, the triangles show the angular distribution obtained after subtraction of the expected \texteta\ Dalitz contribution.
The resulting fit is shown as dashed line and the corresponding anisotropy parameter $B^x$ changes very little. 
In contrast, \pn\ bremsstrahlung is not expected to produce anisotropic distributions (\mycf Sec.\,\ref{sec:sis-bevalac})). 
Hence it might be justified to conclude that in the $0.15<M<0.5$ mass region bremsstrahlung is of minor importance and is dominated by the \textDelta-like contribution. 
Indeed, the additional strength observed in this mass range (in-medium radiation) can be understood as due to radiation from multiple \textDelta\ resonances produced and decaying during  the dense phase of the collisions. 
Such a conclusion is corroborated by the transport code calculations \cite{Bratkovskaya:2013vx}. 

The very different pattern of angular distribution (isotropic emission) is visible for the highest mass bin and might indicate change of the dominating source. Indeed, in this mass range contribution from the vector meson, especially \textrho ~meson, are expected. Such isotropic emission was indeed measured in URHIC at SPS by NA60 in \inin\ collisions~\cite{Arnaldi:2008gp}.

The observed anisotropies of electron angular distributions and their dependence on the pair invariant mass still awaits a consistent explanation.
The anisotropies measured in the ``\textDelta'' mass bin are apparently in disagreement with calculations for thermal radiation discussed in \cite{Speranza:2018osi} (see \mycf Sec. \ref{sec:thermal-radiation}). 
They  might be a fingerprint of a specific mechanism related to radiation from multiple \textDelta\ generations. 
New results from \auau\ collisions  measured by HADES at smaller energies will shed more light on this interesting phenomenon.  
\subsubsection{SPS}
\label{sec:sps}
\begin{figure}[tb]
  \begin{center}    
    \includegraphics[width=0.5\linewidth]{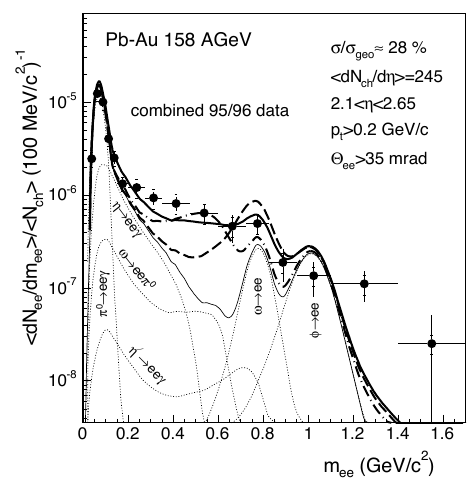}
  \end{center}
\caption{Dielectron invariant mass spectrum obtained by CERES in \pbau\ collisions at maximum SPS energies.  
The data was collected in two runs and is for the 28\% most central fraction of the geometrical cross section (the respective mean charged particle rapidity density is given in the inset together with other conditions applied in the analysis like pair transverse momentum $p_t$ and opening angle $\Theta_{\mreen}$). 
Systematical and statistical errors are given as horizontal and vertical bars, respectively. 
The thin lines depict the mesonic cocktail (with related labels) and the thick lines the sum of the cocktail plus various assumptions for the contributions due to two-body \textrho\ decay \cite{Agakichiev:2005ai}.
}   
    \label{fig:pbau-ceres}
\end{figure}
In the early nineties, second generation experiments\footnote{This notation was used by Hans Specht in his presentation on the occasion of a celebration event ``30 years lead beam at SPS''. It is in distinction to the first experiments operated at SPS with Sulfur and proton beams which where realized by reusing and complementing already existing detector systems.} were installed in the target halls of the SPS at CERN to search for the conjectured new phase of strongly interacting matter, the QGP.
First lead beam in the SPS became available in 1994. 
Dilepton data for lead beams at maximum SPS energy were taken by CERES in 1995 and 1996. 
The dielectron invariant mass spectrum obtained from the combined statistics is shown in Fig.\,\ref{fig:pbau-ceres}~\cite{Agakichiev:2005ai}. 
Although the statistics and resolution of the spectrometer were not sufficient to clearly identify contributions from the vector mesons, the result nevertheless established for the first time a substantial excess yield above contributions from meson Dalitz decays in the invariant mass region $300 < M_{ee}/$(\gevcc )$<700$. 
After the ``anomalous'' excess in proton-induced reactions had been finally explained by incorrectly accounted \texteta\-Dalitz decays (see Sec.\,\ref{sec:elementary_sps}), now an excess in the same mass range was found in collisions of heavy ions. 
This excess could soon be explained as being due to in-medium \textrho ~meson decays. 
Indeed, the most plausible underlying process is continuous production and decay of \textrho ~mesons in the pion-dominated medium as discussed in Sec.\,\ref{sec:thermal-radiation}. 
At each time and location during the evolution of the fireball \textrho ~mesons are expected to be in chemical equilibrium at these beam energies.
This is exactly the situation assumed in calculations of the emissivity of a hadron resonance gas based on the in-medium self-energy of the \textrho\ and strict VDM (\mycf Sec.\,\ref{sec:chiral}). 
The integral yield above the contribution from late meson decays not only depends on the temperature (and to a smaller account on the baryo-chemical potential) of the system, but also scales with the lifetime and volume of the evolving fireball formed in the collision center. 
Since a longer lifetime allows for more generations of \textrho ~mesons, the yield was said to provide a "\textrho\ clock" for fireball evolution.  

Besides the shear amount of excess radiation, also its spectral distribution is of interest.
In comparison with model calculations, shown as thick lines in Fig.\,\ref{fig:pbau-ceres}, it is apparent that the dominant medium effect on the \textrho ~meson is rather a strong broadening than a shift of the \textrho\ pole-mass towards smaller values without broadening, as conjectured in the late eighties~(\mycf Sec. \ref{sec:coldmatter}). 
These findings can be inferred from the full and the dashed--dotted lines,  which represent calculations using a broadening and a dropping scenario, respectively. 
Further support for the explanation of the excess radiation as being due to thermal radiation out of a pion-dominated fireball came from the inspection of the $p_t$ dependence of the excess.
Indeed, the invariant mass distribution obtained for pairs with transverse momenta in excess of 500~\mevc\ shows much less excess radiation and cannot discriminate between calculations with and without medium modification of the  \textrho~\cite{Agakishiev:1997au}. 
This is exactly what is expected in case the \textrho ~mesons are produced in a thermal environment with a temperature around $kT = 150$~\mev . 
Yet, the quality of the data did not allow to draw firm conclusions at the time of their publication.

Later, the CERES experiment was complemented by adding a Time Projection Chamber with a separate, second magnetic field behind the RICH section~\cite{Adamova:2008mk}. 
The TPC not only improved the momentum resolution but also provided additional discrimination power between electrons and pions making use of their specific energy loss in the counting gas.
The new setup was first used for a 40\agev\ run~\cite{Adamova:2002kf}, but with a statistics not reaching close to the one collected for 158\agev .
At lower beam energies the fractional stopping of the incoming baryons is increased while anti-baryon production is generally below the percent level of the stopped nucleons at SPS energies. 
Consequently, more sensitivity to the baryon-driven broadening of the \textrho\ spectral function is expected. 
The invariant mass spectrum obtained for this collision system indeed showed a stronger excess radiation as observed for 158\agev .
The CERES collaboration quantified the excess radiation by subtracting from the the measured yield, integrated in the range $0.2 \leq M_{ee}/(\mathrm{G}e\mathrm{V}/c^2 )\leq 1.0$, the respective yield as calculated with the GENESIS code.
The numbers obtained amount to $2.31 \pm 0.19 (stat) \pm 0.55 (syst) \pm 0.69 (decays)$ and $5.1 \pm 1.3(stat) \pm 1.0 (syst) \pm 1.5 (decays)$  for the measurements at $158$\agev\  and $40$\agev , respectively. 
Moreover, the spectral shape appeared to be rather featureless above the \piz\ Dalitz region, decreasing monotonously by a factor two in yield when going from 200\mevcc\ to 1\gevcc .  
These observations were taken as strong evidence for the dominance of baryonic effects over temperature as leading factor for medium modifications of the \textrho~\cite{Adamova:2002kf}.
Data obtained in a third run with this 158\agev\ Au on Pb reproduced the earlier findings. 
The CERES results demonstrated for the first time the existence of dilepton radiation out of a hot and dense fireball, but the data could not answer if the radiating medium was dominantly of hadronic or partonic character. 
A possible contribution to the excess from a partonic stage could have likely been outshone by radiation from the hadronic stage. 
A separation of the two contributions in the LMR is even more intricate as the spectral distribution of radiation emitted from medium-modified \textrho ~mesons resembles properties of the radiation of a QGP, once the temperature of the emitting matter approaches the critical temperature.
This not so surprising finding has its origin in the quark-hadron duality (\mycf Sec.\,\ref{sec:thermal-radiation}). 

A break through in the search for thermal partonic dilepton radiation was achieved with the NA60 dimuon spectrometer.
It grew from NA50, which had taken data for \pbau\ already but with a strong focus on the IMR and HMR regions.
An important finding of NA50 was the suppression of charmonium production appearing stronger as more central event classes are selected. 
Such a signature was conjectured after it had been realized that the presence of a QGP would prevent the formation of charmonium states out of c$\bar{\text{c}}$ pairs produced in a hard scattering due to color screening~\cite{Matsui:1986dk}.
Attention was also put to the continuum around the pole masses of the charmonium states.
As discussed in Sec.\,\ref{sec:hard-processes}, both, Drell-Yan pairs and correlated semi-leptonic open charm decays are the dominant sources of continuum radiation at invariant masses beyond 1\gevcc .
However, thermal radiation out of a deconfined phase at temperatures in excess of $T_c$ will also contribute to this invariant mass region.
Yet, a final answer to a possible existence of thermal radiation could not be given on the basis of the NA50 data, chiefly because of insufficient knowledge about charm production in heavy-ion collisions in this energy regime.  

\begin{figure}[tb]
  \begin{center}    
    \includegraphics[width=0.46\linewidth]{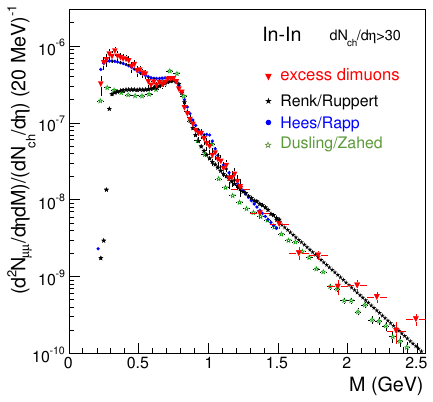}
\includegraphics[width=0.44\linewidth]{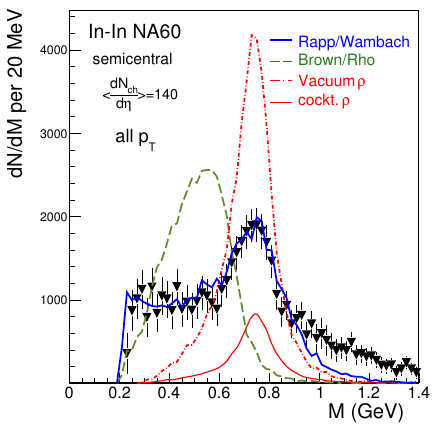}
  \end{center}
\caption{Left panel: Invariant mass distribution of excess dimuons obtained by NA60 in In+In collisions at $158$\agev\ kinetic beam energy~\cite{Arnaldi:2008er}.  
The data shown (excess dimuons) is corrected for efficiency, extrapolated to full phase space and for all collisions with at least 30 charged tracks in one unit of pseudo-rapidity (\myie leaving out very peripheral events only).
Also shown are model predictions assuming thermal rates calculated for an expanding fireball and including partonic and hadronic phases.
For further explanations see text.
(With kind permission of the European Physics Journal (EPJ)) 
Right panel: Invariant mass distribution of excess dimuons reconstructed in the acceptance of the NA60 detector for semi-central In+In collisions at 158\agev in comparison to models \cite{Arnaldi:2008fw}. 
The lines represent expected yields based on various assumptions:  Cocktail \textrho\ (thin solid red), unmodified \textrho\ (dashed--dotted red), in-medium broadened \textrho\  (thick solid blue), shifted \textrho\ (dotted green). 
The latter calculations were normalized to the data in $M<0.9$\gevcc .}
\label{fig:na60-inin}
\label{fig:na60-inin-rho}
\end{figure}
This situation changed when in 2008 data was taken with NA60 for the collision system In+In. 
To improve momentum resolution and pointing accuracy close to the interaction vertex, the existing muon spectrometer was complemented by a high-resolution silicon tracking system placed between the interaction region and the hadron absorber. 
A total of 16 planes of radiation-hard silicon pixel detectors~\cite{Keil:2005vh} were positioned inside a 2.5~Tesla magnetic dipole field.
Muon candidates were found by matching track segments in the vertex tracker and in the muon tracking section both in position and momentum.  
This strategy improved the momentum resolution to about 1\% for muons with one G$e$V momentum and also provided sufficient resolution to discriminate emission from the primary vertex (prompt) from displaced tracks (secondary), characteristic for muons from weak decays. 
In about four weeks of running at interaction rates of around 130 kHz, a total of $2.3\,10^8$ dimuon events were recorded for the collision system \inin\ at 158\agev. 
Two different magnetic field settings were used to shift emphasis from the IMR to the LMR.

The high statistics data, providing an integral of 440.000 signal pairs over a wide range of centralities, allowed for a multi-differential analysis of the dilepton signal for the first time. 
All conventional sources, except the contributions from \textrho ~meson decay, were studied in detail and finally subtracted to obtain a spectral distribution of excess muons, which are expected to be emitted solely from the hot and dense stage of the fireball evolution.  
To the conventional sources belong dimuons from hard processes, which contribute mostly to the IMR, and decays of mesons after freeze-out (\mycf Secs.\,\ref{sec:freezeout} and \ref{sec:cocktail-phenom}).
From inspection of the muon tracks associated with pairs in the IMR, the contribution from correlated open charm could be quantified.
It turned out that the slope of the dimuon invariant mass distribution from open-charm decays falls in line with the slope for the excess radiation while the yields of the two are about the same \cite{Arnaldi:2008er}.
The relative yield of Drell-Yan and open charm dimuons was estimated based on results of \pA\ collisions at a higher energy (450\agev ) obtained by NA3 and NA50, scaled down in energy using the excitation function for Drell-Yan and open charm as calculated with the properly adjusted event generator PHYTIA~\cite{Arnaldi:2008er}.
It was found that the spectrometer suppresses contributions from correlated open charm decay due to an implicit constraint on the muon decay topology, quantified by taking for each pair the emission angle $\theta_\mathrm{CS}$ of the positive muon relative to the beam axis in the \nnnn\ center-of-mass frame (Collins--Soper frame).
Only (pseudo) pairs with $\cos \theta_\mathrm{CS}<0.5$ were accepted in the analysis to minimize edge effects in the reconstruction procedure (fiducial volume). 
The ratio of the two contributions was then kept fixed for all centralities and the Drell-Yan yield was fixed by using the measured J/\textpsi\ yield as reference for the Drell-Yan cross section.
A fit to the high mass region of the measured dimuon continuum, \myie\ in a region where Drell-Yan dominates over open charm, was not pursued due to the limited statistics.
On the other hand, uncertainties in the anomalous suppression of J/\textpsi\ production introduced a 10\% systematic uncertainty of this procedure.
After the subtraction of the so defined contribution from hard scattering processes, the remaining yield was attributed to thermal excess radiation shown as ``excess dimuons'' at invariant masses beyond 1.2\gevcc\ in Fig.\,\ref{fig:na60-inin}.
Indeed, this excess radiation can be saturated scaling both the Drell-Yan and open charm contributions with factors $1.26 \pm 0.09$ and $2.61 \pm 0.2$, respectively. 
This observation (with different factors) was also made in data taken with NA50 for the heavier collision system Pb(158\agev )+Au. 
An important additional asset of the NA60 was the very much improved vertexing capability. 
It helped to rule out charm production as possible explanation of the observed excess yield.
The likelihood of emission from the primary vertex was quantified by the distance of the reconstructed muon tracks to the primary vertex, measured in the transverse plane at the position of the primary vertex.
It could be demonstrated that the excess yield is indeed chiefly emitted from the primary vertex and consequently prompt radiation. 
The most remarkable result is the near exponential decay of the excess radiation as it is expected for thermal (black body) radiation (\mycf Fig.\,\ref{fig:na60-inin}).
We will come back to this topic after the discussion of the LMR dimuons.

The striking feature of the LMR is the structure appearing at 770\mevcc , the location of the $\rho_0$ pole.
This is the structure expected from VDM; while the general trend is still exponential (phase space like), the yield is modulated by the spectral properties of the in-medium \textrho ~meson. 
The excess radiation in the LMR emerges above contributions from decays of the long-lived mesons $\eta, \eta\prime$ and \textomega\, as discussed above.
The $\rho^0$ meson does explicitly not belong to that class and therefore its contribution is kept as part of the excess yield.
The strength of the contributions from late meson decays in this experiment could largely be determined by inspection of the dimuon spectral distribution itself, thanks to the excellent mass resolution of 20-30\mevcc\ in the vector meson pole region (\mycf Sec. \ref{sec:elementary_sps}).
Using precise line shapes obtained from full Monte Carlo simulations including the detector response functions, the pole mass region was fitted assuming a smooth underlying excess yield and the two line shapes for the $\omega \rightarrow$\mumu and $\phi \rightarrow$\mumu .
The contributions from \texteta\-Dalitz decay were determined by the conservative ansatz that the LMR yield at invariant masses around $200$\mevcc\ is fully saturated by Dalitz decays~\cite{Arnaldi:2006jq}. This implies the excess to vanish at very low mass, by construction.

The data set presented in~Fig.\,\ref{fig:na60-inin} constitutes a landmark in dilepton spectroscopy. 
Three essential observations characterizing the properties of the fireball created in In+In collisions at a beam energy of 158\agev could be made.
We first address the change of properties of mesons in an hadronic environment and the role of meson baryon coupling herein. 
The large statistics and the minimum bias dimuon trigger used in the experiment allowed to study the \textrho\ line shape as function of centrality of the collision. 
Indeed, inspecting the spectral distribution of the excess yield for different centrality classes revealed that the bump structure of the \textrho ~meson in the (almost) minimum bias distribution depicted in Fig.\,\ref{fig:na60-inin} originates dominantly from peripheral collisions. 
For that the excess radiation in the LMR was analyzed in 12~bins of centrality in \cite{Damjanovic:2006bd}, with mean multiplicities ranging from $4 \leq \mathrm{d}n_{\mathrm{ch}}/\mathrm{d}\eta < 10$ to $190 \leq \mathrm{d}n_{\mathrm{ch}}/\mathrm{d}\eta < 240$.
Invariant mass distributions of the excess radiation for different centrality classes show two distinct features; a broad distribution ranging from the two-muon cutoff at low invariant masses to beyond $1$\gevcc\ (\myie beyond the poles of the low-mass vector mesons), and a bump structure  centered around the \textrho\ pole.
This separation into continuum and bump may be viewed as due to continuum radiation and dileptons due to decay of quasi-free \textrho ~mesons after chemical freeze-out or throughout in the periphery of the expanding fireball (corona).
The relative yield of the two contributions vary in a characteristic way suggesting that in central collisions thermal dileptons emitted throughout the collision shine out contributions from the freeze-out stage. 
This interpretation is further supported if the invariant mass spectra are accumulated for dimuons with transverse momenta below 0.5\gevc . 
In that case the fraction attributed to the so-called freeze-out \textrho\ is even reduced. 
This effect would likely be even more prominent if the acceptance of NA60 dimuon spectrometer would not drop strongly for low-mass, low-$p_t$ muon pairs.  

A theoretical interpretation of the dimuon invariant mass spectra measured in the acceptance of the NA60 spectrometer is shown in Fig.\,\ref{fig:na60-inin-rho}.
As a consequence of the subtraction scheme the spectrum is assumed to contain essentially only thermal radiation and any contribution from \textrho\ decay other than thermal, most prominently from \textrho ~mesons at chemical freeze-out.  
The spectral distribution is successfully modeled assuming contributions from a hadronic, pion dominated thermalized fireball. 
The respective yield is calculated by integrating the emissivity of a hadron gas at variable temperature and baryo-chemical potential over a conjectured space-time evolution of the collision zone.  
The latter is adjusted so as to describe properly a number of hadronic observables like pion rapidity densities and transverse momentum spectra (see \cite{vanHees:2007th} for a detailed description of the model). 
The results which incorporate the in-medium spectral function of the \textrho\, obtained from hadronic many-body calculations and discussed in Sec.\,\ref{sec:hadronic-model}, is shown in Fig.\,\ref{fig:na60-inin} as solid (blue) curve (right panel, labeled Rapp/Wambach).
Also shown are three further curves to illustrate the discrimination power of the data against other hypotheses: expected yield assuming free (unmodified) \textrho\ decay with \textrho\ multiplicities derived from $m_t$ scaling (cockt.\ \textrho), spectral distribution assuming free \textrho\ decays but with a yield normalized to the measured yield in the $M<0.9$\gevcc mass range (Vacuum \textrho) and \textit{ditto} but assuming a shift of the \textrho\ pole mass according to a conjecture (Brown/Rho \textrho).   
It is important to mention here again that the spectrum is shown in the acceptance of the spectrometer. 
This is important as it was demonstrated that the acceptance filter of the NA60 spectrometer roughly cancels the Boltzmann term in the evaluation of emissivity based on VDM (see \myeg \ref{eqn:rad-rate-bol} ).
Hence, the result essentially represents the spectral distribution of the in-medium \textrho. 
The yield below masses of 0.6\gevcc\ therefore reflects \textrho\ strength dominantly caused by a coupling of the propagating \textrho\ to baryon-resonance hole states. 
We will come back to this observation in Sec.\,\ref{sec:status}. 
The obvious mismatch of the shifted \textrho\ mass distribution with data has been taken as clear evidence for the absence of a pole mass shift due to a depletion of the chiral condensate as suggested by Brown-Rho scaling. 

Lastly, concerning the discussion of the NA60 data, we would like to touch polarization effects observable through angular correlations of the  emitted leptons \mywrt\ the collision axis. 
The electron distributions in the helicity and Collins--Soper frame were analyzed and no significant anisotropies were found and concluded to be in agreement with those expected for thermal source \cite{Arnaldi:2008gp}.  
On the other hand theoretical calculations \cite{Speranza:2018osi}, described in Sec. \ref{sec:thermal-radiation},  show that effects of only a few percent modulation of the lepton angular distribution are to be expected,  depending on the virtual photon transverse momentum, the invariant mass and flow. 
Hence, the observed isotropy of the angular distribution of NA60 strictly speaking is not a proof of thermalization. 
\subsubsection{RHIC and LHC}
Highest center-of-mass energies in collisions of heavy ions are reached at colliders. 
At the LHC, the three major experiments ALICE, ATLAS and CMS have full capability for dilepton physics, though the latter two with limited sensitivity for low-$p_t$ pairs.
Yet, the inclusive dimuon spectrum of CMS, reaching to invariant masses of TeV \cite{Khachatryan:2016zqb}, and showing signals for all vector meson ground states and several excited states, is certainly a landmark of dilepton spectroscopy. 
The first results on dielectron production in heavy-ion collisions have recently been reported by ALICE \cite{Acharya:2018nxm}. 
The dielectron invariant mass spectrum in the range $0<M_{e^+e^-}<3.5$ GeV/c$^2$ has been measured in ($0-10\%$) \pbpb\ collisions at $\sqrt(s)=2.76$ \tev .
It was found that the total electron pair signal exceeds the cocktail in the mass range $0.15<M_{e^+e^-}<0.7$ \gevcc\ by about 40\%. 
The signal to cocktail ratio, excluding the \textrho\ contribution, was extracted to be $1.4\pm0.28(stat)\pm0.27(cocktail)$. 
The measured excess has been found to be in agreement with models including thermal dielectron production from both partonic and hadronic phases with in-medium spectral function \textrho. 
In the IMR, dominated by heavy flavor decays, the data could be described by contributions obtained from PYTHIA calculations scaled with the number of binary collisions assuming either no or full correlation in the charm decays.
The systematic uncertainties of the measurement prevent any conclusion on the effects of interactions between heavy quarks and other partons in the medium as both versions of the hadronic cocktail are consistent with the data. 
Measurements with significantly improved precision are planned after LHC shutdown (in Run 3) with the new TPC and vertex detector \cite{Abelevetal:2014cna}.    

%
\begin{figure}[tb]
  \begin{center}    
    \includegraphics[width=0.5\linewidth]{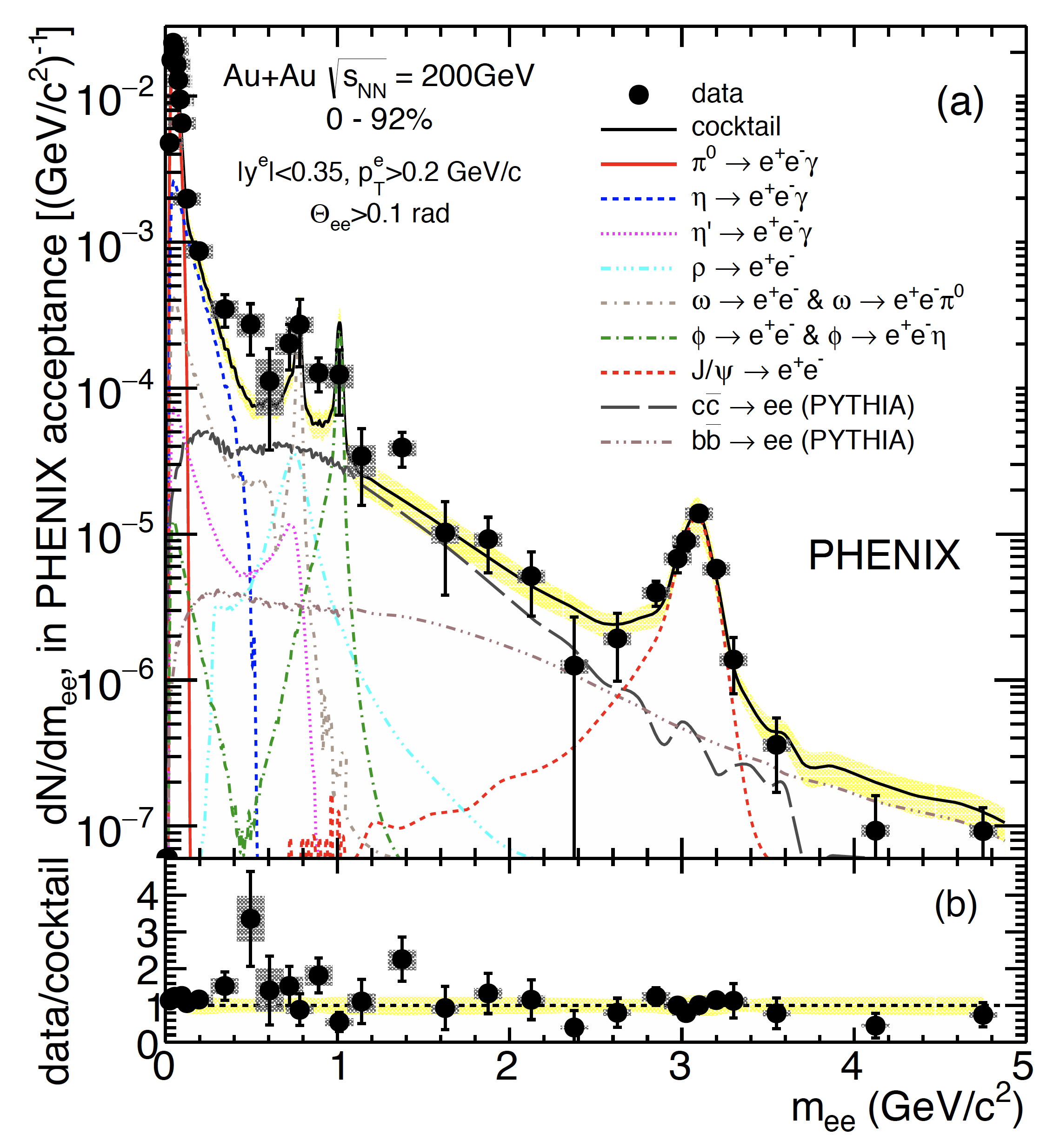} 
  \end{center}
    \caption{(Color online) Invariant mass distribution of dielectrons reconstructed in the acceptance of the PHENIX detector for minimum-bias $Au+Au$ collisions at 
    $\sqrt{s_{NN}} = 200 \, \mathrm{G}e\mathrm{V}$~\cite{Adare:2015ila} (a). The solid (black) lines represent  the sum of all cocktail sources including thermal. The yellow band gives the uncertainty in the cocktail estimation. Thermal radiation should appear above the cocktail, \mycf the ratio depicted in (b).
    }   
\label{fig:phenix-auau-hbd-inv}
\end{figure}
%
At RHIC, two experiments have measured dilepton spectra, PHENIX and STAR.
A historic outcome of the early heavy-ion runs at RHIC has been the discovery of a strong suppression of neutral pions with high transverse momentum, relative to the distribution observed in collisions of protons scaled by the number participant nucleons. 
It is by now widely accepted that this quenching of high-$p_t$ \piz\ mesons (and other hadrons) is caused by quenching of hard partons in the QGP formed nearly instantaneously.   
Many complementary observables have since been addressed by PHENIX and STAR to further support this conjecture and even characterize the properties of this state of matter, in particular also the expected thermal dilepton radiation in the LMR and IMR.
Yet, as will become evident soon, a direct determination of the early temperature of the fireball by means of slope of the IMR dileptons, like it has been achieved by NA60, is difficult at collider energies. 

Both, PHENIX and STAR have addressed continuum dilepton radiation at central rapidity by means of electrons. 
The muon detectors of both experiments are located at large pseudo-rapidities where the hadron absorbers do not hamper particle identification for charged hadrons and muons do have sufficient punch.  
At 200\agev\ \auau\ collisions the fireball freezes out at around the pseudo-critical temperature $T_c \approx$ 160 \mev\ and at close to vanishing baryo-chemical potential ($\mu_B$). 
The measured \piz\ rapidity density in central collisions ($A_{part} = 350$) reaches 300~\cite{Andronic:2005yp}. 
Consequently, substantial combinatorial background arises due to the abundant electrons and positrons from \piz\ Dalitz decays (\myie\ about three pairs per central event and unit rapidity), from incompletely rejected electron and positrons from external pair conversion in detector material or the beam pipe and due to remaining hadron contaminations of the electron sample.

Figure~\ref{fig:phenix-auau-hbd-inv} shows the invariant mass distribution of dielectrons in the acceptance of the spectrometer obtained by PHENIX in minimum-bias Au+Au collisions at \sqsnn{200}~\cite{Adare:2015ila}.
The data has been taken in the year 2010 with the HBD detector in place (\mycf Sc.:\,\ref{sec:electrons}.
The invariant mass distributions for dielectrons show a steep decline in the \piz\ Dalitz region and a prominent signal from J/\textpsi\ decay. 
All expected contributions from processes other than thermal radiation are depicted by a full line. 
This line also prominently shows peak structures in the low-mass vector meson mass region were also the data shows respective structures. 
The accordance of the cocktail with the data can be judged in the lower panel where the ratio of the two is depicted. 
Good agreement is shown throughout the invariant mass range, besides the region below the vector meson poles where the data points deviate more than one sigma from the cocktail.

The spectral distribution observed for the IMR is in accordance to the data taken in 2004 in a run without the HBD.
The yield in the IMR drops nearly exponentially by about an order of magnitude over one\gevcc\ in invariant mass. 
In the IMR, the cocktail contains contributions from open charm decay and from Drell-Yan processes.
Correlated semi-leptonic decay of charmed mesons (\DD\ pairs) dominate this region.
The cocktail yield in the IMR region was estimated based on the dielectron yield measured in \pp\ reaction at the same collision energy per nucleon pair (see \mycf{Sec \ref{sec:elementary-rhic}}). 
From the measured yield in \pp\ the hadronic cocktail (meson decays contributing to the LMR) was subtracted.
The remaining continuum distribution was extrapolated to zero invariant mass in order to obtain the total \textit{prompt} dilepton cross section. 
The cross section and invariant mass distribution obtained in this way were found to be consistent with PYTHIA\footnote{Code version 6.319 with cteq5l parton distribution functions. Details of the settings are reported in~\cite{Adare:2015ila}.} 
Binary-collision scaling was then used to convert the measured \pp\ cross section to an estimate for dielectron production from correlated open charm/bottom decay in \auau\ collisions.
A systematic uncertainty arises from this procedure as it is a priori not known to what extent the transverse momentum distribution of the open-charm mesons is modified by the fireball.
The PHENIX collaboration investigated \myie\ the effect of open-charm thermalization on the spectral distribution of the respective dielectron invariant mass distribution.
Generally, quasi-correlated \DD\ pairs from thermalized matter show a softer dielectron mass distribution since their average opening angle and transverse momenta are smaller. 
The correlation \DD\ mesons are expected to be modified in the medium. 
PHENIX estimated this effect by assuming two extreme situations, namely a full thermalization of the \DD\ mesons and a fully unmodified distribution as obtained from initial hard scattering. Details of the procedure are reported in \cite{Adare:2009qk}.  
From the comparison of data to cocktail it appears that for this minimum-bias sample no evidence for medium radiation was found in IMR.

In the invariant mass range $0.2 < M_{ee}< 1$\gevcc\ all data points lie above the cocktail and signal contributions from thermal radiation. 
Integration of the excess reveals that the dilepton yield is significantly higher than the cocktail.
The PHENIX collaboration derives an Enhancement Factor (ratio of the integrated yields) of $1.7 \pm 0.3\,(\mathrm{stat})\pm 0.3\,(\mathrm{sys})\pm 0.2\,(\mathrm{cocktail})$ and $2.3 \pm 0.4\,(\mathrm{stat})\pm 0.2\,(\mathrm{sys})\pm 0.2\,(\mathrm{cocktail})$, depending on the assumption made for the production cross section for open charm.    
Stronger contributions of thermal radiation are expected if more central collisions are selected. 
Indeed, for the 0--10~\% most central collisions the Enhancement Factor rises to $2.3 \pm 0.7\,(\mathrm{stat})\pm 0.5\,(\mathrm{sys})\pm 0.2\,(\mathrm{cocktail})$ and $3.2 \pm 1.0\,(\mathrm{stat})\pm 0.7\,(\mathrm{sys})\pm 0.2\,(\mathrm{cocktail})$, respectively. 
The quoted errors for the measurement are a combination of statistical and systematic uncertainties as well as uncertainties due to model assumptions and are as large as 40\% (quadratic average).

\begin{figure}[tb]
  \begin{center}    
    \includegraphics[width=0.9\linewidth]{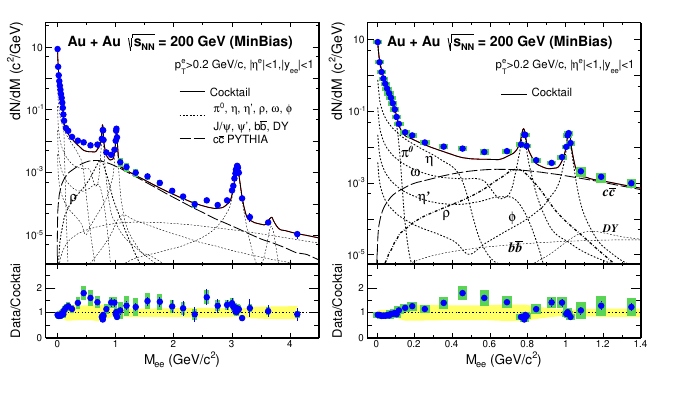} 
  \end{center}
\caption{(Color online) Invariant mass distribution of dielectrons reconstructed in the acceptance of the STAR detector for minimum-bias \auau\ collisions at $\sqrt{s_{NN}} = 200 \, \mathrm{G}e\mathrm{V}$ in comparison to hadronic cocktail (left)  \cite{Adamczyk:2015lme}. The LMR region is zoomed in the right panel. 
The vertical bars show statistical error while the green boxes display systematic uncertainty. The ratio of data to the cocktail is shown in the lower panel.(figure from APS)
}   
\label{fig:star-auau-cocktail}
\end{figure}
The dielectron invariant mass distribution measured by the STAR collaboration for \sqsnn{200} minimum bias collisions is shown in Fig.\,\ref{fig:star-auau-cocktail} together with the hadronic cocktail~\cite{Adamczyk:2015lme}. 
In the LMR one can see also see an access which amounts to $1.76 \pm 0.06\,(\mathrm{stat})\pm 0.26\,(\mathrm{sys})\pm 0.29\,(\mathrm{cocktail})$ when integrated in the $0.3-0.76$\gevcc mass region (without inclusion of the vacuum \textrho\ contribution). 
The data and the estimated cocktail are in accordance with the measurement of the PHENIX collaboration, yet showing higher significance.
STAR has used a different variant of PYTHIA\footnote{Code version 6.419 with cteq5l parton distribution functions. Details of the settings are reported in~\cite{Adamczyk:2015lme}} and appears to lie slightly below the data. 
In the STAR acceptance, which does not put a strong bias on the IMR spectral distribution at mid-rapidity, the effect of thermal radiation amounts to about 10-20\% of the cocktail for minimum bias collisions. 
The yellow band in the lower panels depicts the uncertainty of the cocktail determination which appears to be of same size than the thermal radiation relative to the hadronic cocktail, \myie\ in the yield and shape of the contributions from correlated open heavy-flavor decay (dominantly charm). 
The situation can be improved if a substantial fraction of the open charm contribution is rejected making use of the displaced vertex of semi-leptonic decays of charmed mesons.
To achieve sufficient separation power for the secondary vertices, a dedicated micro vertex detector needs to be employed.  
STAR has taken data in 2016 with the Heavy Flavor Tracker (HFT) installed in the STAR barrel.
The analysis has been ongoing by the time this report was published. 
%
%
\section{Status and Open Questions}
\label{sec:status}
A capital goal of studying dilepton emission in heavy-ion collisions is to ``trace'' the evolution of the expanding fireball in the QCD phase diagram.
Approaching this goal requires to unambiguously understand dilepton emission from thermalized QCD matter at given temperature and chemical potential and, by this, to establish dilepton emission as {\em standard candle} of QCD matter.
As discussed in Sec.\,\ref{sec:theory}, a proper description of the in-medium \textrho\ meson propagator is key to approaching this challenge.
Strict VDM provides a good basis for the theoretical modeling of the emissivity of hadronic matter (Sec.\,\ref{sec:VDM}.
According to VDM, the dominant fraction of dilepton emission in the LMR is emitted via intermediary \textrho\ mesons. 
They are produced in annihilations of pions (\pip\pim ) or in decays of hadron resonances like \myeg $\mathrm{a}_1 \rightarrow$ \textpi \textrho\ or N$^{*} \rightarrow$ N\textrho.
The second important and independent goal of dilepton spectroscopy is to explore the limits of hadronic existence by studying the in-medium \textrho\ propagator.
It can provide insight into the role of a (partial) restoration of the dynamically broken chiral symmetry on the hadron properties (Sec.\,\ref{sec:chiral}).
Thermal radiation from a partonic medium dominates the thermal radiation in the IMR and is an important observable as well. 
In contrast to emission from a hadronic medium, the modeling of partonic processes is more straight forward as composite objects are not involved.
In particular, at vanishing baryo-chemical potential it can directly be computed by lattice QCD. 
The current-current correlator is structureless and hence the extraction of the source temperature straight forward (Sec.\,\ref{sec:thermal-radiation}).
However, a challenge of dilepton spectroscopy in the IMR is to determine with high enough precision the contribution from correlated open charm (Sec.\,\ref{sec:hard-processes}).
In the following section we compare experimental results to model calculations and focus on multi-differential dilepton observables. 
For this comparison we have chosen two theoretical approaches, which both provide a description of dilepton emission for all collision energies studied so far.

\vspace{2ex}
\noindent
\textit{Thermal expansion models}\\[1ex]
The first category of models assumes an expanding, (locally) thermalized fireball and consequently focuses on describing the thermal radiation only.
Respective dilepton spectra are calculated through a four-dimensional (space-time) integration of the emissivity defined for equilibrated QCD matter at given temperature and chemical potential~\mycf Sec.\,\ref{sec:thermal-radiation}. 
For energy densities, at which the system is expected to be in a hadronic state, the most essential theoretical ingredients are the in-medium spectral functions of vector mesons, most notably the \textrho\ meson.
To address in particular \rhoa\ mixing,  invariant masses beyond 1\gevcc\ are to be investigated where multi-pion processes contribute.
The reference model for calculating emissivities of hadronic matter is the one developed by Rapp and Wambach providing evolution of the vector current correlator as a function of the temperature and  chemical potential (\mycf Sec.\,\ref{sec:hadronic-model}). 
We remind that the evolution of the spectral function is governed by the temperature and effective baryon density, not the net-baryon density. 
For the radiation from the QGP phase annihilation of $q\bar{q}$ into dileptons is taken into account.

The second essential ingredient for such calculations, besides the emissivity, is the description of the fireball evolution in the T--\mub\ plane.
The reference calculations to describe the expansion of the collision zone in reactions at ultra-relativistic energies are solutions of relativistic hydro-dynamics with suited definitions of the initial thermalized state. 
Such calculations are very involved and CPU-hungry and go well beyond simple descriptions to study thermal dilepton radiation~(for details see \myeg \cite{Gale:2013da,Kurkela:2018wud}). 
For systematic studies of thermal radiation, over a wide range in beam energy, a common (simplified) strategy is to define an initial volume of cylindrical shape (following the concept of boost invariance) with given initial energy density and chemical potentials. 
The dynamics of the fireball is then obtained by implying isentropic expansion and using a partonic as well as a hadronic equation-of-state for the QGP and hadronic phase, respectively. 
Details of such an expansion model can be found \myeg in~\cite{vanHees:2007th}). 

%
\begin{figure}[tbh]
  \begin{center}
  \begin{minipage}{1.0\textwidth}
  \includegraphics[width=0.44\textwidth,height=0.36\textwidth]{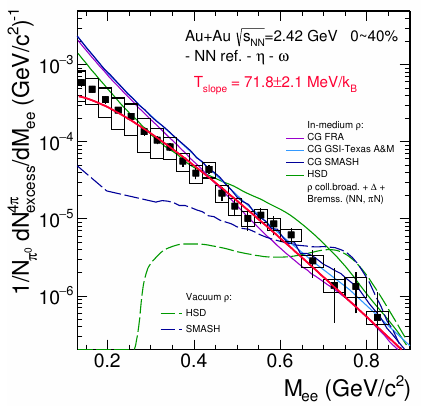}
  \includegraphics[width=0.44\textwidth,height=0.36\textwidth]{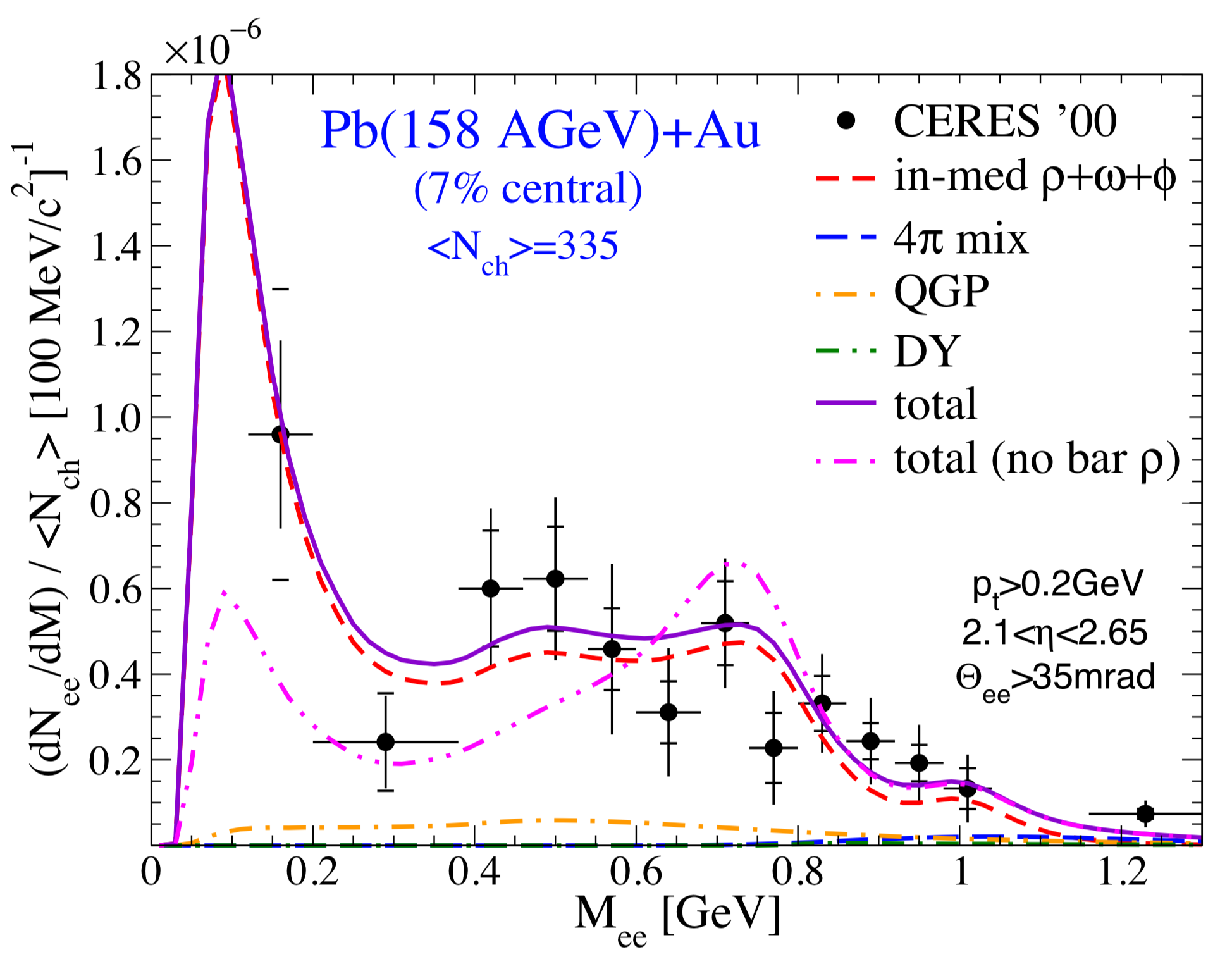}
  \includegraphics[width=0.49\textwidth,]{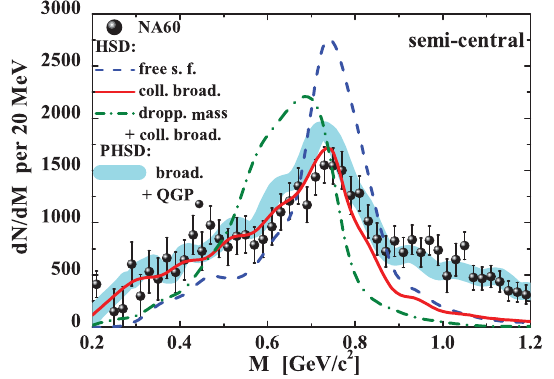}
  \hfill
  \includegraphics[width=0.5\textwidth]{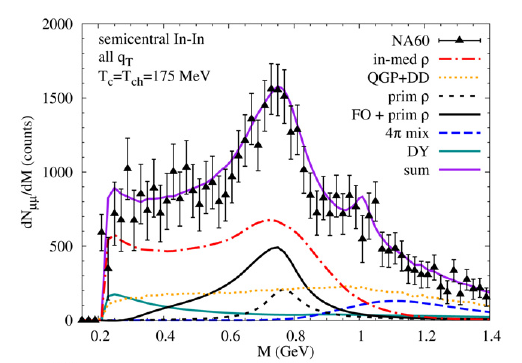}
  \end{minipage}
\caption{Dilepton invariant mass distributions of the excess radiation in the LMR, obtained for various beam energies and collisions systems, in comparison to model calculations described in more detail in the text: upper left: \auau~ at \sqsnn{2.4} \gev ( data HADES coll.) \cite{Adamczewski-Musch:2019byl}, upper right:  \pbau\ at \sqsnn{17.3} \cite{Agakichiev:2005ai} (data CERES coll.) compared to thermal model \cite{vanHees:2007th}, bottom: \inin\ at \sqsnn{17.3} (data NA60 coll.) compared to PHSD (left)\cite{Linnyk:2011hz} and to a calculation assuming isentropic expansion and respective emissivities \cite{vanHees:2007th}.
} 
   \label{fig:excess-theory}
  \end{center}
\end{figure}
%
%
An alternative method to model the fireball expansion is realized by coarse-graining microscopic transport calculations (\mycf  Secs.\,\ref{sec:stages}, \ref{sec:thermal-radiation}).
It bridges between the above described simplified approach and fully microscopic calculations addressed below.  
In this approach, temperatures and chemical potentials are derived via microscopic transport models by sampling the partition function in space-time cells over a large ensemble of events simulated for the exactly same event class. 
In a second step, the intensive parameters $T$ and $\mu$ are calculated in each cell from the particle abundance, their phase space distribution or using an appropriate EOS.
An important aspect in this procedure is to evaluate the degree on thermalization for a given cell. 
The assumption of thermalization enters the emissivity calculation in two aspects: The calculation assumes thermal (grand-canonical) particle densities and respective momentum distributions when degrees of freedom are integrated out.
To allow application of the thermal rates the respective criteria should be approximately fulfilled, a strict thermalization condition is not applied in the calculations.
Remaining effects have to be taken as model uncertainties as it is also not a priori clear, that the true phase space distribution is realized in microscopic transport calculations based on a quasi-classical approach.
In the coarse-grain calculations discussed here, thermalization is evaluated inspecting the pressure isotropy in each cell. 
To derive the density in each cell, the respective sampled particle distribution of each cell is boosted in its (individual) center-of-momentum frame.
Different procedures have been used to determine the respective temperatures: In \cite{Galatyuk:2015pkq} (GSI-TAMU group) the pion spectra in the local rest frame were fitted with a Boltzmann distribution, while in \cite{Endres:2014zua} (Frankfurt group) the temperatures and baryo-chemical potential where derived from the local baryon density and energy density using a hadron gas equation of state.  
Finally, the thermal radiation rates are applied to each cell and integrated of the full space-time evolution. 
An advantage of this approach is that it allows to apply thermal emissivities also in case of collisions at low beam energies where the assumption of a globally thermalized fireball is not justified (\mycf Sec.\,\ref{sec:initial}). 
Comparisons of the two approaches discussed above for RHIC/SPS energies provide consistent results on dilepton emission rates, despite some significant differences in the modeling of expansion process like for example the time evolution of the pion and kaon fugacities.
Therefore, the authors conclude that dilepton spectra are not very sensitive to details of the space-time evolution~(for details see \cite{,Endres:2014zua,Galatyuk:2015pkq,Endres:2015fna,Endres:2015egk,Endres:2016tkg}).

\vspace{2ex}
\noindent
\textit{Microscopic transport models}\\[1ex]
The second category is microscopic transport calculations.
Such models incorporate non-equilibrium effects by propagating individual quasi-particles, including mean-fields and two-body interactions, by solving the Boltzmann--Ueling--Ulenbeck equation~(\mycf Sec.\,\ref{sec:stages}). 
Here we take as an example the Parton--Hadron String Dynamics (PHSD, equivalent to HSD in the hadronic phase)~\cite{Bratkovskaya:2007jk,Linnyk:2011vx,Linnyk:2011hz} and the recently developed SMASH code.
A comprehensive comparison of dilepton spectra obtained with the Gie{\ss}en BUU transport model to HADES data can be found in~\cite{Larionov:2020fnu}.
Emphasis is laid on a proper treatment of the collisional \textrho\ broadening based on the collision integral and using off-shell transport. 
Furthermore, the contribution from \nnnn\ bremsstrahlung has been optimized in the calculation by adjusting the code to HADES data on \pp\ and \np\ reactions.

Transport calculations do not require the assumption of local thermalization and take into account modifications of the spectral distribution of the quasi-particles due to the surrounding medium in an empirical way. 
In PHSD, off-shell particles are propagated using a method inspired by the kinetic theory developed by Kadanoff and Baym, as discussed in Sec. \ref{sec:mass-transport}.
Spectral distributions for hadrons can be explicitly broadened and/or shifted in mass depending on the local density.
The former resembles collisional broadening while the latter has been introduced to mimic possible effects on the mass distribution originally proposed by Brown and Rho  (\mycf Sec.\,\ref{sec:thermal-radiation}).
The dynamics in the partonic phase is calculated within the Dynamical Quasi-Particle Model (DPM), where partons have finite masses with spectral functions which in turn depend \myeg on temperature. 
The parametrization of the the spectral density is adjusted to fit the lattice EOS and the running coupling constant. 
The finite width of partons is essential to reproduce in this model a small ratio or the shear viscosity to entropy ratio which results in hydrodynamical evolution of the partonic phase (for details for the model see \cite{Cassing:2008nn}).
With this microscopic ansatz the full evolution of the fireball, from initial hard scatterings of partons, through a dense medium including a possible QGP hadron-gas phase transition up to the transition to a non-interacting gas of hadrons  can be simulated.
The coupling of the electromagnetic currents of the hadrons to dileptons is usually calculated in a perturbative approach. 
It means that dilepton emission is calculated in a ``second round'' by inspecting the full collision history of an event and calculating the dilepton emission based on known (or estimated) branching ratios for leptonic or semi-leptonic decays of hadrons.
The dynamics of the hadrons is not modified.  
In addition, dileptons are also radiated from  nucleon-nucleon elastic collisions ("quasi-elastic" bremsstrahlung) using the Soft Photon Approximation. 
Dilepton production in the partonic phase include, besides Born terms, also interactions with gluons (\myeg $q\bar{q}\rightarrow\gamma^*+g$ and $q(\bar{q})g\rightarrow\gamma^*+q(\bar{q})$). 
The most striking influence of finite parton masses on dilepton production is visible in the invariant mass distribution where a mass threshold in the infrared occurs and the production cross section at low masses increases \mywrt\ results obtained from perturbative calculations.    
\subsection{In-medium properties of the rho meson}
The investigation of the in-medium \textrho\ spectral function has been the central motivation of dilepton spectroscopy at moderate beam energies.
While results on medium-modifications of vector mesons in cold nuclear matter are still not fully conclusive, particularly  at low energies  (\mycf Sec.\,\ref{sec:coldmatter}), dilepton data obtained in heavy-ion reactions paint a rather clear picture.
In all experiments, from \sqsnn{2.4} to \sqsnn{200}, contributions to the dilepton yield attributed to \textrho\ decay show a strong broadening such that no conclusion on a possible shift of its pole mass can be drawn. 
The consistent theoretical description of the LMR excess radiation in the full energy range, based on the thermal average of the in-medium \textrho\ propagator, provides strong evidence for substantial modification of the \textrho\ through its coupling to the hadronic heat bath.
Additional \textrho\ strength, well below the \textrho\ pole mass, develops through coupling to baryon-resonance hole states.
This feature of the in-medium \textrho\ propagator provides an explanation why the broadening is strongest for the lowest beam energy.
In such collisions, the incoming nucleons are stopped in the Center of Mass to a high degree while pionic excitations are present but not dominating the dynamics.  
In case of such high baryonic densities pionic excitations are substantially modified and a clear separation between pion nucleon and baryon-resonance (hole) states does not exist anymore. 

\begin{figure}[tbh]
  \begin{center}
  \includegraphics[width=0.7\textwidth, height=16pc]{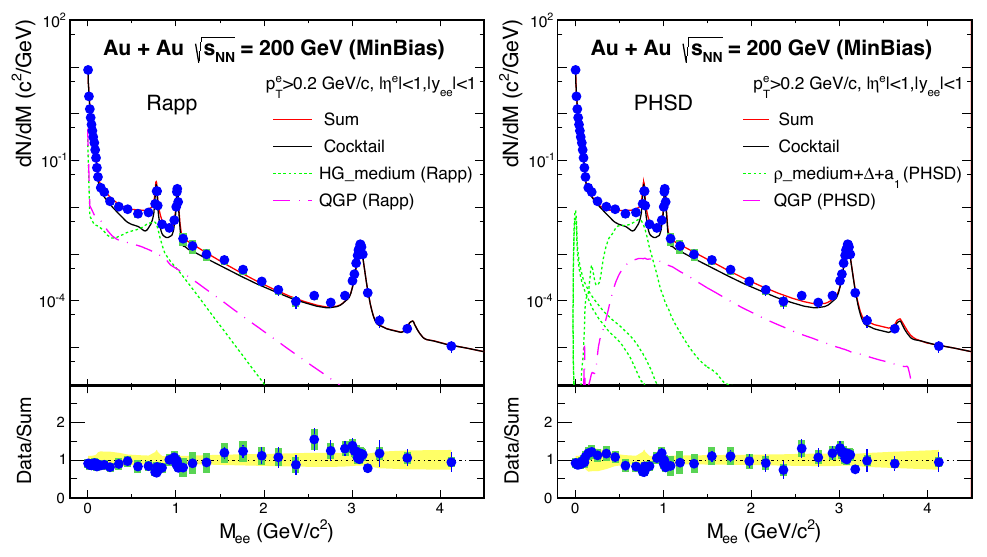}
    \caption{Dielectron invariant mass distribution of the excess radiation from \auau\ collisions at \sqsnn{200} \gev\ obtained by the STAR collaboration in comparison to model calculations based on the isentropic expansion model and PHSD calculations~\cite{Adamczyk:2015lme}.
    }
    \label{fig:STAR_model}
  \end{center}
\end{figure}

The spectra shown in Fig.\,\ref{fig:excess-theory} were obtained, in clock-wise order, by HADES for \auau\ collisions at \sqsnn{2.4} \cite{Adamczewski-Musch:2019byl}, by CERES for \pbau\ collisions \cite{Agakichiev:2005ai}, and NA60 for \inin\ collisions \cite{Arnaldi:2006jq}, both at \sqsnn{17.3}, and in Fig. \ref{fig:STAR_model}  by STAR for \auau\ collisions \cite{Adamczyk:2015lme}. 
Note that all spectra, but the one from NA60, were obtained for dielectrons and that the HADES spectrum is shown in logarithmic representation owing to the low average temperature resulting in a strong enhancement of low-mass dileptons

The comparison of data with model calculations based on the coarse grained approach and microscopic in-medium spectral functions shows a good agreement with the data as well in shape as in absolute yields, except for some excess visible at small invariant masses measured by HADES and displayed in Fig.\,\ref{fig:excess-theory} (consistently for various flavors of coarse grained calculations: violet line CG FRA \cite{Endres:2015fna}, blue line CG GSI-Texas \cite{Galatyuk:2015pkq}, dark blue CG SMASH \cite{Staudenmaier:2017vtq} ).
Indeed, the HADES spectrum, when compared to results of model calculations using the coarse grained approach, reveals the most striking observation:
The spectral distribution is falling of exponentially and monotonously. 
In view of the underlying emissivity this occurs if the weighted in-medium spectral function of the \textrho\ $\Im D_{\rho}/M_{\ell\ell}^2$ is approximately constant, or only slowly varying, as function of invariant mass (\mycf ~Eq. (\ref{eqn:rad-rate-bol}) and (\ref{eqn:ratio-hadron})). 
Indeed, a fit of the Boltzmann function $f_B$ (solid red line) converges for a temperature of 72~M$e$V/$k_{\mathrm{B}}$.
Further evidence for a broadening of the \textrho\ provides the comparison to microscopic transport calculations using vacuum spectral functions for the \textrho\. 
The respective results are shown as dashed curves for HSD and SMASH.
While HSD uses a cut-off in the free spectral function at $M = 2 m_{\pi}$, SMASH includes coupling to baryon resonance (see discussion in \mycf Sec.\,\ref{sec:sis-bevalac}) and consequently the mass spectrum is cut off by the dielectron phase space only. 
Both calculations reveal a structure around the \textrho\ meson pole mass, not supported by data, and do not explain the observed yield towards small invariant masses.
In such microscopic calculations, additional contributions to the excess radiation are expected from bremsstrahlung and from multiple $\Delta$ generation in the dense phase of the collisions, as already suggested for explanation of the \arkcl\ data (see \mycf Sec.\,\ref{subsec:heavy_ion}). 
The  sum of the respective contributions is shown for the HSD case. 
Summing the contributions, the low-mass region of the data is well described, too. However, the bump structure is only partly reduced if a strong collisional broadening is additionally assumed in the calculation~\cite{Linnyk:2011hz}.
It is important to underline that the calculation of the emissivity based on the in-medium \textrho\ propagator includes consistently also graphs with bremsstrahlung and the $\Delta$ resonance in the calculations of the in-medium self energy. Hence, additional treatment of these processes in case of coarse grained approaches would mean to double count contributions. 

%
%

How does the situation change for excess radiation investigated at ultra-relativistic collision energies?
Continuing clockwise in Fig.\,\ref{fig:excess-theory}, the respective mass distributions are depicted for collisions at SPS at an beam energy of \sqsnn{158} for the case of dielectrons (CERES) and dimuons (NA60), where the latter data is compared with two different types of model calculations (shown in the bottom row). 
The CERES spectrum is compared to a calculation based on thermal emissivities~\cite{vanHees:2007th} using the isentropic expansion model.
While the statistical precision of the CERES spectrum is just about good enough to discriminate between a scenario with (labeled as ``total'') and without including coupling to baryons (``no bar \textrho ''), yet the data point at 150~\mevcc\ seems to support the observation of HADES of a strong excess yield at low-masses due to the \textrho N$\,\leftrightarrow\,$N$^*$ loop in the \textrho\ in-medium self energy.
Contributions from Drell-Yan and QGP radiation are not subtracted here but seem to contribute little as can be appraised from the respective contributions also shown in the theory cocktail (DY, QGP).
The model calculation also suggests little contributions due two multi-pion annihilation ($4\pi$ mix).

More information about the origin of the radiation in LMR and IMR can be extracted from the high-statistics NA60 data.
In the lower right panel, the respective data are compared to the same Rapp/Wambach model, yet with a more elaborate tuning of the fireball expansion model.
This has been achieved by addressing at the same time dimuon invariant mass and transverse momentum spectra.
Most importantly, the expansion velocity of the fireball has been increased as to better match the slope of the $p_t$ distributions. 
For details about the expansion model see~\cite{vanHees:2007th}.
Compared to the calculation shown in Fig.\,\ref{fig:na60-inin-rho}, which is based on the same emissivity but uses an earlier parameter set for the expansion model and focusing on \textrho\ contributions only, here also the high-mass region of the experimental distribution is well reproduced. 
The successful description of the dimuon yield above 1\gevcc\ is due to including contributions from open-charm decay, QGP radiation (labeled ``QGP+DD'') and multi-pion decays in a consistent way. 
The latter is an interesting contribution since the yield might be affected by the degree of \rhoa\ mixing.
Note also that the bump at the \textrho\ pole mass is explained as being due to primordial \textrho\ mesons produced in first chance collisions and escaping the fireball as well as due to decays of \textrho\ mesons after chemical freeze-out, the so-called cocktail \textrho\. 
Overall, thermal radiation makes up only about 50\% of the total IMR excess yield. 
As discuss already before, the peak and the low-mass side of the spectral distribution are well described by the \textrho\ contributions with evidence for \textrho\ baryon coupling. 
The lower-left panel shows again the NA60 data, \myie for  semi-central collisions but here compared to results obtained with HSD and PHSD.
The experimental line shape around the \textrho\ pole mass and below is here obtained by explicitly broadening the free \textrho\ spectral function (collisional broadening). 
The degree of broadening is adjusted as to best reproduce the data. 
Also the HSD calculation for hadronic contributions falls short of saturating the yield in the high mass regime. 
Only after inclusion of radiation from a plasma phase, included in the calculations, the yield at 1\gevcc\ and beyond is explained. 
Note, however, that details about the composition of this part of the radiation are not given.  
More details of the calculation can be found in~\cite{Linnyk:2011hz}.
The comparison of the NA60 data to both model calculations clearly disfavors a dropping-mass scenario as already discussed in the previous section (\mycf Sec.\,\ref{sec:sps}).
Since the emitting source (fireball) has non-vanishing baryon densities at all energies, and  since the  \textrho\ mesons couple to baryons and antibaryons in the same way, the \textrho\ spectral distribution acquires also at SPS a significant broadening (\mycf Sec.\,\ref{sec:sps}), yet with less emphasis of the low-mass tail of the spectral function compared to the case at SIS18 energies (HADES).
In the latter case, \textrho\ like virtual states in the vicinity of baryons (cloud) decay to virtual photons.
This is the picture behind VDM in baryonic transitions, for which a thorough proof has still to be provided.
Hence, the very low-mass region is of particular interest for studies of baryon-driven effects.

%
\begin{figure}[tb]
  \begin{center}    
\includegraphics[width=0.5\linewidth]{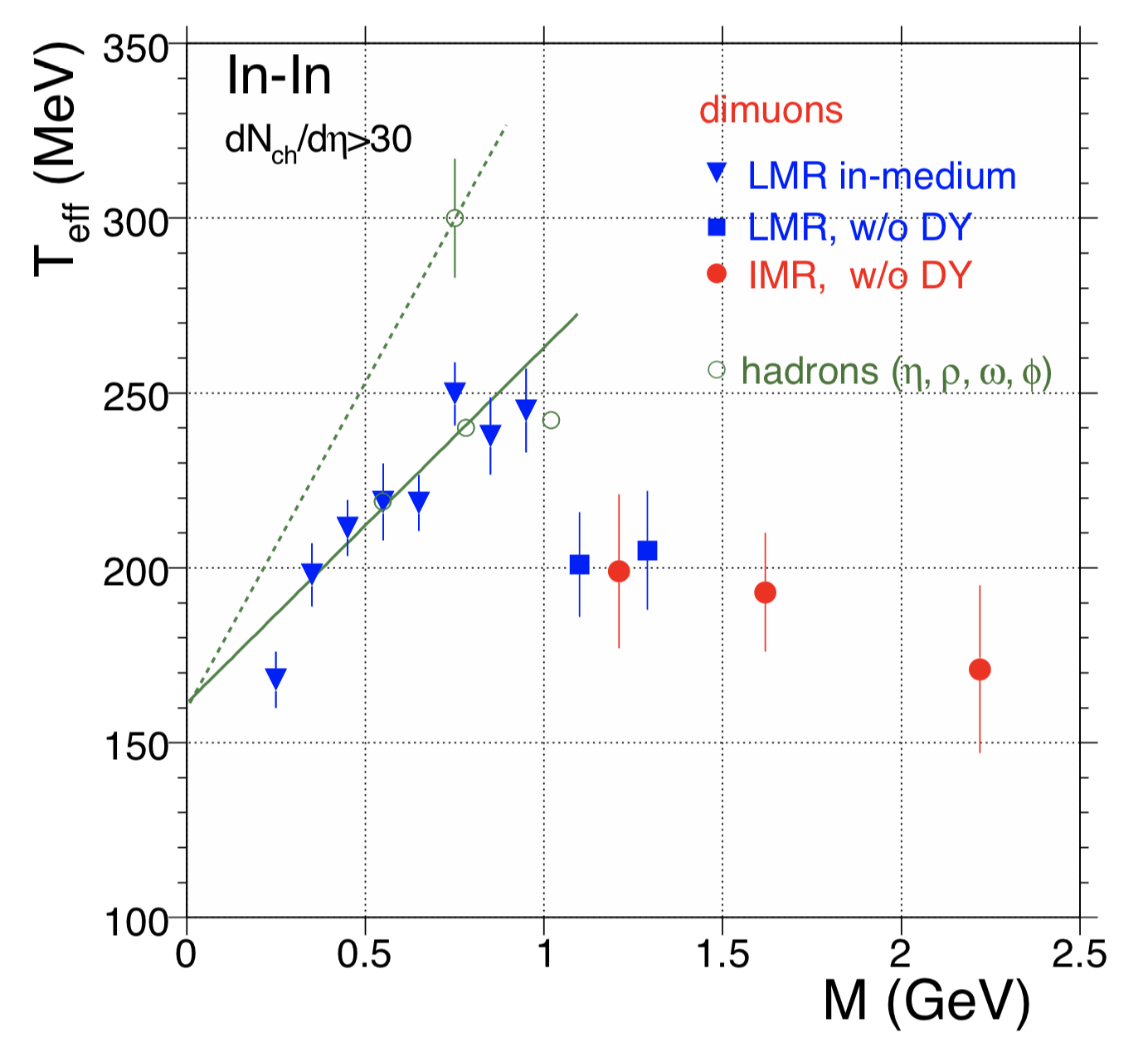} 
   \end{center}
\caption{Systematics of slope parameters ($T_{eff}$) extracted from the $p_t$ distribution of dimuons originating from different sources as function of the invariant mass $M$~\cite{Arnaldi:2007ru}. 
Full symbols correspond to the systematics of the excess radiation and is presented for bins in invariant mass. 
Open circles depict slopes connected to identified mesonic dilepton sources. See text for further explanations.}
\label{fig:na60-inin-teff}
\end{figure}
%
Last, in Fig.\,\ref{fig:STAR_model} the centrality integrated dielectron yield obtained by STAR for \sqsnn{200} \auau\ collisions is compared to both the Rapp/Wambach model (left panel) and to PHSD (right panel) \cite{Adamczyk:2015lme}.
While both calculations achieve a satisfactory agreement with data concerning the integral yield, the underlying contributions are assessed quite differently.
This is in particular true for the radiation from the partonic phase.
The dielectron excess in the LMR is well described based on the thermal emission rates and the expanding fireball model used for SPS energies (with adjusted initial parameters for RHIC).
Also for this collision system, where the highest initial temperatures are reached, the spectral distribution of the access supports a broadening of the \textrho\ driven by baryonic effects.  
These findings provide striking evidence that the dominant role in the emission of dileptons from the fireball, as expected, is played by radiative decays of in-medium \textrho\ mesons throughout the full energy range. 
A similar agreement with data is achieved also with PHSD, despite the completely different approach (right panel).
Striking is here the cut-off for thermal partonic radiation at low invariant masses related to the fact that partons are treated in PHSD as massive quasi-particles with temperature dependent spectral functions.
Generally, the contribution from thermal partonic radiation to the IMR is coming out to be stronger as compared to the Rapp/Wambach model.
Data is not precise enough to scrutinize details of the treatment of the thermal hadronic (in-medium \textrho ) radiation.  
Note also that in PHSD dielectrons from $\Delta$ Dalitz decays are explicitly treated. 
It appears that details that the in-medium \textrho\ meson are sufficiently well approximated by the empirical ansatz of collisional broadening. 
Similar conclusions can be also derived for more differential comparisons of the transverse momenta distributions. 
Future dilepton experiments are designed to provide data quality which will help to address this important issue in great detail and differentiate between the models (see the following section). 
%
\subsection{Multi-differential dilepton observables}
\label{sec:MultiDiff}
%
A second important observation made by NA60 is based on the evaluation of excess-pair transverse-momentum distributions for different regions in invariant mass.  
For a total of 10 invariant-mass bins, spanning from 0.2 to 1.2~\gevcc , acceptance corrected transverse-mass distributions were reconstructed and fitted with an exponential form according to 
\begin{equation}
 \frac{\mathrm{d}M}{m_\perp \mathrm{d}m_\perp} = e^{\frac{-m_\perp}{kT_{\mathrm{eff}}}}
 \, .    
\end{equation}
The respective inverse slope parameters $T_{\mathrm{eff}}$ are depicted  in Fig.\,\ref{fig:na60-inin-teff} and reveal a non-monotonic trend as function of invariant mass. 
Up to an invariant mass of around 1\gevcc\ they rise continuously and then drop by about 50\mev\ to stay constant within errors.  
A similar exercise has also been done for pairs in the IMR and their inverse slope parameters turn out to be  consistent with $T=200$\mev\ within errors.  
This is a remarkable observation as the drop in effective temperatures occurs exactly beyond the low-mass vector meson region. 
It is conceivable, as argued in \cite{Arnaldi:2007ru}, that the rise of the inverse slope parameter is due to the collective expansion of the emitting source forcing shallower slopes as the masses grow.
Having in mind that the flow is built up in the course of the fireball evolution, lower inverse slop parameters for a given invariant mass would signal generally earlier emission, \myie before flow reaches its maximum (\mycf Sec~\ref{sec:stages}). 
\begin{figure}[tb]
  \begin{center}    
\includegraphics[width=0.8\linewidth]{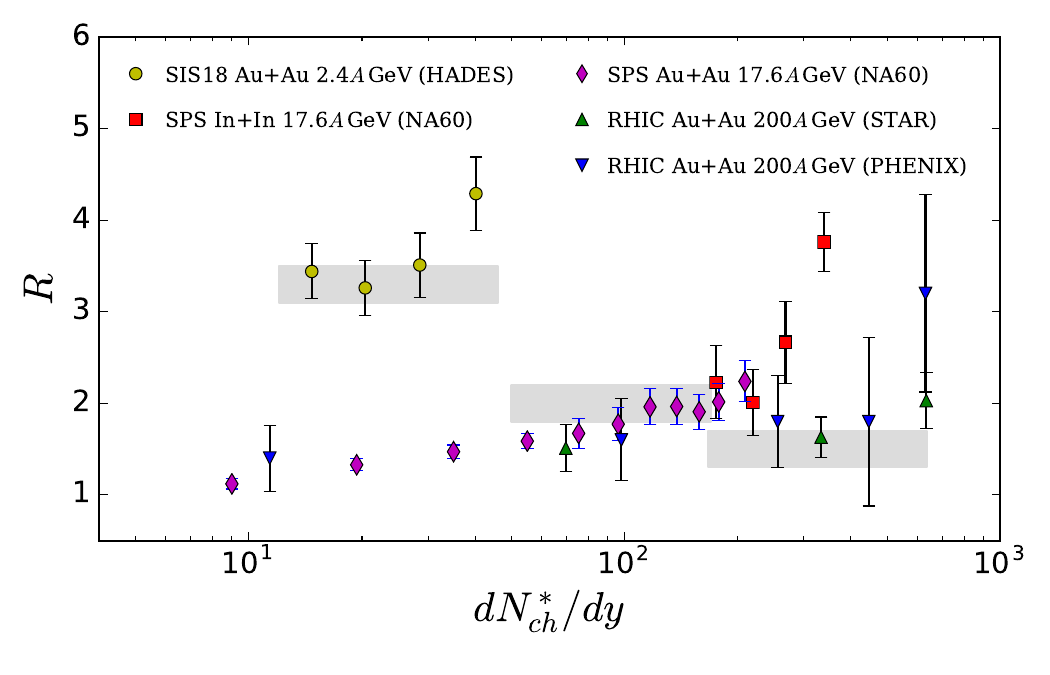} 
   \end{center}
\caption{
Systematic of excess radiation. 
Plotted is the ratio $R$ of the dilepton signal over the cocktail yield versus the number of produced charged particles in the reaction $\mathrm{d}N_{\mathrm{ch}}^*/\mathrm{d}y$. 
SIS18: $R$ is computed as ratio of the signal yield in the (large) acceptance of the HADES spectrometer around mid-rapidity divided by the respective cocktail yield.
Other: Signal yields over cocktail at mid-rapidity in one unit of rapidity are plotted.
For details see text.
}   
\label{fig:excess-syst}
\end{figure}
%
A similar observation and conclusion~\cite{vanHecke:1998yu} has been made inspecting slope parameters extracted from hadronic final states measured with NA44~\cite{Bearden:1996dd} and also dilepton mesonic sources in NA60 data (open symbols in Fig.\,\ref{fig:na60-inin-teff}).

The interpretation of reduced mean transverse momenta (at given invariant mass) in terms of an emission at earlier stages can be further scrutinized by turning back to the fully acceptance corrected invariant mass distribution shown in Fig.\,\ref{fig:na60-inin}.
The third observation is the perfectly smooth falloff in the mass region above the eminent \textrho\ bump.
Other than the $p_\perp$ distributions, the invariant mass distribution is not subject to blue (or red) shifts since the invariant mass is a Lorentz-scalar. 
The NA60 has fitted a relativistic Planck distribution of shape
$\mathrm{d}N/\mathrm{d}M \propto M^{3/2} \exp(-M/T_{\mathrm{th}})$
to the distribution to extract a space-time averaged emission temperature of $T_{\mathrm{th}} = 200\pm 12$\mev .
This temperature is significantly above the pseudo-critical temperature extracted from lattice QCD which it turn coincides with the limiting temperature derived from particle multiplicities assuming emission from a thermalized sources. 
Taking all together, the data provides substantial support for the interpretation that a deconfined phase has been created in HI collisions at the SPS.  

The penetrating nature of dilepton radiation is also reflected in the centrality dependence of the excess yield in the LMR. 
While the observed hadron multiplicities are understood to scale with the volume of the fireball, for given freeze-out temperature and chemical potentials, the dilepton yield is also determined by the lifetime of the radiating fireball. 
Moreover, the emission rate is strongly depending on the temperature of the emitting source which of course changes in the course of fireball evolution. 
A simple evolution of the excess radiation as a function of centrality is therefore not to be expected. 
Fig.\,\ref{fig:excess-syst} shows an attempt to establish the systematics of the ratio $R$ of signal over cocktail obtained in various experiments. 
The collision energies span two orders of magnitude while mostly heavy collision system are investigated, besides the medium size \inin\ system explored by NA60. 
Because of the wide range in collision energy, the definition of the ``excess'' is not strictly unique. 
The "low energy" case is represented by the HADES \auau\ data taken at \sqsnn{2.4}. 
Four classes of centrality are shown, $0-10$\%, $10-20$\%, $20-30$\% and $30-40$\% of the total geometrical cross section. The centrality classes are obtained from fits of Glauber MC simulations to the measured charged particle multiplicities in the HADES acceptance.
Details of the centrality selection can be found in~\cite{Adamczewski-Musch:2017sdk}.
For the systematics presented here, $R$ is taken to be the signal yield integrated in the mass range $200$\mevcc\ $\le M_{\mreen} \le 600$\mevcc\ 
divided by the respective integral over the sum of meson cocktail and the \nnnn\ reference spectrum in the same mass range.
The yields are derived in the acceptance which, in the case of HADES, covers a large fraction of the phase space around mid-rapidity.
The ratio is plotted as a function of the multiplicity of {\em produced} charged particles, in this case pions. 
We have chosen to do so since at this energy the final state is dominated by the stopped baryons and hence not representing a measure for the event activity.
Hence, the data points for HADES are placed at the respective charged pion multiplicity in full phase space. 
Pions are the only abundantly {\em produced} hadrons at the SIS18 energy domain. 

The ``medium energy'' is represented by the \inin\ data obtained by the NA60 collaboration.
Note that this is the only medium size collision system included in the systematics.
For this case $R$ was computed from the ratios published in~\cite{Arnaldi:2008fw} where the continuum yield, normalized to a so-called cocktail \textrho\ is presented for 12 centrality bins. 
To include these data to our systematics, the continuum yield was assumed to be constant\footnote{Indeed the continuum yield measured in the NA60 acceptance appears nearly constant due to the particular acceptance filter of the spectrometer.} 
and accounted for in the $200-600$ \gevcc\ invariant mass region (for details see \myeg \cite{Damjanovic:2008ta}).
The ratios originally presented refer to the excess yield (in acceptance) without \textrho\ peak in the $200<M_{\mu^+\mu^-}<1000$~\mevcc\ relative to the cocktail \textrho\ yield.
The latter is calculated from the measured $\omega$ yield assuming that the \textrho\ signal is as strong as the $\omega$ signal.  
The respective cocktail yield has been obtained from the centrality integrated spectrum accounting for $\eta$ and $\omega$-Dalitz decays in the  $200-600$ \gevcc\ mass range. 
The cocktail was assumed to scale linearly with the charged particle multiplicity as reported in~\cite{Arnaldi:2008fw}.
The data points (denoted by diamond symbols) are plotted as function of the charged particle multiplicity density. 
The data collection from SPS is complemented with the data obtained by the CERES collaboration for \pbau\ collisions also at \sqsnn{158}~\cite{Agakichiev:2005ai}.
The ratios $R$ have been derived as signal over cocktail yield integrated from 200 to 600 \mevcc\ and are shown (squares) for four centrality classes spanning the 40\% most central collisions according to the trigger conditions used in the experiment.

The third energy is represented by the STAR and PHENIX data (triangles) taken from \cite{Adamczyk:2015lme} and \cite{Adare:2015ila}, respectively.
Both data sets were obtained at maximum RHIC energy of \sqsnn{200}. 
Also here $R$ represent the integral signal  
yield normalized to the hadronic cocktail, however here integrated from 300 to 760 \mevcc .
They are as well plotted for the respective charged particle rapidity densities of the respective centrality classes.
The fact that the largest fraction of excess radiation is observed at the lowest collision energy can be traced back to two effects.
First, the fireball freezes out at moderate temperatures of around $kT \simeq 60$\mev . 
Consequently, there are only moderately many excited hadron states which can contribute to the cocktail yield. 
The second reason is the relatively long life time of the dense phase of the collisions (up to $15\;\mathrm{fm}/c$).
Note that the absolute size of the fireball does not matter as the hadronic cocktail scales linearly with the freeze-out volume as also the emissivity.

With upcoming high-statistics data on dilepton production it will become possible to extract flow patterns through higher harmonics in the azimuthal angular distribution relative to the event plane (\mycf Eq.\,\ref{eq:higher-harmonics}). 
The elliptic flow signal ($v_2$) in particular will provide a more quantitative analysis of the actual emission time of the radiation. 
Very interesting will be the comparison of the elliptic anisotropy of the radiation emitted in the LMR and IMR.
Certainly, one would expect that the the strength is diminished at higher mass as it is suggested by the result depicted in Fig.\,\ref{fig:na60-inin-teff}.
Elliptic flow has been observed in the direct photon signal \cite{Adare:2011zr, Abelev:2013cva}  and interpreted using a full expansion simulation of the fireball including various scenarios \cite{Vujanovic:2016anq}.
It was found that the flow signature also encodes bulk properties of the expanding medium like shear viscosity for real photons and consequently also for dileptons.
Such measurements are challenging as flow signatures are best observed when centrality selections are applied. 
Hence, such measurements not only represent triple differential analyses but also have to be corrected for possible non-flow contributions.   
%
%
\section{Outlook}
\label{sec:outlook}
Since the pioneering dilepton experiments in heavy-ion collisions, the existence  of thermal radiation of virtual photons has been clearly identified and successfully described on the basis of thermal averages of in-medium hadronic and partonic current-current correlators.
Meanwhile, dileptons are recognized as valuable tool to study the microscopic properties of QCD matter under extreme conditions and to search for landmarks in the strong-interaction matter phase diagram. 
Significant measurements of thermal dilepton radiation need high statistics data and excellent discrimination of background. 
Only new, state-of-the art detector systems can cope with this demand. 

To explore the complete region of the QCD phase diagram accessible in heavy-ion reactions therefore require to ``go back'' in beam energy and to operate respective detector systems at accelerators providing collision energies spanning from \sqsnn{2} to energies beyond \sqsnn{100}.
Several such projects are on their way.
ALICE will continue dilepton spectroscopy with emphasize on thermal radiation in run 3 with the upgraded ALICE detector  \cite{Abelevetal:2014cna}.
The spectrometer will operate at 50~kHz interaction rate and include two upgrades essential for this task: the GEM-based TPC read-out chambers enabling continuous operation and the new Inner Tracking System (ITS) based on thinned monolithic CMOS sensors. 
There are even plans to further replace inner layers of the recently developed ITS of 2nd generation in LS3 with possibly then available ultra-thin, flexible pixel detector. 
The boost in detection capability of low momentum particle is expected to provide even better signal to background  ratio as compared to the new ITS. High statistics measurements in LMR and IMR region are scheduled for Run3 with the main goal to access in-medium $\rho$ meson spectral function and  thermal radiation in the region above vector meson poles .  
The STAR detector will continue its Beam Energy Scan and will provide dielectron spectra with good statistics down to collision energies of around \sqsnn{20} and possibly even lower, depending on the luminosity finally achievable by utilizing improved beam cooling in the RHIC.
Also the SPS physics program is looking forward to the approval of a next generation NA60 experiment termed NA60++ \cite{Agnello:2018evr}.  It will stay with the concept of combining two independent spectrometers, one in front of and one behind the hadron absorber. 
However, new detectors are foreseen for installation throughout.

The exploration of the high-\mub\ region using dilepton observables is the realm of new accelerator facilities currently under construction. 
Indeed, the exploration of the QCD phase diagram in the region of high \mub\ has recently gained a lot of attention.
Several new experiments are proposed or being constructed to measure relevant observables with unprecedented precision.
In most of the cases dilepton observables are a prime goal of the experiments.
With these detectors becoming operational, high statistics data on dilepton radiation can then be taken down to SIS18 energies where HADES is currently operational. 
The CBM  experiment will be installed at the new FAIR accelerator complex and feature both techniques with same rapidity coverage, dielectron and dimuon spectroscopy.
It is designed to be operated at interaction rates beyond one MHz. 

For the first time
the high-$\mu_B$ region of the phase diagram will be explored with penetrating probes in the STAR BES-II run. 
Statistics, however, will still be limited for the lowest beam energies under consideration. 
The third facility, also currently under construction, is NICA at the Joint Institute of Nuclear Research. 
This facility will operate both, a fix-target and s collider experiment. 
For a more detailed discussion of the upcoming facilities dedicated to high-$\mu_B$ QCD matter we refer to~\cite{Galatyuk:2019lcf}.

Indeed, thermal dileptons are a promising tool for the search for yet undiscovered landmarks in the QCD phase diagram.
The prospect to establish excitation functions of thermal dilepton emission in the LMR and IMR has very much inspired the community in recent years.
A high-precision reconstruction of the excess radiation in the invariant mass range around 1\gevcc\ will allow to study the effects of spontaneous chiral symmetry breaking through combined measurements of in-medium $\rho$ meson spectral function and \textrho--$\text{a}_1$ mixing, possibly a more clear indication for the onset of its restoration.
The direct extraction of a mean temperature from the slope of the invariant mass distribution will give access to the ``caloric curve'' of QCD matter.
It is expected that a non-monotonicity in the excitation function might occur once the phase space evolution of the fireball transits through a first order deconfinement--confinement transition. 
Such a landmark could also leave non-monotonic structure in the excitation function of integrated excess yield. 
Roughly spoken, the excess radiation is emitted from the fireball in a period of less than 10\,ns. 
Any ``slowing down''of the fireball expansion due to the  conversion the latent heat would be visible if the ``delay'' would be significant, \myie reach several 10\,\% of the total radiation time.    

Till today, the observation of thermal IMR radiation by NA60 and the extraction of a source mean temperature $T = 200\pm12$~\mev\  is one of the most direct proofs that a partonic medium is created in ultra-relativistic heavy-ion collisions.  In the low energy domain in the HADES measurement temperature of the  of above 70\,\mev\ is extracted from radiation of hadronic medium with constituents whose properties have been strongly modified. Upcoming experiments in colliders and using stationary targets will provide more precise data.
In this review we have addressed the challenges and strategies to accomplish this ambitious goal based on a precise understanding of the emissivity of strong-interaction matter over full phase space diagram.  
To establish the ``standard candle'' of strong-interaction matter significant reference measurements with pion and proton beams, using in particular also exclusive channels, are very important. 
Particularly, successful description of LMR with in-medium $\rho$  meson spectral function based on hadronic models requires further constraints to scrutinize mason-baryon interactions which drive the medium effects. 

A second challenge is to provide clear evidence for a (partial) restoration of the spontaneously broken symmetry.
Reactions, where multi-pion contributions to the dilepton yield dominate over QGP radiation, and provided that contributions from open charm and Drell-Yan are under control with sufficient precision, seem to be promising for studies of V--A mixing and the vector mass modifications. 
Beam energies in the SIS100 range provide favorable conditions since hard processes are sufficiently suppressed. 
Such measurements are planned with the upgraded HADES detectors already during Phase-0 of FAIR and later with CBM. 
At SPS energies, a new NA60$^+$ detector will be sensitive to detect effect of the chiral mixing in IMR with the anticipated precision of vertex detectors \cite{Agnello:2018evr}.   
At the high-energy frontier, the upgrade of the ALICE detector is planned to be finished for Run~3 and will provide excellent vertex resolution, sufficient to separate the correlated charm from prompt thermal dileptons. 
Moreover, the rate capability will reach 50\,kHz recorded events per second due to the TPC upgrade.
Precision measurements in the LMR and IMR will then be in reach~\cite{Abelevetal:2014cna}. 
It is expected that about 60\,\% of the radiation in the LMR region will originate from a hadronic medium with temperatures slightly below $T_{\mathrm{pc}}$, the measurement will be sensitive to effects related to chiral symmetry restoration close to the hadron gas-QGP transition.    
The near future will show how close the community will come to this ambitious goal.
%
%
\vspace{10ex}
\newline
\noindent
{\bf Acknowledgement}\\[1ex]
We are very grateful for many stimulating and inspiring discussions with our colleagues Bengt Friman, Tetyana Galatyuk, Stefan Leupold and Ralf Rapp. 
We also like to thank two of them and Peter Senger for careful reading of the manuscript. 
Last but not least, we would like to acknowledge the very fruitful collaboration with our HADES colleagues.  
We acknowledge also support from grants of the National Science Foundation (Poland)  2017/25/N/ST2/00580 and from GSI, Darmstadt (Germany). 
%
%

\newpage
This text is added to avoid that the label ``lastpage'' is not written to the aux file
\end{document}